\documentclass[a4paper,11pt]{article}

\usepackage{lineno,hyperref}
\usepackage{tikz}
\usepackage{tikz-cd}
\usepackage{dsfont}
\usepackage{amsfonts}
\usepackage{amsmath}
\usepackage{slashed}
\usepackage{graphics}
\usepackage{amssymb}
\usepackage[compat=1.1.0]{tikz-feynman}
\usepackage{setspace}
\usepackage{float}
\usepackage{cleveref}
\usepackage{units}
\usepackage{amscd}
\usepackage[left=1in, right=1in, top=1in, bottom=1in]{geometry}
\usepackage{nicefrac}
\usepackage{subcaption}
\usepackage{changepage}
\usepackage{setspace}
\usepackage{comment}
\usepackage{amscd}
\usepackage{xcolor}
\usepackage{comment}

\usepackage[T1]{fontenc}
\usepackage[utf8]{inputenc}
\usepackage{booktabs}
\usepackage{tabularx}
\usepackage{array}
\usepackage{graphicx}
\usepackage{url}
\usepackage{upgreek}
\usepackage{microtype}
\usepackage{caption}
\ifdefined\extralength\else\newlength{\extralength}\fi
\newcolumntype{L}{>{\raggedright\arraybackslash}X}
\newcolumntype{C}{>{\centering\arraybackslash}X}
\providecommand{\hl}[1]{#1}
\providecommand{\textls}[2][0]{#2}

\providecommand{\PublishersNote}[1]{}

\providecommand{\funding}[1]{\par\bigskip\noindent\textbf{Funding:} #1\par}
\providecommand{\dataavailability}[1]{\par\medskip\noindent\textbf{Data Availability Statement:} #1\par}
\providecommand{\acknowledgments}[1]{\par\medskip\noindent\textbf{Acknowledgments:} #1\par}
\providecommand{\conflictsofinterest}[1]{\par\medskip\noindent\textbf{Conflicts of Interest:} #1\par}

\modulolinenumbers[5]

\providecommand{\keywords}[1]{%
  \vspace{0.5cm}
  \noindent\textbf{Keywords: } #1
  \vspace{0.5cm}
}

\title{\textbf{Gravitational-Wave Signatures of Compact Extra Dimensions and Their Detectability with Third-Generation Interferometers}}

\author{
  M.D.C.~Torri$^{12}$\thanks{Corresponding author: marco.torri@unimi.it, marco.torri@mi.infn.it}\;,\\
  \textit{$^1$Dipartimento di Fisica, Università degli Studi di Milano,} \\
  \textit{$^2$INFN Milano,} \\
  \textit{via Celoria 16, I-20133 Milano, Italy} \\
}

\begin{document}
\maketitle


\begin{abstract}We investigate gravitational-wave dynamics in theories of gravity formulated in $D=4+N$ dimensions, with matter localized on a four-dimensional brane or coupled to the effective four-dimensional metric obtained after compactification. We consider four representative geometries, namely toroidal and Calabi--Yau compactifications, and the warped Randall--Sundrum I and II models, which cover the principal discrete and continuous Kaluza--Klein spectra. We derive the modifications of binary dynamics, propagation, quasinormal modes and polarization content, and use them as a feasibility screening for third-generation interferometers, with particular attention to the Einstein Telescope. The screening shows that the direct effects of the heavy Kaluza--Klein towers are strongly suppressed and are expected to be inaccessible to third-generation detectors. Indeed, the effective-field-theory corrections scale as $(\hbar\omega/m_1c^2)^2$$\sim$$10^{-23}$--$10^{-22}$, while the explicit conservative inspiral-phase correction of RS2 remains below the ET measurability threshold over the region allowed by laboratory bounds. Nevertheless, the screening identifies additional phenomenological channels that may remain testable: a tidal-charge deformation of the black-hole background, which shifts the QNM frequencies and damping times; a continuum-inspired phase deformation; and a radion-induced scalar polarization. We treat these channels as phenomenological benchmarks and formulate a Fisher-matrix Monte Carlo forecast based on the Earth-fixed geometry of the candidate sites, comparing the triangular and double-L Einstein Telescope configurations within an idealized common-noise setting.\end{abstract} 

\keywords{compactification of extra dimensions; gravitational waves; binary merger}

%


\section{Introduction}
\label{introduction}
General relativity (GR) has successfully passed a wide range of experimental and observational tests, including Solar-System measurements and gravitational-wave (GW) detections by ground-based interferometers~\cite{Virgo,Middleton}. Nevertheless, theoretical and cosmological questions, such as the search for a quantum theory of gravity and the nature of dark energy, motivate tests of GR in regimes where deviations may emerge~\cite{COST,Whitepaper}. Compact-binary coalescences, observed through GW astronomy~\cite{Blair,Berti,Ando,Arimoto,Sagunski}, provide a powerful laboratory for probing gravity in the strong-field, highly dynamical regime and constraining extensions of~GR.

Extra spatial dimensions are among the best-motivated such extensions. A~generic $D=4+N$-dimensional theory produces, after~compactification, a~massless graviton zero mode together with Kaluza--Klein (KK) \hl{excitations}
~\cite{Han,Overduin,Appelquist}. The~KK spectrum depends on the geometry, topology, boundary conditions, and~stabilization mechanism of the extra-dimensional space. Beyond~standard compactifications, recent GW literature has also explored braneworld models with higher-curvature bulk terms, such as Gauss--Bonnet corrections, which modify black-hole (BH) horizons and perturbation potentials without introducing ghost instabilities, as~well as scenarios with dynamical bulk fields, fluxes, or~energy leakage into massive bulk gravitons. Multimessenger observations, notably GW170817 and GRB170817A, further constrain propagation speeds and time lags in warped geometries~\cite{Datta}.

We consider four representative geometries, chosen to cover the main spectral structures relevant for GW phenomenology. Flat toroidal compactifications, $M_4\times T^N$, provide the simplest discrete KK spectrum, with~masses fixed by the compactification radii~\cite{Appelquist}. Ricci-flat compactifications, $M_4\times X_N$, including the string-motivated Calabi--Yau subclass, yield spectra determined by the eigenvalues of the appropriate Laplacian or Lichnerowicz operator~\cite{Greene}. Finally, the~Randall--Sundrum I (RS1) model, with~a finite warped interval bounded by two branes, supports a discrete KK tower~\cite{Randall1}, whereas Randall--Sundrum II (RS2), with~a noncompact warped extra dimension, contains a normalizable massless graviton and a continuum of massive KK modes~\cite{Randall2}. These cases allow for a systematic comparison of discrete and continuous spectra and of flat, curved, and~warped~geometries.

The observable consequences depend on the ratio between the KK mass scale $m_\alpha$ and the GW frequency $\omega$. Modes with $m_\alpha<\omega$ can propagate as asymptotic waves and may produce dispersive propagation, additional radiation channels, energy leakage into the bulk, or~independent waveform components. For~the phenomenologically realistic case $m_\alpha\gg\omega_{\rm GW}$, considered throughout this work, massive KK modes are not propagating degrees of freedom in the detector band and can be integrated out. Their effects may nevertheless survive through virtual corrections to the source dynamics, modifications of the brane-to-brane response, continuum-induced nonlocality, zero-mode--KK mixing, changes in the BH background, or~light radions and other moduli. These mechanisms can affect binary dynamics, GW propagation, quasinormal modes (QNMs), damping times, and~polarization~content.

We adopt a two-stage framework: a feasibility screening followed by a phenomenological sensitivity study, rather than a measurement of microscopic compactification parameters. First, we derive the GW signatures analytically and assess their accessibility to third-generation detectors~\cite{LIGO,Virgo2,KAGRA}, with~particular attention to the Einstein Telescope \mbox{(ET) \cite{Branchesi,Maggiore,Punturo}.} The~screening follows the full chain from microscopic parameters to the waveform. Heavy KK tensor towers affect the waveform only through effective-field-theory corrections suppressed by $(\hbar\omega/m_1c^2)^2$$\sim$$10^{-23}$--$10^{-22}$, with~coefficients determined by the toroidal lattice sums and by brane-induced zero-mode--KK mixing in RS1. In~RS2, the~conservative inspiral-phase correction, derived from the brane-to-brane Green function through the weak-field potential, remains below the ET measurability threshold throughout the region allowed by laboratory bounds, as~confirmed by an absolute detector-independent Fisher bound. Direct KK propagation in the detector band is also excluded by laboratory bounds on the compactification size; dispersive propagation would require an effectively ultralight graviton mass. Each derivation is highlighted in the text, and~the status of all channels is summarized in \cref{tab:channel_map}.

The channels that survive this screening are not suppressed by the KK gap: a tidal-charge deformation of the BH background, parametrized by an effective quantity $\Upsilon$ that shifts the QNM frequencies and damping times; a continuum-inspired phase deformation; and a radion-induced scalar polarization. Since their normalizations are not fixed by a specific compactification, they are treated as phenomenological benchmarks. The~Fisher uncertainties therefore quantify the detectability of the assumed waveform deformations, not constraints on compactification~parameters.

Second, we perform a Fisher-matrix Monte Carlo forecast for these channels, using a parametrized post-Einsteinian description of the phase deformations. The~forecast includes a continuum-inspired phase deformation, a~tidal-charge phase amplitude, QNM frequency and damping-time shifts, and~a scalar-polarization amplitude. Its purpose is to compare the response of the two ET configurations currently under consideration to deformations of a given phase, QNM, or~polarization form, independently of their microscopic amplitude. The~configurations are a triangular layout of three co-located underground interferometers, each with 10 km arms and internal angles of $60^\circ$, and~a double-L ($2L$) layout consisting of two L-shaped detectors at sites separated by approximately 1000 km~\cite{Branchesi,Maggiore,Punturo}.

Previous studies of third-generation networks, including the triangular and $2L$ ET designs, have mainly evaluated their performance within GR and general astrophysical population assumptions~\cite{Virgo2,LIGO,KAGRA,Branchesi,Maggiore,Punturo}. Ref.~\cite{Branchesi} compares the two designs systematically for standard science targets. Our analysis is complementary, as~it assesses their relative performance for compactification-inspired waveform deformations. The~detector geometry is taken from the real Earth-fixed positions of the candidate sites, and~Earth rotation is included for long inspiral signals. Results are reported both at the natural signal-to-noise ratio (SNR) of each design and at fixed network SNR, separating network sensitivity from geometric effects. Since the comparison uses an idealized common-noise setting, it characterizes the adopted waveform parametrizations rather than providing an engineering ranking of the final ET designs. The~contribution of this work is therefore twofold: a quantified assessment of the channels that cannot be probed by ET, and~an evaluation of how detector geometry affects the channels that~can.

This paper is organized as follows. In~\cref{sec2}, we introduce the theoretical framework for the compactification scenarios considered in this work, namely flat toroidal compactification, Ricci-flat compactification, RS1, and~RS2. In~\cref{sec3,sec4}, we derive the linearized dynamics and the brane-to-brane response, including the exactly solvable RS1 example with brane-induced mixing and the RS2 Green function. In~\cref{sec7}, we organize the observable signal by the part of the waveform that is modified, and~we state which channels are derived from the microscopic parameters. The~channels are then treated by macro-sector, with~source generation in \cref{sec5}, which includes the heavy-KK effective theory, the~brane-induced mixing, and~the RS2 conservative inspiral phase, propagation in \cref{sec:prop}, the~ringdown phase of a BH remnant in \cref{sec5ring}, and~the scalar sector, polarizations, and~moduli in \cref{sec:scalar}. The~scope of the analytic framework is stated in \cref{sub:pheno-scope}. In~\cref{sec:numerics}, we evaluate these results numerically for each model and establish their accessibility to ET. In~\cref{sec8}, we introduce the Fisher-matrix forecasting formalism for the surviving channels and compare the proposed detector configurations. Our conclusions, together with the limitations of the analysis and the planned extensions, are presented in \cref{sec:conclusions}.

\section{Theory~Formulation}
\label{sec2}

We consider a $D=4+N$ spacetime with warped extra dimensions, described by the bulk \hl{action} 
\begin{equation}
\label{eq:bulk_action}
S_D=\frac{M_D^{D-2}}{2}\int d^4x\,d^Ny\,\sqrt{-g_D}\,
\left(R_D-2\Lambda_D\right)+S_{\rm brane}+S_{\rm matter}\,.
\end{equation}

\hl{For} 
 models with localized gravitational dynamics, one may also include the brane-induced Einstein term
\begin{equation}
\label{eq:induced_gravity}
S_{\rm ind}=\frac{M_b^2}{2}\int_{\rm brane}d^4x\,\sqrt{-\hat g}\,R_4[\hat g]\,,
\end{equation}
which is distinct from the four-dimensional Einstein term generated by dimensional reduction of the bulk~action.

The general warped background is taken as
\begin{equation}
\label{eq:general_warped_metric}
d\bar s_D^2=e^{-2A(y)}\bar g_{\mu\nu}(x)\,dx^\mu dx^\nu
+\bar G_{ab}(y)\,dy^ady^b\,.
\end{equation}

If matter is localized on a brane at $y=y_b$, it does not propagate into the bulk and experiences the pulled-back metric
\begin{equation}
\hat g_{\mu\nu}(x)=g_{\mu\nu}(x,y_b)\,,
\end{equation}
which inherits the warp factor $e^{-2A(y_b)}$.

The relation between the fundamental and effective Planck scales depends on the geometry. For~an unwarped compactification,
\begin{equation}
\label{eq:Planck_unwarped}
M_{\rm Pl}^2 \sim M_D^{D-2}V_N\,,
\end{equation}
where $V_N$ is the internal volume. In~warped models, this relation is modified by an integral over the warp~factor.

The four scenarios studied below correspond to specific choices of $A(y)$ and $\bar G_{ab}(y)$ in \mbox{\cref{eq:general_warped_metric}}. The~toroidal and Ricci-flat compactifications are unwarped, with~$A(y)={\rm const}$, so \mbox{\cref{eq:Planck_unwarped}} applies and the induced-gravity term plays no role. The~Randall--Sundrum models instead have a nontrivial warp factor sourced by $\Lambda_D$ and the brane tensions in $S_{\rm brane}$; the Planck mass receives the corresponding warped correction, and~\mbox{\cref{eq:induced_gravity}} becomes relevant when brane-localized gravitational dynamics are retained. In~all cases, the~KK spectrum follows from the linearized fluctuations about the background \mbox{\cref{eq:general_warped_metric}}. Its discrete or continuous, gapped or gapless structure determines the propagating-versus-heavy distinction introduced in \cref{introduction}.

\subsection{Flat Toroidal~Compactification}
\label{sub:toroidal}
The simplest case is the flat compactification $M_4\times T^N$. For~a rectangular torus with radii $R_a$, the~KK masses are
\begin{equation}
\label{eq:toroidal_masses}
m_{\vec n}^2=\sum_{a=1}^N\frac{n_a^2}{R_a^2}\,,
\qquad n_a\in\mathbb{Z}\,.
\end{equation}

The zero mode corresponds to $\vec n=0$. This model provides the simplest explicit discrete tower and serves as a reference for the heavy-mode effective theory. In~the phenomenologically relevant regime $R_a^{-1}\gg\omega_{\rm GW}$, the~tower can be integrated out, leaving the zero mode as the only propagating tensor degree of~freedom.

\subsection{Ricci-Flat~Compactification}
\label{sub:ricciflat}
A more general unwarped compactification is $M_4\times X_N$, where $X_N$ is compact and Ricci-flat, $R_{ab}[X_N]=0$. Calabi--Yau manifolds form a particularly well-motivated string-inspired subclass; if their complex dimension is $n_{\rm CY}$, then $N=2n_{\rm CY}$.

For the four-dimensional tensor sector $h_{\mu\nu}$, the~internal dependence is governed by the scalar Laplacian,
\begin{equation}
\label{eq:CY_spectrum}
-\Delta_{X_N}Y_\alpha=\lambda_\alpha Y_\alpha\,,
\qquad m_\alpha^2=\lambda_\alpha\,,
\end{equation}
in units $c=\hbar=1$ and with the internal-metric normalization adopted in \cref{eq:general_warped_metric}. Under~an overall rescaling of $X_N$, $\lambda_\alpha$ scales as an inverse length squared, i.e.,\ as a mass squared. Unlike the toroidal case, the~eigenvalues are generally not known analytically. Calabi--Yau compactifications also generically contain complex-structure and K\"ahler moduli, whose light representatives provide additional scalar degrees of freedom of the type anticipated in \cref{introduction}.

\subsection{Randall--Sundrum~\hl{I} 
}
\label{sub:rs1}
The RS1 model contains a warped finite interval $0\leq y\leq L$ bounded by two branes. In~the minimal five-dimensional realization, the~background metric may be written as
\begin{equation}
ds^2=e^{-2k|y|}\eta_{\mu\nu}dx^\mu dx^\nu+dy^2\,,
\end{equation}
in a suitable coordinate~gauge.

The tensor spectrum is discrete,
\begin{equation}
\label{eq:RS1_masses}
m_0=0\,,
\qquad
m_n \simeq x_n k e^{-kL}\,,
\qquad n=1,2,\ldots\,,
\end{equation}
where the $x_n$ are fixed by the boundary conditions. For~standard tensor boundary conditions, they are related to zeros of a Bessel function, with~$x_1\simeq3.83$.

RS1 may also contain a radion, namely the scalar mode associated with fluctuations in the interbrane separation. Its stabilization determines its mass and couplings. The~radion can source the scalar polarization included in the Fisher forecast of \cref{sec8}. A~caveat concerns BH binaries: in scalar--tensor gravity, a~stationary, asymptotically flat BH carries no static scalar charge at leading order. Thus, for~BH--BH systems, the~leading static radion coupling is suppressed relative to systems containing neutron stars or other compact objects with scalar charge. This statement concerns the leading static multipole and does not exclude subleading dynamical or tidal contributions during the inspiral. We therefore regard the scalar-polarization channel as primarily relevant for binaries containing at least one non-BH compact~object.

\subsection{Randall--Sundrum~II}
\label{sub:rs2}
The RS2 model contains a single brane and a noncompact warped extra dimension, $y\geq0$. Its tensor spectrum consists of a normalizable zero mode and a continuum,
\begin{equation}
\label{eq:RS2_spectrum}
m_0=0\,,
\qquad
m\in(0,\infty)\,.
\end{equation}

RS2 therefore differs qualitatively from the toroidal and RS1 cases: it has neither an isolated first massive KK mode nor a discrete mass gap. The~continuum response is the origin of the continuum-induced nonlocality anticipated in \cref{introduction}.

\section{Linearized Dynamics and KK~Decomposition}
\label{sec3}

GW observations probe the weak-field regime, in~which the metric deviates only slightly from a fixed background solution of the bulk theory introduced in \cref{sec2}. Extracting the four-dimensional detector-frame dynamics therefore requires two steps: linearization of the higher-dimensional field equations about the background \cref{eq:general_warped_metric}, followed by a KK decomposition that separates the internal dependence from the four-dimensional propagation. The~resulting internal eigenvalue problem is discrete for the toroidal, Ricci-flat, and~RS1 cases and continuous for RS2. Its spectrum fixes the masses $m_\alpha$ entering the four-dimensional wave equation and hence the propagating-versus-heavy distinction discussed in \cref{introduction}. The~formalism of this section is common to all four geometries; the model dependence is encoded in the internal operator $\mathcal L_y$, its spectrum, the~KK wavefunctions, boundary conditions, and~brane~couplings.

We perturb the metric at linear order,
\begin{equation}
g_{MN}=\bar g_{MN}+h_{MN}\,,
\end{equation}
where $\bar g_{MN}$ is the background metric of \cref{eq:general_warped_metric}. Capital Latin indices $M,N=0,\ldots,3+N$ run over the full $D=4+N$ spacetime.

The linearized Einstein equations are, in~a suitable gauge,
\begin{equation}
\label{eq:linearized_general}
\delta G_{MN}[h]=0
\end{equation}
in vacuum, with~the source terms added in \cref{eq:4D_KK_equation} below. For~a Ricci-flat background and a transverse-traceless perturbation, when such a gauge is available, \cref{eq:linearized_general} reduces to
\begin{equation}
\label{eq:Lichnerowicz}
\bar\Box_Dh_{MN}+2\bar R_{MANB}h^{AB}=0\,,
\end{equation}
where $\bar\Box_D$ is the background d'Alembertian and $\bar R_{MANB}$ the background Riemann tensor. The~Riemann term cannot be discarded merely because the background is Ricci-flat:
\begin{equation}
\bar R_{MN}=0
\quad\not\Rightarrow\quad
\bar R_{MANB}=0\,,
\end{equation}
as illustrated by the Schwarzschild and Kerr~spacetimes.

The perturbation decomposes under the $4+N$ split into a four-dimensional tensor sector $h_{\mu\nu}$, a~vector sector $h_{\mu a}$, and~an internal scalar/moduli sector $h_{ab}$, with~Greek indices $\mu,\nu=0,\ldots,3$ and internal indices $a,b=1,\ldots,N$. In~this work, we focus on the tensor sector directly probed by interferometric GW observations and adopt the \mbox{phenomenological truncation}
\begin{equation}
h_{\mu a}=0\,,
\qquad
h_{ab}=0\,.
\end{equation}

The vector and scalar/moduli sectors are governed by the corresponding internal differential operators and are treated in \cref{sec:scalar}. This is a phenomenological truncation rather than a general proof of dynamical decoupling: cross-couplings may arise from the linearized equations and from the internal~geometry.

For the tensor sector, the~KK decomposition reads
\begin{equation}
\label{eq:KK_decomposition}
h_{\mu\nu}(x,y)=\sum_\alpha u_\alpha(y)H_{\mu\nu}^{(\alpha)}(x)
\end{equation}
for a discrete spectrum, where $u_\alpha(y)$ are internal mode functions and $H_{\mu\nu}^{(\alpha)}(x)$ are four-dimensional tensor fields. When a continuum is present, as~in RS2, the~sum is \mbox{replaced by}
\begin{equation}
\label{eq:continuum_decomposition}
h_{\mu\nu}(x,y)=u_0(y)H_{\mu\nu}^{(0)}(x)
+\int_0^\infty d\mu(m)\,u_m(y)H_{\mu\nu}^{(m)}(x)\,,
\end{equation}
where $d\mu(m)$ is the spectral measure and $u_0$, $H_{\mu\nu}^{(0)}$ denote the normalizable zero~mode.

The internal modes satisfy
\begin{equation}
\label{eq:internal_eigenproblem}
\mathcal L_yu_\alpha=m_\alpha^2u_\alpha\,,
\end{equation}
where $\mathcal L_y$ is the appropriate Laplace, Sturm--Liouville, or~Lichnerowicz operator for the geometry considered. With~the sign convention
\begin{equation}
\Delta_Lh_{MN}
=
-\bar\Box_Dh_{MN}
-2\bar R_{MANB}h^{AB}\,,
\end{equation}
the vacuum equation \cref{eq:Lichnerowicz} becomes $\Delta_Lh_{MN}=0$. After~projection onto a KK mode, the~four-dimensional field satisfies
\begin{equation}
\label{eq:4D_KK_equation}
\left[\Delta_L^{(4)}-m_\alpha^2\right]H_{\mu\nu}^{(\alpha)}=
J_{\mu\nu}^{(\alpha)}\,,
\end{equation}
where $\Delta_L^{(4)}$ is the four-dimensional Lichnerowicz operator on $\bar g_{\mu\nu}(x)$ and $J_{\mu\nu}^{(\alpha)}$ is the effective source obtained by projecting the higher-dimensional stress tensor onto $u_\alpha(y)$. The~zero mode has $m_0=0$, while massive modes have $m_\alpha>0$.

\section{Brane-to-Brane Dynamics and Microscopic~Matchings}
\label{sec4}

The KK decomposition of \cref{sec3} determines how a tensor perturbation propagates in the bulk once it is sourced. In~warped scenarios with matter confined to a brane, however, the~observable quantity is the response measured on the brane. A~signal generated by a source on one brane, or~on the same brane in RS2, must therefore be reconstructed from the full KK tower evaluated at the brane location. This defines the brane-to-brane~propagation.

The coupling of brane-localized matter to gravity is described by
\begin{equation}
\label{eq:brane_interaction}
S_{\rm int}=\frac{1}{2}\int d^4x\sqrt{-\hat g}\,
T^{\mu\nu}\delta\hat g_{\mu\nu}\,,
\end{equation}
where $T^{\mu\nu}$ is the stress tensor of the brane-localized matter, such as a compact binary, and~$\hat g_{\mu\nu}$ is the induced metric at $y=y_b$ defined in \cref{sec2}.

Evaluating \cref{eq:KK_decomposition,eq:continuum_decomposition} at $y=y_b$ shows that each mode couples to the brane with a strength set by its wavefunction value at the brane,
\begin{equation}
\label{eq:KK_coupling}
g_\alpha\propto u_\alpha(y_b)\,.
\end{equation}

The induced-metric perturbation is therefore
\begin{equation}
\label{eq:delta_metric_discrete}
\delta\hat g_{\mu\nu}=g_0H_{\mu\nu}^{(0)}
+\sum_{\alpha>0}g_\alpha H_{\mu\nu}^{(\alpha)}
\end{equation}
in the discrete case, and~\begin{equation}
\label{eq:delta_metric_continuum}
\delta\hat g_{\mu\nu}=g_0H_{\mu\nu}^{(0)}
+\int_0^\infty d\mu(m)\,g_mH_{\mu\nu}^{(m)}
\end{equation}
in the continuum case relevant for RS2, with~$d\mu(m)=\rho(m)\,dm$.

Projecting \cref{eq:brane_interaction} onto a KK mode gives the source in \cref{eq:4D_KK_equation},
\begin{equation}
\label{eq:J_alpha_coupling}
J_{\mu\nu}^{(\alpha)}=g_\alpha J_{\mu\nu}\,,
\end{equation}
where $J_{\mu\nu}$ is the brane-localized four-dimensional source obtained from $T_{\mu\nu}$. This is a source coupling, not a dynamical mixing: all modes are sourced by the same brane matter, but~no off-diagonal kinetic term is generated. If~$\mathcal M_{0\alpha}$ denotes the off-diagonal zero-mode--massive-mode kinetic mixing, introduced explicitly below, then
\begin{equation}
\label{eq:no_dynamical_mixing}
J_{\mu\nu}^{(\alpha)}\neq0\quad\not\Rightarrow\quad
\mathcal M_{0\alpha}\neq0\,.
\end{equation}

Let $G_\alpha(p)$ denote the four-dimensional propagator of the canonically normalized field $H_{\mu\nu}^{(\alpha)}$, obtained from \cref{eq:4D_KK_equation}. The~brane-to-brane propagator, namely the two-point function of $\delta\hat g_{\mu\nu}$ at the brane, is obtained by summing over the tower:
\begin{equation}
\label{eq:RS1_propagator}
G_{\rm br}^{\rm RS1}(p)=g_0^2G_0(p)+\sum_{n\geq1}g_n^2G_n(p)
\end{equation}
for RS1, and~\begin{equation}
\label{eq:RS2_propagator}
G_{\rm br}^{\rm RS2}(p)=g_0^2G_0(p)+\int_0^\infty dm\,\rho(m)g_m^2G_m(p)
\end{equation}
for RS2. These expressions describe independent mode exchange: each KK mode contributes additively to the brane response, with~no interference between~modes.

A genuine zero-mode self-energy, i.e.,\ a modification in the zero-mode propagation itself, requires off-diagonal mixing between the zero mode and the massive tower. At~quadratic order, this is represented by
\begin{equation}
\label{eq:mixing_term}
S_{\rm mix}^{(2)}=\sum_\alpha H^{(0)}\mathcal M_{0\alpha}H^{(\alpha)}\,.
\end{equation}

Eliminating the massive sector through its equations of motion gives an effective equation for the zero mode,
\begin{equation}
\label{eq:effective_operator}
\left[\mathcal D_0^P-\Sigma_P\right]\Psi_0^P=S_{\rm eff}^P\,,
\end{equation}
where $P$ labels the propagating polarization channel, tensor or, in~RS1, the~additional scalar channel associated with a radion. Here, $\mathcal D_0^P$ is the free zero-mode kinetic operator, $\Psi_0^P$ the corresponding effective four-dimensional field, $S_{\rm eff}^P$ the effective source obtained from \cref{eq:J_alpha_coupling}, and~\begin{equation}
\label{eq:Sigma_RS1}
\Sigma_P^{\rm RS1}(p^2)=\sum_{n\geq1}\mathcal M_{0n}^P(p^2)
G_n^P(p^2)\mathcal M_{n0}^P(p^2)
\end{equation}
for RS1, while
\begin{equation}
\label{eq:Sigma_RS2}
\Sigma_P^{\rm RS2}(p^2)=\int_0^\infty dm\,\rho(m)\,\mathcal M_{0m}^P(p^2)
G_m^P(p^2)\mathcal M_{m0}^P(p^2)
\end{equation}
for RS2. In~a minimal separable background, with~no zero-mode--massive-mode mixing,
\begin{equation}
\label{eq:no_mixing}
\mathcal M_{0\alpha}=0\,,\qquad\Sigma_P=0\,,
\end{equation}
the zero mode propagates as a free four-dimensional field, and~the extra-dimensional effect is confined to the source coupling $g_0$ of \cref{eq:KK_coupling}.

\subsection{RS1 Wavefunction Profiles and Brane~Couplings}
\label{sub:rs1_derivation}
For the standard RS1 tensor spectrum, the~massive KK masses are approximately
\begin{equation}
m_n=x_nke^{-kL}\,,
\end{equation}
where $x_n$ are the positive zeros of the Bessel function $J_1(x)$. After~canonical normalization, the~massive modes couple to matter on the infrared brane with approximately universal strength $g_n=1/\Lambda_\pi$, where $\Lambda_\pi=\overline M_{\rm Pl}e^{-kL}$. The~low-energy sum over virtual massive modes is then
\begin{equation}
\label{eq:RS1_sum}
\sum_{n\geq1}\frac{g_n^2}{m_n^2} \simeq\frac{e^{2kL}}{8\Lambda_\pi^2k^2}\,.
\end{equation}

This result uses $\sum_{n\geq1}x_n^{-2}=1/8$ and assumes that the effective RS1 description can be extended over the modes included in the sum. In~a strict EFT treatment, the~tower must be truncated at the warped cutoff, so the coefficient should be regarded as cutoff-dependent. \Cref{eq:RS1_sum} is therefore used only as a parametric estimate of virtual massive-mode exchange, assuming the standard RS1 boundary conditions and the absence of additional brane-localized kinetic~terms.

\subsection{An Exactly Solvable Example with Brane-Induced KK Kinetic~Mixing}
\label{subsec:exactRS1}
To realize explicitly an off-diagonal zero-mode--KK coupling, consider RS1 supplemented by a four-dimensional Einstein--Hilbert term localized on the infrared brane. Expanding transverse-traceless tensor perturbations in the eigenfunctions of the unmodified RS1 problem,
\begin{equation}
h_{\mu\nu}(x,y)=\sum_{n=0}^{\infty}u_n(y)H_{\mu\nu}^{(n)}(x)\,,
\end{equation}
the localized curvature term contributes the quadratic interaction
\begin{equation}
\Delta S^{(2)}=\frac{1}{8}\sum_{m,n}\Delta K_{mn}
\int d^4x\,H_{\mu\nu}^{(m)}\Box H^{(n)\mu\nu}\,,
\end{equation}
with
\begin{equation}
\label{eq:DeltaK}
\Delta K_{mn}=M_{\rm IR}^2e^{-2kL}u_m(L)u_n(L)\,.
\end{equation}

For a nonzero localized curvature coefficient, the~off-diagonal element $\Delta K_{0n}$ is nonzero and generates the momentum-dependent mixing kernel
\begin{equation}
\mathcal M_{0n}(p^2)=p^2\Delta K_{0n}\,.
\end{equation}

Separating the zero mode from the massive sector, the~quadratic inverse propagator takes the block form
\begin{equation}
\mathcal K(p^2)=
\begin{pmatrix}
p^2(1+\Delta K_{00}) & p^2\boldsymbol{\Delta K}_{0H} \\
p^2\boldsymbol{\Delta K}_{H0} & \mathcal K_{HH}(p^2)
\end{pmatrix}\,,
\end{equation}
where $[\mathcal K_{HH}]_{mn}=(p^2-m_n^2)\delta_{mn}+p^2\Delta K_{mn}$ for $m,n\geq1$. Integrating out the massive modes by the Schur complement gives
\begin{equation}
\label{eq:Keff_00}
\mathcal K_{{\rm eff},00}(p^2)=p^2(1+\Delta K_{00})-\Sigma_0(p^2)\,,
\qquad\Sigma_0(p^2) \simeq p^4\sum_{n\geq1}\frac{(\Delta K_{0n})^2}{p^2-m_n^2}\,.
\end{equation}

For momenta below the lightest massive threshold, $|p^2|\ll m_1^2$, the~zero-mode pole remains at $p^2=0$. This protects the massless four-dimensional tensor mode from a mass shift and from ghost instabilities. The~observable quantity is therefore the full modified spectral response and the brane-to-brane Green function $G_{\rm br}(p)$, not the basis-dependent mixing coefficient alone. The~low-momentum expansion of \cref{eq:Keff_00} and the resulting waveform correction are derived in \cref{sub:mixing_chain}.

\subsection{RS2 Brane-to-Brane Green~Function}
\label{sub:rs2_propagator}
We consider the minimal RS2 model~\cite{Randall2}, with~a single positive-tension brane in a five-dimensional anti-de Sitter bulk. In~conformal coordinates,
\begin{equation}
ds^2=\frac{\ell^2}{z^2}\left(\eta_{\mu\nu}dx^\mu dx^\nu+dz^2\right)\,,
\qquad z\geq\ell\,,
\end{equation}
\textls[-25]{where $\ell$ is the AdS curvature radius. We use the bulk normalization $S_{\rm EH}=M_5^3\int d^5x\sqrt{-g}\,R_5/2$} and the standard $Z_2$ orbifold~convention.

For transverse-traceless tensor perturbations,
\begin{equation}
h_{\mu\nu}(x,z)=\epsilon_{\mu\nu}e^{ip\cdot x}u_p(z)\,,
\end{equation}
the Euclidean bulk equation is
\begin{equation}
\left[\partial_z^2-\frac{3}{z}\partial_z-p^2\right]u_p(z)=0\,.
\end{equation}

The solution regular at the AdS horizon is $u_p(z)\propto z^2K_2(pz)$, where $K_\nu$ is the modified Bessel function of the second kind. Using $K_2'(x)=-K_1(x)-2K_2(x)/x$, one finds:
\begin{equation}
\partial_zu_p(z)\propto2zK_2(pz)+z^2pK_2'(pz)=-pz^2K_1(pz)\,.
\end{equation}

Thus:
\begin{equation}
\frac{u_p(\ell)}{\partial_zu_p(\ell)}=-\frac{K_2(p\ell)}{pK_1(p\ell)}\,.
\end{equation}

Imposing the linearized junction condition at the brane gives the scalar coefficient of the transverse-traceless brane-to-brane propagator,
\begin{equation}
G_{\rm TT}^{E}(p)=\frac{1}{2M_5^3p}\frac{K_2(p\ell)}{K_1(p\ell)}\,.
\label{eq:rs2_exact_green}
\end{equation}

The full tensor response to a conserved brane source is obtained by contracting this coefficient with the appropriate tensor projectors; possible brane-bending contributions are not included in this TT~coefficient.

Using $K_2(x)=K_0(x)+2K_1(x)/x$, \cref{eq:rs2_exact_green} becomes
\begin{equation}
G_{\rm TT}^{E}(p)=\frac{1}{M_5^3\ell p^2}+\frac{1}{2M_5^3p}
\frac{K_0(p\ell)}{K_1(p\ell)}\,.
\label{eq:rs2_pole_continuum}
\end{equation}

The first term is the normalizable massless graviton, while the second encodes the continuum of massive KK modes. Equation \cref{eq:rs2_pole_continuum} is therefore the explicit realization of \cref{eq:RS2_propagator} in the no-mixing case \cref{eq:no_mixing}.

For $p\ell\ll1$, using $K_0(x)\simeq-[\ln(x/2)+\gamma_E]$ and $K_1(x)\simeq1/x$, one obtains
\begin{equation}
G_{\rm TT}^{E}(p)=\frac{1}{M_{\rm Pl}^2p^2}\left[1-\frac{(p\ell)^2}{2}
\left(\ln\frac{p\ell}{2}+\gamma_E\right)
+\mathcal O\!\left((p\ell)^4\ln^2(p\ell)\right)\right]\,,
\label{eq:rs2_low_energy}
\end{equation}
where $M_{\rm Pl}^2=M_5^3\ell$ in the minimal RS2 construction. The~logarithmic term is a nonanalytic contribution generated by the KK continuum. It does not shift the pole of the localized massless graviton; therefore, the~Euclidean momentum-space correction cannot be identified directly with a frequency-dependent propagation phase for an on-shell GW. Its observable consequences are the static response and the associated conservative inspiral correction derived in \cref{sub:rs2_conservative_phase}.

\section{From the KK Dynamics to the Detector~Waveform}
\label{sec7}

Having established the KK spectra of the four compactifications (\cref{sec2}), the~linearized dynamics (\cref{sec3}), and~the brane-to-brane response (\cref{sec4}), we now assemble these ingredients into the observable gravitational-wave signal. The~following sections organize the physical mechanisms according to the part of the signal that they modify: source generation (\cref{sec5}), propagation (\cref{sec:prop}), black-hole response and ringdown (\mbox{\cref{sec5ring}}), and~additional polarizations and moduli (\cref{sec:scalar}). \Cref{sub:pheno-scope} states which effects and models are not covered. All numerical evaluations, detector thresholds, experimental bounds, and~accessibility statements are collected in \cref{sec:numerics}.

\subsection{General Waveform~Decomposition}
\label{sub:pheno-decomp}

The detector output may be written schematically as
\begin{equation}
\label{eq:detector_waveform}
\tilde h_I(f)=\mathcal R_I(f)\,\mathcal P(f)\,\tilde h_{\rm em}(f)\,.
\end{equation}

Here, $\tilde h_{\rm em}(f)$ is the emitted source-frame tensor waveform, $\mathcal P(f)$ is the propagation transfer function encoding dispersive, dissipative, or~nonlocal modifications between emission and arrival, and~$\mathcal R_I(f)$ is the response of detector $I$, including its antenna pattern and projection onto the incoming polarization basis. The~black-hole structure and its QNM response belong to the source sector and are therefore part of $\tilde h_{\rm em}(f)$ or of the corresponding ringdown waveform, not of $\mathcal R_I(f)$.

We organize deviations from GR by physical origin as
\begin{equation}
\label{eq:phenomenological_decomposition}
\delta\tilde h=\delta\tilde h_{\rm src}+\delta\tilde h_{\rm prop}
+\delta\tilde h_{\rm bg/QNM}+\delta\tilde h_{\rm sc}\,.
\end{equation}

The first three terms denote, respectively, corrections to source generation, propagation, and~the black-hole background or QNM spectrum. They contribute to the tensor channel and are added coherently in the waveform. The~fourth term denotes an additional non-tensor polarization channel, treated in \cref{sec:scalar}; it enters the detector output alongside the tensor signal rather than as a correction internal to \cref{eq:detector_waveform}.

\Cref{eq:phenomenological_decomposition} is therefore a bookkeeping classification of physically distinct channels, not a unique linear expansion of a single observable. The~terms are not universally present in every compactification, and~their relative importance is assessed model by model in \cref{sec:numerics}.

\subsection{Status of the~Derivations}
\label{sub:derivation_status}

The physical mechanisms map onto \cref{eq:phenomenological_decomposition} as summarized in \cref{tab:channel_map}, which also records how far each channel is derived from the microscopic parameters. We distinguish three levels. A~\emph{\hl{complete chain}
} means that the passage from the microscopic or model parameters to a definite waveform deformation is carried out analytically under explicitly stated assumptions. A~\emph{\hl{parametric}} channel has a derived scaling or a literature-based parametrization, but~its coefficient depends on model details not computed here. A~channel that is \emph{\hl{not derived}} is identified and discussed but not~quantified. 

\begin{table}[h]
\centering
\small
\caption{\hl{Mapping} between the physical mechanisms of \cref{sec5,sec:prop,sec5ring,sec:scalar} and the terms of \cref{eq:phenomenological_decomposition}, together with the status of their derivation from microscopic~parameters.}
\label{tab:channel_map}
\begin{tabularx}{\textwidth}{LlLL}
\toprule
\textbf{Mechanism} & \textbf{Term} & \textbf{Main Observable} & \textbf{Derivation Status} \\
\midrule
Virtual heavy-KK exchange (torus) & $\delta\tilde h_{\rm src}$ & inspiral phase & complete chain to the scaling; coefficient model dependent (\cref{sub:eft_derivation}) \\\midrule
Virtual exchange with brane-induced mixing & $\delta\tilde h_{\rm src}$ & inspiral phase & complete analytic chain (\cref{sub:mixing_chain}) \\\midrule
RS2 conservative potential & $\delta\tilde h_{\rm src}$ & inspiral phase at 2PN & complete chain (\cref{sub:rs2-theory}) \\\midrule
Real KK/bulk emission & $\delta\tilde h_{\rm src}$ & chirp rate, flux & parametric (\cref{sub:rs2_conservative_phase}) \\\midrule
Tidal Love numbers & $\delta\tilde h_{\rm src}$, $\delta\tilde h_{\rm bg/QNM}$ & inspiral phase at 5PN & phenomenologically introduced (\cref{sub:rs2_conservative_phase}) \\\midrule
Massive-mode dispersion & $\delta\tilde h_{\rm prop}$ & phase $\propto m^2/f$ & complete chain (\cref{sub:pheno-prop}) \\\midrule
Cumulative bulk leakage & $\delta\tilde h_{\rm prop}$ & $d_L^{\rm GW}\neq d_L^{\rm EM}$ & parametrized from the literature (\cref{sub:pheno-leak}) \\\midrule
Propagation speed, bulk shortcuts & $\delta\tilde h_{\rm prop}$ & GW--EM arrival-time difference & complete for a massive mode (\cref{sub:pheno-speed}) \\\midrule
QNM shift from KK self-energy & $\delta\tilde h_{\rm bg/QNM}$ & $\delta\hat\omega$, $\delta\hat\tau$ & parametric (\cref{sub:qnm_selfenergy}) \\\midrule
QNM shift from background deformation & $\delta\tilde h_{\rm bg/QNM}$ & $\delta\hat\omega$, $\delta\hat\tau$ & complete chain from $\Upsilon$ (\cref{sub:qnm_background}) \\\midrule
Late-time tails & $\delta\tilde h_{\rm bg/QNM}$ & power-law tail & not derived (\cref{sub:pheno-tails}) \\\midrule
Radion, moduli, extra polarizations & $\delta\tilde h_{\rm sc}$ & longitudinal mode, dipole flux & scaling derived; phenomenological couplings (\cref{sec:scalar}) \\
\bottomrule
\end{tabularx}
\end{table}
\unskip

\section{Source~Generation}
\label{sec5}

Corrections to the emitted waveform $\tilde h_{\rm em}(f)$ arise from virtual exchange of massive KK modes, which modifies the conservative orbital dynamics, and~from real emission of massive modes, which opens additional radiative channels. We first treat the heavy-KK regime through an effective-field-theory expansion. We then derive two cases in which the correction can be followed from microscopic parameters to the signal phase: the brane-induced mixing of \cref{subsec:exactRS1} and the RS2 weak-field~potential.

\subsection{Heavy-KK Effective~Dynamics}
\label{sub:heavyKK}

The brane-to-brane self-energy $\Sigma_P$ derived in \cref{sec4} is generally nonlocal in four-dimensional spacetime, as~will be made explicit for the radial representation in \mbox{\cref{eq:QNM_nonlocal_equation}}. For~a discrete and gapped KK spectrum, however, $\Sigma_P$ admits a local expansion in powers of $\omega/m_1$, where
\begin{equation}
m_1=\min_{\alpha>0}m_\alpha
\end{equation}
is the lightest massive KK mass, i.e.,\ the KK gap. This applies to the toroidal, Calabi--Yau, and~RS1 compactifications, whose spectra Equations (\ref{eq:toroidal_masses}), (\ref{eq:CY_spectrum}) and (\ref{eq:RS1_masses}) are discrete. It does not apply to RS2, whose continuum spectrum \cref{eq:RS2_spectrum} has no gap and must be treated through the nonlocal continuum expressions \cref{eq:Sigma_RS2,eq:rs2_pole_continuum}.

We therefore assume the heavy-KK regime
\begin{equation}
m_1\gg\omega_{\rm GW}\,,
\end{equation}
so that the massive modes are off-shell. With~the mostly plus signature and the convention $p^2=-\omega^2+\mathbf p^2$, the~propagator of a massive mode introduced in \cref{sec4} admits the low-energy expansion
\begin{equation}
G_\alpha(p)
=
-\frac{1}{m_\alpha^2}
-\frac{p^2}{m_\alpha^4}
+\mathcal O(m_\alpha^{-6})
=
-\frac{1}{m_\alpha^2}
+\frac{\Box}{m_\alpha^4}
+\mathcal O(m_\alpha^{-6})\,,
\label{eq:heavy_propagator}
\end{equation}
where $\Box$ is the four-dimensional d'Alembertian on $\bar g_{\mu\nu}(x)$.

\subsection{Regulated Toroidal EFT Matching and Cutoff~Dependence}
\label{sub:eft_derivation}

We illustrate the matching for a two-dimensional torus, with~KK masses $m_{\vec n}^2=|\vec n|^2/R^2$, $\vec n\in\mathbb Z^2\setminus\{0\}$, in~units $c=\hbar=1$. As~a scalar toy calculation, summing the low-energy propagators \cref{eq:heavy_propagator} over the tower with a universal brane coupling $g$ gives
\begin{equation}
\label{eq:lattice_sum_expansion}
\sum_{\vec n\neq0}g^2G_{\vec n}
=
-g^2R^2S_2
+g^2R^4S_4\Box
+\mathcal O(R^6\Box^2)\,,
\qquad
S_s\equiv\sum_{\vec n\neq0}\frac{1}{|\vec n|^s}\,.
\end{equation}

The two lattice sums behave differently. The~coefficient $S_2$ is logarithmically divergent. With~$K=\Lambda_{\rm UV}R$,
\begin{equation}
\label{eq:S2_regulated}
S_2(K)
=
\sum_{0<|\vec n|<K}\frac{1}{|\vec n|^2}
=
2\pi\ln K+C_{\rm latt}+o(1)\,.
\end{equation}

The coefficient $S_4$ is instead convergent for $N<4$ and equals an Epstein zeta function at $s=2$,
\begin{equation}
\label{eq:S4_epstein}
S_4
=
4\zeta(2)\beta(2)
=
\frac{2\pi^2}{3}G_{\rm C}
 \simeq6.03\,,
\end{equation}
where $\beta$ is the Dirichlet beta function and $G_{\rm C}$ is Catalan's constant. For~$N\ge4$, the~corresponding sums acquire power-law cutoff~dependence.

The $S_2$ term is momentum-independent and represents a contact interaction between sources. Its coefficient is regulator- and matching-scheme-dependent, so it cannot be assigned a universal value from the unregulated tower alone. Since it carries no frequency dependence, it does not by itself generate a phase deformation. The~frequency-dependent correction is carried by the finite, regulator-independent $S_4$ term.

Integrating out the massive tower order by order gives the effective action
\begin{equation}
\label{eq:EFT_action}
S_{\rm eff}
=
S_{\rm GR}
+\sum_i\frac{c_i}{\Lambda_i^{d_i-4}}
\int d^4x\sqrt{-g}\,\mathcal O_i\,,
\end{equation}
where $S_{\rm GR}$ is the four-dimensional Einstein--Hilbert action recovered in the decoupling limit, consistently with \cref{eq:Planck_unwarped}; $\mathcal O_i$ are higher-dimension curvature and stress-tensor operators generated by \cref{eq:heavy_propagator}; $d_i$ is the mass dimension of $\mathcal O_i$; $\Lambda_i$$\sim$$ m_1$; and $c_i$ are dimensionless Wilson coefficients. The~latter depend on the microscopic action, KK masses, wavefunction overlaps \cref{eq:KK_coupling}, brane couplings, and~tensor~structure.

The leading frequency-dependent source correction scales as
\begin{equation}
\frac{\delta h}{h}
 \sim 
c_i\left(\frac{\omega}{m_1}\right)^2\,.
\label{eq:EFT_scaling}
\end{equation}

This motivates the parametrization
\begin{equation}
\label{eq:emitted_waveform}
\tilde h_{\rm em}(f)
=
\tilde h_{\rm GR}(f)
\left[
1+\delta A_{\rm em}(f)
+i\delta\Psi_{\rm em}(f)
\right]\,,
\end{equation}
where $\delta A_{\rm em}$ and $\delta\Psi_{\rm em}$ are the fractional amplitude and phase corrections induced by the heavy-KK operators. They are polynomial in $\omega/m_1$ and describe virtual off-shell effects, not delayed propagation of a massive KK~wave.

For the toroidal case, the~derivation proceeds from $R$ to the masses $m_{\vec n}$, sums the heavy-mode propagators over the lattice, evaluates the regulated sums \cref{eq:S2_regulated,eq:S4_epstein}, and~extracts the Wilson coefficients and scaling \cref{eq:EFT_scaling}. The~scalar calculation fixes the structure of the expansion but not the full gravitational Wilson coefficients: a complete spin-2 matching requires canonically normalized tensor propagators, polarization projectors, brane-source couplings, and~a specified operator basis. The~gravitational coefficients are therefore treated as model-dependent effective parameters, with~the scale set by \cref{eq:S4_epstein}.

\subsection{Brane-Induced Mixing and the Heavy-KK~Coefficient}
\label{sub:mixing_chain}

The exactly solvable example of \cref{subsec:exactRS1} allows one to follow a genuine zero-mode--KK mixing down to an explicit correction of the brane response. For~$|p^2|\ll m_1^2$, the~self-energy \cref{eq:Keff_00} expands as
\begin{equation}
\Sigma_0(p^2)
=
-p^4\sum_{n\geq1}
\frac{(\Delta K_{0n})^2}{m_n^2}
\left[
1+\frac{p^2}{m_n^2}+\cdots
\right]\,.
\end{equation}

The effective zero-mode inverse propagator becomes
\begin{equation}
\mathcal K_{{\rm eff},00}(p^2)
=
p^2(1+\Delta K_{00})
\left[
1+c_{\rm mix}\frac{p^2}{m_1^2}
+\mathcal O\!\left(\frac{p^4}{m_1^4}\right)
\right]\,,
\label{eq:cmix}
\end{equation}
with
\begin{equation}
c_{\rm mix}
\equiv
\frac{1}{1+\Delta K_{00}}
\sum_{n\geq1}
(\Delta K_{0n})^2\frac{m_1^2}{m_n^2}\,.
\end{equation}

The zero-mode pole remains at $p^2=0$, while $1+\Delta K_{00}$ renormalizes the effective gravitational coupling. The~frequency-dependent correction to the brane response is
\begin{equation}
\frac{\delta G_{00}}{G_{00}}
=
-c_{\rm mix}\frac{p^2}{m_1^2}\,,
\end{equation}
which has the scaling of \cref{eq:EFT_scaling} with $c_i\to c_{\rm mix}$. This is a virtual effect: it modifies the propagator away from the pole and does not change the on-shell dispersion of the emitted~wave.

For RS1 with brane-induced mixing, the~chain proceeds from $M_{\rm IR}$, $k$, $L$, and~the mode functions $u_n$ to $\Delta K_{0n}$ (\cref{eq:DeltaK}), then to $\Sigma_0(p^2)$ (\cref{eq:Keff_00}), $c_{\rm mix}$ (\cref{eq:cmix}), and~finally to $\delta G_{00}/G_{00}$. The~coefficient is fixed by the brane-localized curvature term and the RS1 mode functions, up~to the spin-2 projector structure that multiplies the scalar-like expression~above.

\subsection{Virtual Exchange, Source Action and Energy~Balance}
\label{sub:pheno-source}

Virtual KK exchange modifies the source action according to
\begin{equation}
\label{eq:source_effective_action}
\Delta S_{\rm source}
=
-\frac{1}{2}
\sum_{\alpha>0}
g_\alpha^2
J_{\mu\nu}
G_\alpha^{\mu\nu\rho\sigma}
J_{\rho\sigma}\,,
\end{equation}
where $J_{\mu\nu}$ is the brane source of \cref{eq:J_alpha_coupling} and $G_\alpha^{\mu\nu\rho\sigma}$ is the tensor propagator of mode $\alpha$. This correction modifies the conservative orbital dynamics and hence the relation between orbital energy and frequency. Real emission of massive KK modes, radions, or~other bulk fields instead contributes an additional dissipative~flux.

The energy balance may therefore be written as
\begin{equation}
\label{eq:energy_balance}
\frac{dE_{\rm orb}}{dt}
=
-\mathcal F_{\rm GR}^{\rm corr}
-\mathcal F_{\rm real}\,,
\end{equation}
where $\mathcal F_{\rm GR}^{\rm corr}$ is the leading quadrupole flux evaluated on the conservatively corrected orbit, and~$\mathcal F_{\rm real}$ collects the radiative channels associated with real bulk emission. To~linear order, the~source-level waveform correction is
\begin{equation}
\label{eq:source_waveform}
\tilde h_{\rm em}(f)
=
\tilde h_{\rm GR}(f)
\left[
1+\delta A_{\rm src}(f)
+i\delta\Psi_{\rm src}(f)
\right]\,,
\end{equation}
where $\delta A_{\rm src}$ and $\delta\Psi_{\rm src}$ are the source-generation contributions to $\delta A_{\rm em}$ and $\delta\Psi_{\rm em}$ in \mbox{\cref{eq:emitted_waveform}}. Inspiral phase corrections are especially relevant because they accumulate coherently over many~cycles.

For a heavy KK tower, the~virtual-exchange correction is governed by the EFT of \cref{sub:heavyKK} and scales as
\begin{equation}
\label{eq:eps_KK_def}
\epsilon_{\rm KK}
 \simeq
c_i\left(\frac{\hbar\omega}{m_1c^2}\right)^2\,,
\end{equation}
with $c_i$ given by the toroidal coefficient of \cref{sub:eft_derivation} or by $c_{\rm mix}$ of \cref{eq:cmix}.

\subsection{Randall--Sundrum II: Weak-Field Potential and Conservative~Inspiral}
\label{sub:rs2-theory}

In the low-energy effective description, RS2 retains distinctive signatures through two complementary routes: a conservative weak-field inspiral correction controlled by the AdS curvature scale $\ell$, derived below, and~a strong-field merger-ringdown modification controlled by the effective tidal-charge parameter
$\Upsilon$.

\subsubsection{Static Response and Weak-Field~Potential}
\label{sub:rs2_potential}

A concrete low-energy observable follows from the static response of the
brane-to-brane Green function derived in \cref{sec4}. For~a static source,
$q\equiv|\mathbf q|$ replaces $p$, and~the Fourier transforms
\begin{equation}
\int\frac{d^3q}{(2\pi)^3}\frac{e^{i\mathbf q\cdot\mathbf r}}{q^2}=
\frac{1}{4\pi r}\,,\qquad\int\frac{d^3q}{(2\pi)^3}
e^{i\mathbf q\cdot\mathbf r}\ln q=-\frac{1}{4\pi r^3}
\end{equation}
turn the logarithmic term of the low-energy RS2 propagator into a relative correction $\ell^2/(2r^2)$ for a scalar~structure.

Including the tensor structure, the~zero-mode exchange carries the trace
coefficient $1/2$, whereas the continuum exchange carries $1/3$. For~static sources, the~relative weight is therefore:
\begin{equation}
\frac{1-\frac13}{1-\frac12}=\frac43\,,
\end{equation}
so the coefficient $\frac12$ becomes $\frac23$. Including this tensor structure and the brane-bending contribution, the~weak-field potential between nonrelativistic brane sources is:
\begin{equation}
V(r)=-\frac{G_Nm_1m_2}{r}\left[1+\frac{2\ell^2}{3r^2}
+\mathcal O\!\left(\frac{\ell^4}{r^4}\right)\right]\,,\qquad
r\gg\ell\,.
\label{eq:rs2_potential}
\end{equation}

Thus, the~minimal RS2 model predicts the definite fractional correction
\begin{equation}
\frac{\delta V}{V_N}=\frac{2}{3}\frac{\ell^2}{r^2}\,,
\label{eq:rs2_fractional_potential}
\end{equation}
which provides a microscopic connection between the AdS curvature radius and the conservative gravitational interaction. The~following inspiral calculation uses this weak-field correction as an effective conservative interaction. It does not assume that the complete five-dimensional compact-binary dynamics are determined by the static potential~alone.

\subsubsection{Conservative Correction to the Compact-Binary Inspiral~Phase}
\label{sub:rs2_conservative_phase}

We now derive an explicit inspiral-phase correction from the weak-field RS2 potential. The~calculation assumes a quasi-circular binary with orbital separation $r$, a~conservative two-body interaction obtained from the static potential, and~leading-order quadrupole radiation. It therefore isolates a definite conservative contribution rather than the complete radiative dynamics of a five-dimensional~binary.

Let
\begin{equation}
M=m_1+m_2\,,\qquad\mu=\frac{m_1m_2}{M}\,,\qquad\eta=\frac{\mu}{M}
\end{equation}
be the total mass, reduced mass, and~symmetric mass ratio. The~effective
potential is:
\begin{equation}
V(r)=-\frac{G_N\mu M}{r}\left(1+\frac{2\ell^2}{3r^2}\right)\,.
\end{equation}

For a circular orbit, the~radial force balance:
\begin{equation}
\mu\Omega^2r=\frac{dV}{dr}=G_N\mu M\left(\frac{1}{r^2}+\frac{2\ell^2}{r^4}\right)
\end{equation}
gives the modified Keplerian relation:
\begin{equation}
\Omega^2=\frac{G_NM}{r^3}\left(1+\frac{2\ell^2}{r^2}\right)\,.
\label{eq:rs2_kepler}
\end{equation}

The orbital energy is:
\begin{equation}
E=\frac12\mu\Omega^2r^2+V(r)\,,
\end{equation}
namely
\begin{equation}
E(r)=-\frac{G_N\mu M}{2r}+\frac{G_N\mu M\ell^2}{3r^3}\,.
\label{eq:rs2_energy_r}
\end{equation}

Introduce the dimensionless parameters:
\begin{equation}
x=\left(\frac{G_NM\Omega}{c^3}\right)^{2/3}\,,\qquad
\alpha_{\rm RS2}=\left(\frac{\ell c^2}{G_NM}\right)^2\,,
\end{equation}
and
\begin{equation}
u\equiv\frac{G_NM}{c^2r}\,,
\end{equation}
so that $\ell^2/r^2=\alpha_{\rm RS2}u^2$. From~\cref{eq:rs2_kepler},
\begin{equation}
x^3=u^3\left(1+2\alpha_{\rm RS2}u^2\right)\,.
\end{equation}

Inverting to first order in $\alpha_{\rm RS2}x^2$ gives:
\begin{equation}
u=x\left[1-\frac{2}{3}\alpha_{\rm RS2}x^2\right]\,,
\end{equation}
or equivalently
\begin{equation}
r=\frac{G_NM}{c^2x}\left[1+\frac{2}{3}\alpha_{\rm RS2}x^2\right]
+\mathcal O(\alpha_{\rm RS2}^2)\,.
\end{equation}

Substituting into
\begin{equation}
E=\mu c^2\left[-\frac{u}{2}+\frac13\alpha_{\rm RS2}u^3\right]
\end{equation}
gives
\begin{equation}
E(x)=-\frac{\mu c^2x}{2}\left[1-\frac{4}{3}\alpha_{\rm RS2}x^2\right]
+\mathcal O(\alpha_{\rm RS2}^2)\,.
\label{eq:rs2_energy_x}
\end{equation}

Therefore:
\begin{equation}
\frac{dE}{dx}=-\frac12\mu c^2\left(1-4\alpha_{\rm RS2}x^2\right)\,.
\end{equation}

For the radiative sector, we adopt the leading four-dimensional quadrupole flux evaluated on the corrected circular orbit:
\begin{equation}
\mathcal F_{\rm quad}=\frac{32}{5}\frac{G_N\mu^2r^4\Omega^6}{c^5}\,.
\end{equation}

Using $\Omega=c^3x^{3/2}/(G_NM)$ and the corrected $r(x)$ above, one finds:
\begin{equation}
r^4\Omega^6=\frac{c^{10}x^5}{(G_NM)^2}\left[1+\frac83\alpha_{\rm RS2}x^2
\right]\,,
\end{equation}
and hence
\begin{equation}
\mathcal F_{\rm quad}(x)=\frac{32}{5}\frac{c^5}{G_N}\eta^2x^5
\left[1+\frac83\alpha_{\rm RS2}x^2\right]
+\mathcal O(\alpha_{\rm RS2}^2)\,.
\label{eq:rs2_flux}
\end{equation}

This flux is a controlled quadrupole-model assumption; possible additional five-dimensional radiative corrections are not~included.

Using the balance equation:
\begin{equation}
\frac{dE}{dt}=-\mathcal F_{\rm quad}\,,
\end{equation}
i.e.,
\begin{equation}
\dot x=-\frac{\mathcal F_{\rm quad}}{dE/dx}\,,
\end{equation}
and expanding
\begin{equation}
\left(1-4\alpha_{\rm RS2}x^2\right)^{-1} \simeq1+4\alpha_{\rm RS2}x^2\,,
\end{equation}
one obtains
\begin{equation}
\frac{dx}{dt}=\frac{64}{5}\frac{c^3}{G_NM}\eta x^5
\left[1+\frac{20}{3}\alpha_{\rm RS2}x^2\right]
+\mathcal O(\alpha_{\rm RS2}^2)\,,
\label{eq:rs2_chirp}
\end{equation}
where $\frac83+4=\frac{20}{3}$.

The dominant GW frequency is
\begin{equation}
f=\frac{\Omega}{\pi}\,.
\end{equation}

Within the stationary-phase approximation,
\begin{equation}
\frac{d^2\Psi}{df^2}=\frac{2\pi}{\dot f}\,.
\end{equation}

Writing
\begin{equation}
v\equiv\left(\frac{\pi G_NMf}{c^3}\right)^{1/3}=x^{1/2}\,,\qquad
K\equiv\frac{\pi G_NM}{c^3}\,,
\end{equation}
so that $f=v^3/K$ and $df=3v^2dv/K$, one finds:
\begin{equation}
\dot f=\frac{96}{5}\frac{\eta}{K^2}v^{11}
\left[1+\frac{20}{3}\alpha_{\rm RS2}v^4\right]\,,
\end{equation}
and therefore
\begin{equation}
\frac{d^2\Psi}{df^2}=\frac{5K^2}{48\eta}\left[v^{-11}
-\frac{20}{3}\alpha_{\rm RS2}v^{-7}\right]\,.
\end{equation}

A first integration in $f$ gives:
\begin{equation}
\frac{d\Psi}{df}=\frac{5K}{16\eta}\left[-\frac{v^{-8}}{8}
+\frac53\alpha_{\rm RS2}v^{-4}\right]+\text{const}\,,
\end{equation}
and a second integration yields
\begin{equation}
\Psi(f)=\frac{3}{128\eta}v^{-5}-\frac{25}{16\eta}\alpha_{\rm RS2}v^{-1}
+\cdots\,.
\end{equation}

Restoring the integration constants:
\begin{equation}
\Psi(f)=2\pi ft_c-\phi_c-\frac{\pi}{4}+
\frac{3}{128\eta}x^{-5/2}\left[1-\frac{200}{3}\alpha_{\rm RS2}x^2\right]
+\mathcal O(\alpha_{\rm RS2}^2)\,,
\label{eq:rs2_spa_phase}
\end{equation}
where $t_c$ and $\phi_c$ are the coalescence time and phase. Terms constant or
linear in $f$ have been absorbed into $\phi_c$ and $t_c$.

The resulting conservative RS2-inspired phase deformation is therefore
\begin{equation}
\delta\Psi_{\rm RS2}(f)=-\frac{25}{16\eta}\alpha_{\rm RS2}
\left(\frac{\pi G_NMf}{c^3}\right)^{-1/3}\,.
\label{eq:rs2_explicit_phase}
\end{equation}

Since the leading GR term scales as $v^{-5}$ and the correction as $v^{-1}$, the~relative order is $v^4=x^2$, i.e.,\ a relative 2PN phase correction obtained from the RS2 weak-field potential and the adopted quadrupole flux. Its normalization is fixed within these assumptions, unlike an arbitrary phenomenological phase~kernel.

The result should not be interpreted as the complete RS2 compact-binary
waveform, since corrections to the radiative multipoles, bulk energy loss, strong-field two-body dynamics, and~the source-to-observer retarded response require separate calculations. The~approximation is appropriate only when the binary can be described by a weak-field, adiabatic effective interaction and $\ell/r\ll1$.

For the RS2 model, the~derivation proceeds from $\ell$ and $M_5$ to the
brane-to-brane Green function $G_{\rm TT}^E(p)$
(\cref{eq:Sigma_RS2,eq:rs2_pole_continuum}), derives the static response and potential $V(r)$ (\cref{eq:rs2_potential}), computes the orbital energy $E(x)$ and quadrupole flux $\mathcal F_{\rm quad}(x)$
(\cref{eq:rs2_energy_x,eq:rs2_flux}), obtains the frequency evolution $\dot x$ and Fourier phase $\Psi(f)$ (\cref{eq:rs2_chirp,eq:rs2_spa_phase}), and~finally yields the explicit phase deformation $\delta\Psi_{\rm RS2}$ as a function of the curvature radius $\ell$ (\cref{eq:rs2_explicit_phase}). These steps assume the weak-field, adiabatic, quasi-circular regime and the standard four-dimensional quadrupole~flux.

\section{Propagation~Effects}
\label{sec:prop}
Corrections to the propagation transfer function $\mathcal P(f)$ arise when a massive mode propagates over astrophysical distances, when gravity leaks into the bulk along the path, or~when the graviton speed differs from the speed of light. The~Euclidean continuum correction of the RS2 propagator does not belong to this class, since it does not shift the zero-mode pole (\cref{sub:rs2_propagator}).

\subsection{Massive-Mode Dispersion and Evanescent~Modes}
\label{sub:pheno-prop}
For a genuinely propagating massive mode of spatial wavenumber $k$ the dispersion relation is, restoring units,
\begin{equation}
\label{eq:massive_dispersion}
\hbar^2c^2k^2=\hbar^2\omega^2-m^2c^4\,.
\end{equation}

The condition $m<\omega$ (in natural units) is necessary for asymptotic propagation, consistent with the criterion identified in the Introduction. For~$m\gg\omega$, the wavenumber becomes imaginary, $k=i\kappa$ with $\hbar c\,\kappa=\sqrt{m^2c^4-\hbar^2\omega^2}$, and~the mode is evanescent, $\propto e^{-\kappa r}$, so its contribution is exponentially suppressed in the wave zone. This is the heavy-KK regime treated through the effective theory of \cref{sub:heavyKK}, and~not through direct~propagation.

In the light, propagating regime, the group velocity follows from \cref{eq:massive_dispersion},
\begin{equation}
\label{eq:group_velocity}
v_g=\frac{d\omega}{dk}=c\sqrt{1-\left(\frac{mc^2}{\hbar\omega}\right)^2} \simeq c\left[1-\frac{1}{2}\left(\frac{mc^2}{\hbar\omega}\right)^2\right]\,.
\end{equation}

Over a propagation distance $L$ (with negligible redshift), the arrival-time delay with respect to a massless signal is therefore
\begin{equation}
\Delta t(f)=\frac{L}{v_g}-\frac{L}{c} \simeq\frac{L}{2c}\left(\frac{mc^2}{hf}\right)^2\,.
\end{equation}

In the stationary-phase approximation, $d\Psi/df=2\pi t(f)$, so integrating the delay gives
\begin{equation}
\label{eq:dispersion_phase}
\delta\Psi(f)=2\pi\int\Delta t(f)\,df=-\frac{\pi L}{c}\left(\frac{mc^2}{h}\right)^2\frac{1}{f}=-\frac{\pi Lc}{\lambda_g^2}\,\frac{1}{f}\,,\qquad\lambda_g=\frac{h}{mc}\,,
\end{equation}
which reproduces the standard massive-graviton-type phase correction $\delta\Psi\propto m^2/f$. The~proportionality retains the frequency scaling, while, for~a source at cosmological redshift, the~distance $L$ must be replaced by the appropriate redshift-dependent distance integral, which cannot be fixed from the local dispersion relation alone. This scaling must not be applied to a heavy KK tower, whose phase correction instead follows the polynomial-in-$\omega/m_1$ scaling of \cref{eq:EFT_scaling}.

The propagating regime requires a mass gap below the GW photon energy, $m_1c^2<\hbar\omega$. For~a single flat extra dimension of size $R$, one has $m_1c^2$$\sim$$\hbar c/R$, so this condition corresponds to
\begin{equation}
\label{eq:propagation_condition}
R\gtrsim\frac{\hbar c}{\hbar\omega}\,.
\end{equation}

The dispersive channel in the detector band is therefore relevant only for an effective ultralight graviton mass $m_g$, independent of the KK~tower.

For the propagating massive mode, the~derivation proceeds from the mass $m$ to the dispersion relation (\cref{eq:massive_dispersion}), computes the group velocity $v_g$ (\cref{eq:group_velocity}) and the arrival-time delay $\Delta t(f)$, and~finally yields the phase deformation $\delta\Psi(f)$ as an explicit function of the mass and propagation distance (\cref{eq:dispersion_phase}), alongside the kinematic condition on the compactification size (\cref{eq:propagation_condition}).
\subsection{Cumulative Bulk Leakage and the GW Luminosity~Distance}
\label{sub:pheno-leak}
If gravity propagates in $D>4$ dimensions up to a distance scale $R_c$ while photons are confined to the brane, the~GW amplitude is damped relative to the four-dimensional $1/d_L$ law. Parametrizing the transition with a steepness $n$, the~GW and EM luminosity distances are related by~\cite{Pardo}:
\begin{equation}
\label{eq:dL_leakage}
d_L^{\rm GW}=d_L^{\rm EM}\left[1+\left(\frac{d_L^{\rm EM}}{R_c}\right)^{n}\right]^{\frac{D-4}{2n}}\,,
\end{equation}
which reduces to $d_L^{\rm GW}=d_L^{\rm EM}$ for $D=4$ or $d_L\ll R_c$. In~the opposite limit $d_L^{\rm EM}\gg R_c$, \mbox{one finds}
\begin{equation}
d_L^{\rm GW} \simeq d_L^{\rm EM}\left(\frac{d_L^{\rm EM}}{R_c}\right)^{\frac{D-4}{2}}\,,
\end{equation}
i.e.,\ the amplitude falls as $d^{-(D-2)/2}$, the~$D$-dimensional geometric law. For~sources at cosmological redshift, additional redshift-dependent factors must be included~\cite{Corman}. This effect is cumulative along the propagation path and is distinct from the local energy loss at the source discussed in \cref{sub:pheno-source}.

In RS1 and RS2, the four-dimensional zero mode is normalizable and localized on the brane. The~KK continuum modifies the brane-to-brane response only at distances comparable to the curvature radius $\ell$, so no cumulative amplitude damping accumulates over astrophysical distances, and~\cref{eq:dL_leakage} does not apply. It applies to scenarios with quasi-localized gravity, which are discussed in \cref{sub:pheno-scope}. No ET forecast for this channel is computed in the present~work.

\subsection{Propagation Speed and Bulk~Shortcuts}
\label{sub:pheno-speed}
Let $\epsilon_c\equiv(c_{\rm GW}-c)/c$. For~a massive mode with $m<\omega$, \cref{eq:group_velocity} gives $\epsilon_c=-\tfrac{1}{2}\left(mc^2/\hbar\omega\right)^2\le0$, so  a lower bound $\epsilon_c^{\min}$ on the GW speed translates into an upper bound on the mass,
\begin{equation}
\label{eq:mass_from_speed}
mc^2\le\sqrt{2\,|\epsilon_c^{\min}|}\;\hbar\omega\,.
\end{equation}

In warped braneworlds with brane-confined photons, gravitons may additionally propagate along bulk geodesics and arrive earlier than photons~\cite{Chung}. The~magnitude of such a shortcut depends on the bulk geometry and requires the retarded brane-to-brane response, and~it is not computed here. A~model predicting $|\epsilon_c|$ above the experimental bound (\cref{sec:numerics}) is excluded irrespective of the waveform-level sensitivity of \mbox{third-generation~detectors.}

\section{Ringdown and Late-Time~Response}
\label{sec5ring}

We investigate how extra-dimensional warping affects the ringdown following the coalescence of compact objects. The~relevant modifications are encoded in the effective brane dynamics of \cref{sec4} and alter the perturbation equations governing the post-merger signal. Depending on the spin of the perturbing field and on the sector considered, they enter the Regge--Wheeler, Zerilli, and~Teukolsky equations. We first fix the QNM conventions, then write the effective perturbation operator for all sectors, and~finally specialize it to the axial, polar, and~rotating cases. The~two physically distinct sources of QNM shifts, namely the KK self-energy and a deformation of the black-hole background, are treated in the last~subsections.

\subsection{QNM Parametrization and~Conventions}
\label{sub:pheno-qnm}

The ringdown is described as a superposition of damped QNMs,
\begin{equation}
\label{eq:ringdown}
h_{\rm RD}(t)=\sum_{\ell mn}A_{\ell mn}\,e^{-t/\tau_{\ell mn}}
\cos\!\left(\omega_{\ell mn}t+\phi_{\ell mn}\right)\,,
\end{equation}
where $A_{\ell mn}$, $\tau_{\ell mn}$, $\omega_{\ell mn}$, and~$\phi_{\ell mn}$ are the excitation amplitude, damping time, oscillation frequency, and~phase of the $(\ell,m,n)$ mode. The~perturbations of nonrotating and rotating black holes are described by the Regge--Wheeler--Zerilli and Teukolsky formalisms, respectively~\cite{Berti2}.

Equivalently, each mode is characterized by the complex frequency
\begin{equation}
\label{eq:complex_QNM_frequency}
\omega_{\ell mn}^{\rm c}=\omega_{\ell mn}^{R}+i\omega_{\ell mn}^{I}\,,
\qquad\omega_{\ell mn}^{I}>0\,,
\end{equation}
with time dependence $e^{i\omega_{\ell mn}^{\rm c}t}
=e^{i\omega_{\ell mn}^{R}t}e^{-\omega_{\ell mn}^{I}t}$. Hence,
\begin{equation}
\label{eq:QNM_damping_convention}
\tau_{\ell mn}=\frac{1}{\omega_{\ell mn}^{I}}\,.
\end{equation}

Writing
\begin{equation}
\label{eq:complex_QNM_shift}
\omega_{\ell mn}^{\rm c}=\left(\omega_{\ell mn}^{\rm c}\right)^{\rm GR}
+\delta\omega_{\ell mn}^{\rm c}\,,\qquad\delta\omega_{\ell mn}^{\rm c}=
\delta\omega_{\ell mn}^{R}+i\delta\omega_{\ell mn}^{I}\,,
\end{equation}
the real part shifts the oscillation frequency, while the imaginary part changes the damping rate. Varying \cref{eq:QNM_damping_convention} gives
\begin{equation}
\delta\tau_{\ell mn}=-\left(\tau_{\ell mn}^{\rm GR}\right)^2
\delta\omega_{\ell mn}^{I}\,.
\end{equation}

The fractional shifts used as phenomenological parameters are
\begin{equation}
\label{eq:fractional_QNM_shifts}
\delta\hat\omega_{\ell mn}=\frac{\delta\omega_{\ell mn}^{R}}
{\omega_{\ell mn}^{R,{\rm GR}}}\,,\qquad\delta\hat\tau_{\ell mn}=
\frac{\delta\tau_{\ell mn}}{\tau_{\ell mn}^{\rm GR}}\,.
\end{equation}

They are introduced as independent phenomenological deviations from the GR ringdown~spectrum.

\subsection{Effective Perturbation Operator on a Black-Hole~Background}
\label{sub:qnm_operator}

There are two physically distinct sources of QNM shifts: a modification in the perturbation operator at fixed black-hole background, and~a modification in the background itself. In~the first case, the~four-dimensional zero-mode operator is corrected by the KK-induced self-energy of \cref{eq:effective_operator},
\begin{equation}
\label{eq:effective_operator_QNM}
\mathcal D_P(\omega)=\mathcal D_0^P(\omega)-\Sigma_P(\omega)\,,
\end{equation}
where $P\in\{{\rm RW},Z,T\}$ labels the perturbation sector. The~QNM condition is
\begin{equation}
\label{eq:QNM_condition}
\det\mathcal D_P(\omega)=0\,.
\end{equation}

The determinant is taken over the coupled perturbation space. It reduces to a scalar condition for a decoupled master equation and becomes a matrix condition for coupled systems, such as the polar or rotating sectors described by \cref{eq:Z_massive_system,eq:Teukolsky_effective}.

For a general KK tower, the~effective operator is nonlocal in the radial coordinate. In~a Regge--Wheeler-type representation,
\begin{equation}
\label{eq:QNM_radial_operator}
\mathcal D_P(\omega)=\frac{d^2}{dr_*^2}+\omega^2-V_0^P(r)
-\Sigma_P(r_*,r_*';\omega)\,,
\end{equation}
where the last term is an integral operator. The~QNM equation reads
\begin{equation}
\label{eq:QNM_nonlocal_equation}
\left[\frac{d^2}{dr_*^2}+\omega^2-V_0^P(r)\right]\Psi_P(r_*)-
\int dr_*'\,\Sigma_P(r_*,r_*';\omega)\Psi_P(r_*')=0\,.
\end{equation}

This nonlocality is the radial manifestation of continuum-induced nonlocality and is generically present when the integrated-out tower contains a continuum, as~in RS2. Only for a discrete, gapped spectrum can $\Sigma_P$ be expanded locally as
\begin{equation}
\Sigma_P\longrightarrow\delta V_P(r,\omega)
+\delta Z_P(r,\omega)\frac{d^2}{dr_*^2}+\cdots\,,
\end{equation}
where $\delta V_P$ is a local potential correction and $\delta Z_P$ a wave-function renormalization. This is the heavy-KK regime of \cref{sub:heavyKK}; it does not apply to the RS2~continuum.

The black-hole geometry itself may also be modified by the compactification,
\begin{equation}
\label{eq:BH_background_correction}
g_{\mu\nu}^{\rm BH}=g_{\mu\nu}^{\rm GR}+\delta g_{\mu\nu}^{\rm BH}\,,
\end{equation}
so
\begin{equation}
\label{eq:background_operator_correction}
\mathcal D_0^P\longrightarrow\mathcal D_0^P
+\delta\mathcal D_P^{\rm background}\,.
\end{equation}

The total perturbed operator is therefore
\begin{equation}
\label{eq:total_QNM_operator}
\mathcal D_P^{\rm tot}=\mathcal D_0^P+
\delta\mathcal D_P^{\rm background}-\Sigma_P\,.
\end{equation}

Unlike $\Sigma_P$, the~background contribution is not generically controlled by the KK gap. In~braneworld models, the~projection of the bulk Weyl tensor onto the brane can modify the four-dimensional black-hole geometry~\cite{Maartens}.

\subsection{Axial, Polar, and~Rotating~Sectors}
\label{sub:qnm_sectors}

In the axial sector, the~Regge--Wheeler equation describes odd-parity perturbations of a Schwarzschild black hole. The~background metric \cref{eq:general_warped_metric} is specialized to
\begin{equation}
d\bar s_4^2=-f_{\rm S}(r)dt^2+\frac{dr^2}{f_{\rm S}(r)}+r^2d\Omega^2\,,
\qquad f_{\rm S}(r)=1-\frac{2M}{r}\,,
\end{equation}
with tortoise coordinate $dr_*/dr=1/f_{\rm S}$. The~zero mode obeys the standard Regge--Wheeler potential
\begin{equation}
\label{eq:RW_GR_potential}
V_{\rm RW}^{\rm GR}(r)=f_{\rm S}(r)\left[\frac{\ell(\ell+1)}{r^2}
-\frac{6M}{r^3}\right]\,.
\end{equation}

For a massive KK excitation, a~schematic single-channel reduction gives
\begin{equation}
\label{eq:RW_KK}
\frac{d^2\Psi_\alpha^{\rm RW}}{dr_*^2}+\left[\omega^2-V_\alpha^{\rm RW}(r)\right]\Psi_\alpha^{\rm RW}=S_\alpha^{\rm RW}\,,
\end{equation}
with
\begin{equation}
V_\alpha^{\rm RW}(r)=f_{\rm S}(r)\left[\frac{\ell(\ell+1)}{r^2}
-\frac{6M}{r^3}+m_\alpha^2\right]\,,
\end{equation}
where $m_\alpha^2$ is fixed by \cref{eq:internal_eigenproblem} and $S_\alpha^{\rm RW}$ is the axial projection of \cref{eq:J_alpha_coupling}. This equation is only a schematic single-channel approximation: a quantitative QNM calculation requires the constrained massive spin-2 system on the specified background, including the relevant coupled axial and polar sectors. The~term $f_{\rm S}m_\alpha^2$ modifies a massive KK excitation; it does not automatically modify the massless zero-mode potential. If~dynamical mixing is present, $\mathcal M_{0\alpha}\neq0$ as in \cref{eq:mixing_term}, the~zero mode obeys \cref{eq:effective_operator_QNM} with $P={\rm RW}$.

In the polar sector, the~Zerilli equation describes even-parity perturbations. The~zero-mode equation is $\mathcal D_Z^{\rm GR}\Psi_Z=S_Z$, with~\begin{equation}
\mathcal D_Z^{\rm GR}=\frac{d^2}{dr_*^2}+\omega^2-V_Z^{\rm GR}(r)\,,
\end{equation}
and
\begin{equation}
\label{eq:Zerilli_GR_potential}
V_{\rm Z}^{\rm GR}(r)=\frac{2f_{\rm S}(r)}{r^3(\lambda r+3M)^2}
\mathcal N_{\rm Z}(r)\,,
\qquad
\lambda=\frac{(\ell-1)(\ell+2)}{2}\,,
\end{equation}
where
\begin{equation}
\mathcal N_{\rm Z}(r)=\lambda^2(\lambda+1)r^3+3\lambda^2Mr^2
+9\lambda M^2r+9M^3\,.
\end{equation}

A massive KK graviton carries tensor, vector, and~scalar polarization states (\cref{sub:polar_content}), so the polar perturbations of mode $\alpha$ are generally described by a coupled system,
\begin{equation}
\label{eq:Z_massive_system}
\mathbf D_\alpha^Z\boldsymbol\Psi_\alpha^Z=\mathbf S_\alpha^Z\,,
\end{equation}
where $\boldsymbol\Psi_\alpha^Z$ collects the polar amplitudes, $\mathbf D_\alpha^Z$ is the matrix-valued kinetic operator, and~$\mathbf S_\alpha^Z$ is the source vector. After~eliminating the massive sector, the~zero mode obeys \cref{eq:effective_operator_QNM} with $P=Z$. The~number of coupled polarization sectors depends on the dimension and field content of the bulk; the single-vector, single-scalar counting applies to the minimal $N=1$ warped~case.

For a rotating black hole, the~Regge--Wheeler and Zerilli separations rely on spherical symmetry and are no longer available in the same form. The~appropriate framework is the Newman--Penrose formalism, with~spin-weighted curvature scalars $\psi_{\pm2}$ satisfying the Teukolsky equation on Kerr. In~GR,
\begin{equation}
\label{eq:Teukolsky_GR}
\mathcal T_{\pm2}^{\rm GR}\psi_{\pm2}^{(0)}=S_{\pm2}\,,
\end{equation}
where $\mathcal T_{\pm2}^{\rm GR}$ is the Teukolsky operator, $\psi_{\pm2}^{(0)}$ the massless Newman--Penrose scalar, and~$S_{\pm2}$ the projected source. If~the massless sector is dynamically coupled to massive KK modes, $\mathcal M_{0\alpha}\neq0$, then
\begin{equation}
\label{eq:Teukolsky_effective}
\left[\mathcal T_{\pm2}^{\rm GR}-\Sigma_T\right]\psi_{\pm2}^{(0)}=
S_{\pm2}^{\rm eff}\,,
\end{equation}
where $\Sigma_T$ and $S_{\pm2}^{\rm eff}$ are the Teukolsky counterparts of $\Sigma_P$ and $S_{\rm eff}^P$ in \cref{eq:effective_operator}. A~generic $\Sigma_T$ may depend on $r,\theta,\omega,\ell,\ell'$ and mix angular multipoles. Thus, unlike the Regge--Wheeler and Zerilli cases, generic KK mixing need not preserve the separability of the Teukolsky equation~\cite{Cano}. It is the mixing, rather than the black-hole spin itself, that can spoil~separability.

\subsection{QNM Shift from the KK~Self-Energy}
\label{sub:qnm_selfenergy}

QNMs form a non-Hermitian problem with radiative boundary conditions, for~which normalization and boundary terms are essential~\cite{Leung}. Let $\Psi_R$ and $\Psi_L$ denote the right and left modes, i.e.,\ the solutions of the direct and adjoint unperturbed problems, with~a regularized bilinear product $\langle\cdot,\cdot\rangle$. Writing $\mathcal D=\mathcal D_0+\delta\mathcal D$ and expanding $\mathcal D(\omega^{\rm GR}+\delta\omega)\Psi_R=0$ to first order gives
\begin{equation}
\label{eq:QNM_shift_general}
\delta\omega_{\ell mn}=-\frac{\langle\Psi_L,\delta\mathcal D\,\Psi_R\rangle}{\langle\Psi_L,\partial_\omega\mathcal D_0\,\Psi_R\rangle}\,.
\end{equation}

The KK spectra affect the QNMs through the overlap between the zero-mode--KK mixing vertex and the GR QNM,
\begin{equation}
\label{eq:QNM_KK_overlap}
\mu_\alpha^{\ell mn}\equiv\int dr_*\,\Psi_0^{\ell mn}(r_*)\,
\mathcal M_{0\alpha}^P(r_*)\,
\Psi_\alpha^P\!\left(r_*;\omega_{\ell mn}^{\rm GR}\right)\,,
\end{equation}
where $\Psi_\alpha^P$ is the radial profile of the massive mode in channel $P$, introduced for the axial sector in \cref{eq:RW_KK}. Retaining $\Psi_\alpha^P$ is essential: it promotes the single integral \cref{eq:QNM_KK_overlap} to the correct projection of the nonlocal self-energy \cref{eq:QNM_nonlocal_equation} onto a propagating radial mode. The~normalization inherits the brane coupling $g_\alpha\propto u_\alpha(y_b)$ of \cref{eq:KK_coupling}. For~RS1, algebraic denominators such as $(\omega_{\ell mn}^{\rm GR})^2-m_\alpha^2$ appear only after further simplification; on a black-hole background, the~full response involves the radial Green function, potential, and~boundary conditions. For~RS2, the~tower is replaced by a continuum with spectral density $\rho(m)$.

In the heavy-KK limit:
\begin{equation}
\delta\omega_{\ell mn}^{\Sigma}\propto
\frac{g_{\rm eff}^2\left(\omega_{\ell mn}^{\rm GR}\right)^2}{m_1^2}\,,
\end{equation}
so the shift is suppressed by
\begin{equation}
\left(\frac{\hbar\omega_{\ell mn}^{\rm GR}}{m_1c^2}\right)^2\,,
\end{equation}
up to dimensionless overlap factors, consistently with \cref{eq:EFT_scaling}. For~brane-induced mixing, the~coefficient is governed by $c_{\rm mix}$ of \cref{eq:cmix}. This contribution is generically negligible for direct QNM spectroscopy unless the spectrum contains an accidentally light or near-resonant mode. The~explicit radial Green function and overlaps are not evaluated here, so this channel remains~parametric.

\subsection{Effective Tidal-Charge~Background}
\label{sub:tidal_charge_background}

A representative realization of the background channel is provided by the DMPR tidal-charge solution~\cite{Dadhich}. In~a vacuum region of the brane, and~with the effective four-dimensional cosmological constant set to zero, the~gravitational equations contain the projection of the bulk Weyl tensor. In~the usual normalization, they take the form:
\begin{equation}
\label{eq:tidal_effective_vacuum}
G_{\mu\nu}=-\mathcal E_{\mu\nu}\,,
\end{equation}
where $\mathcal E_{\mu\nu}$ is the projected Weyl contribution.
Consequently, a~vacuum brane need not have a Ricci-flat four-dimensional~geometry.

A static, spherically symmetric effective solution has the 
Reissner--Nordstr\"om-like form:
\begin{equation}
\label{eq:tidal_charge_metric}
ds^2=-f(r)dt^2+\frac{dr^2}{f(r)}+r^2d\Omega^2\,,
\end{equation}
with
\begin{equation}
\label{eq:tidal_charge_lapse}
f(r)=1-\frac{2M}{r}+\frac{\Upsilon M^2}{r^2}\,.
\end{equation}

We use geometrized units $G_N=c=1$ in this section. The~dimensional tidal coefficient is therefore $b=\Upsilon M^2$, while $\Upsilon$ is dimensionless. Upon~restoring units, the~last term in the lapse is
$\Upsilon(G_NM/c^2r)^2$.

Although the metric has the algebraic form of the Reissner--Nordstr\"om solution, $\Upsilon$ is not an electromagnetic charge. It parametrizes an effective gravitational contribution associated with the bulk.
Unlike a squared electromagnetic charge, it may have either sign. Negative tidal charge is particularly motivated by the attractive bulk contribution in the original construction. It is not imposed here as a
universal prediction of \mbox{all~compactifications.}

The Schwarzschild limit is recovered at $\Upsilon=0$. For~$\Upsilon\leq1$, the~outer horizon is
\begin{equation}
\label{eq:tidal_charge_outer_horizon}
r_+=M\left(1+\sqrt{1-\Upsilon}\right)\,.
\end{equation}

A negative value enlarges the horizon relative to Schwarzschild, while a positive value reduces its radius. The~deformation changes the background geometry and therefore affects geodesic motion, characteristic orbital frequencies, and~the perturbation problem governing BH~oscillations.

We adopt this metric as a theory-guided effective strong-field ansatz. The~effective brane equations do not provide, by~themselves, a~unique relation between $\Upsilon$, the~bulk curvature radius, and~the source
properties. Moreover, the~tidal-charge lapse is not a globally established replacement for the asymptotic RS2 weak-field~solution. 

Specifying the background metric and specifying
the dynamics of its perturbations are therefore two separate modeling steps. The~metric alone does not determine a universal gravitational master equation. The~effective axial prescription adopted below defines
the perturbative realization used in this~work.

\subsection{QNM Shift from the Black-Hole~Background}
\label{sub:qnm_background}

The background contribution is obtained by replacing the self-energy operator with the perturbation induced by $\delta g_{\mu\nu}^{\rm BH}$. For~a Regge--Wheeler-type operator, $\partial_\omega\mathcal D_0=2\omega$, and~\begin{equation}
\label{eq:QNM_shift_background}
\delta\omega_{\ell mn}^{\rm bg}=-\frac{1}
{2\omega_{\ell mn}^{\rm GR}N_{\ell mn}}
\left\langle\Psi_0^{\ell mn},\delta\mathcal D_P^{\rm background}
\Psi_0^{\ell mn}\right\rangle\,,
\end{equation}
where $N_{\ell mn}=\int dr_*\,[\Psi_0^{\ell mn}]^2$ is the regularized norm. If~the deformation can be represented as a local potential correction at fixed tortoise coordinate, $\delta\mathcal D_P^{\rm background}\to-\delta V_P^{\rm bg}(r_*)$, then
\begin{equation}
\label{eq:local_background_QNM_shift}
\delta\omega_{\ell mn}^{\rm bg}=\frac{1}
{2\omega_{\ell mn}^{\rm GR}N_{\ell mn}}\int dr_*\,
\left[\Psi_0^{\ell mn}(r_*)\right]^2\delta V_P^{\rm bg}(r_*)\,.
\end{equation}

This formula holds only if the map $r\mapsto r_*$ is unchanged by the deformation, which is not the case for a tidal-charge~background.

A representative example is the DMPR tidal-charge BH, whose metric has a Reissner--Nordstr\"om-like form~\cite{Dadhich} introduced in
\cref{eq:tidal_charge_metric} and \cref{eq:tidal_charge_lapse}.
Here, $\Upsilon$ is a dimensionless tidal-charge parameter. It is not an electromagnetic charge, but~parametrizes the influence of the bulk geometry on the black-hole solution. The~expected sign in RS models is $\Upsilon<0$. This deformation is used as a theory-guided phenomenological ansatz; a top-down derivation of $\Upsilon$ from a specific compactification remains an open~problem.

The background correction scales as
\begin{equation}
\delta\omega_{\ell mn}^{\rm bg} \sim \mathcal O(\Upsilon)\,
\omega_{\ell mn}^{\rm GR}\,,
\end{equation}
up to mode-dependent coefficients, and~is not subject to the KK-gap suppression of the previous~subsection.

For the axial sector, an~effective tidal-charge-like Regge--Wheeler potential is
\begin{equation}
\label{eq:RW_tidal_charge_potential}
V_{\rm RW}(r;\Upsilon)=f(r)\left[\frac{\ell(\ell+1)}{r^2}
-\frac{6M}{r^3}+\frac{4\Upsilon M^2}{r^4}\right]\,.
\end{equation}

Expanding to first order in $\Upsilon$ at fixed $r$ gives
\begin{equation}
\label{eq:RW_tidal_charge_expansion}
V_{\rm RW}(r;\Upsilon)=V_{\rm RW}^{\rm GR}(r)
+\Upsilon\,\delta V_{\rm RW}^{\rm bg}(r)+\mathcal O(\Upsilon^2)\,,
\end{equation}
with
\begin{equation}
\label{eq:RW_tidal_charge_correction}
\delta V_{\rm RW}^{\rm bg}(r)=\frac{M^2}{r^4}
\left[\ell(\ell+1)+4-\frac{14M}{r}\right]\,.
\end{equation}

The corresponding master equation is
\begin{equation}
\label{eq:RW_tidal_charge_equation}
\frac{d^2\Psi_{\rm RW}}{dr_*^2}+\left[\omega^2-V_{\rm RW}(r;\Upsilon)\right]\Psi_{\rm RW}=0\,,
\end{equation}
with
\begin{equation}
\label{eq:tidal_charge_tortoise}
\frac{dr_*}{dr}=\frac{1}{1-\dfrac{2M}{r}+\dfrac{\Upsilon M^2}{r^2}}\,.
\end{equation}

Because $dr_*/dr$ depends on $\Upsilon$, the~perturbation cannot be represented by $\delta V_{\rm RW}^{\rm bg}$ at fixed $r$ alone.

Writing the equation in the Schwarzschild tortoise coordinate and defining
\begin{equation}
\varepsilon(r)\equiv\frac{\Upsilon M^2}{r^2f_{\rm S}}\,,
\end{equation}
one obtains, to~first order in $\Upsilon$,
\begin{equation}
\frac{d^2\Psi}{dr_*^2}+\left[\omega^2-V_{\rm RW}^{\rm GR}\right]\Psi
+\left[2\varepsilon\frac{d^2\Psi}{dr_*^2}+\varepsilon'\frac{d\Psi}{dr_*}
-\Upsilon\delta V_{\rm RW}^{\rm bg}\Psi\right]=0\,,
\end{equation}
where a prime denotes $d/dr_*$. The~bracket is $\delta\mathcal D\,\Psi$. Inserting it into \cref{eq:QNM_shift_background} and integrating by parts, with~boundary terms removed by the regularization, gives
\begin{equation}
\label{eq:RW_tidal_charge_QNM_shift}
\begin{split}
\delta\omega_{\ell mn}^{\rm bg,RW}={}&\frac{\Upsilon}
{2\omega_{\ell mn}^{\rm GR}N_{\ell mn}}\int dr_*\,
\Bigg\{\left[\Psi_0^{\rm RW}\right]^2\delta V_{\rm RW}^{\rm bg}\\
&+\frac{M^2}{r^2f_{\rm S}}\left[\left(\frac{d\Psi_0^{\rm RW}}{dr_*}\right)^2-\Psi_0^{\rm RW}\frac{d^2\Psi_0^{\rm RW}}{dr_*^2}\right]
\Bigg\}\,.
\end{split}
\end{equation}

The second term arises from the $\Upsilon$ dependence of $dr_*/dr$ and is absent from \cref{eq:local_background_QNM_shift}. Its numerical relevance is assessed in \cref{sec:numerics}. The~full $\Upsilon$-dependent problem is evaluated with a third-order WKB expansion, defining
\begin{equation}
\label{eq:RW_tidal_charge_coefficients}
\frac{\delta\omega^{R}}{\omega^{R}}\equiv
\mathcal C_\omega\Upsilon\,,\qquad\frac{\delta\tau}{\tau}
\equiv\mathcal C_\tau\Upsilon\,,
\end{equation}
for the axial $\ell=2$, $n=0$ mode, with~$\mathcal C_\omega$ and $\mathcal C_\tau$ given in \cref{sec:numerics}. For~$\Upsilon<0$, the~signs of the induced shifts are opposite to those of the~coefficients.

\Cref{eq:RW_tidal_charge_correction} is an effective tidal-charge-like Regge--Wheeler correction, not a universal potential for all braneworld models. A~complete treatment requires deriving the master equation from the linearized field equations of the specific model. For~the tidal-charge background, the~chain proceeds from $\Upsilon$ to $f(r)$, then to $V_{\rm RW}(r;\Upsilon)$, the~corrected tortoise coordinate, the~QNM shift \cref{eq:RW_tidal_charge_QNM_shift}, and~the fractional shifts \cref{eq:RW_tidal_charge_coefficients}. The~relation between $\Upsilon$ and microscopic compactification parameters is not~derived.

For the polar and rotating sectors, replacing $f_{\rm S}\to f$ in \cref{eq:Zerilli_GR_potential} gives only a minimal metric-only approximation,
\begin{equation}
\label{eq:Zerilli_effective_tidal_charge}
V_{\rm Z}^{\rm eff}(r;\Upsilon)=\frac{2f(r)}{r^3(\lambda r+3M)^2}
\mathcal N_{\rm Z}(r)\,,
\end{equation}
whose first-order metric contribution at fixed $r$ is
\begin{equation}
\label{eq:Zerilli_effective_correction}
\delta V_{\rm Z}^{\rm eff}(r)=\frac{2M^2\mathcal N_{\rm Z}(r)}
{r^5(\lambda r+3M)^2}\,.
\end{equation}

As in the axial sector, the~full first-order shift contains the tortoise-coordinate contribution. The~complete polar correction is
\begin{equation}
\label{eq:Zerilli_complete_correction}
\delta V_{\rm Z}^{\rm bg}=\delta V_{\rm Z}^{\rm eff}
+\delta V_{\rm Z}^{\rm dyn}\,,
\end{equation}
where $\delta V_{\rm Z}^{\rm dyn}$ includes modifications of the gauge-invariant master variable, constraints, additional degrees of freedom, and~bulk backreaction. It cannot be determined from the background metric alone. In~particular, axial and polar sectors need not remain isospectral once the perturbation dynamics are~modified.

For a rotating black hole, the~background deformation adds to the Teukolsky operator:
\begin{equation}
\label{eq:modified_Teukolsky_operator}
\left[\mathcal T_s^{\rm GR}+\delta\mathcal T_s^{\rm bg}
-\Sigma_T\right]R_s=0\,.
\end{equation}

Its detailed form and separability are model-dependent. A~generic modification of the rotating background need not preserve the separability of the Kerr perturbation equations; a modified Teukolsky treatment requires the explicit rotating background, perturbation action, boundary conditions, and~KK~couplings.

\subsection{Summary of the Two QNM~Channels}
\label{sub:qnm_summary}

The two channels have distinct parametric scalings:

  \begin{table}[H]
\centering

  \begin{tabularx}{\textwidth}{CC}

\toprule
\textbf{\hl{Channel} 
} & \textbf{Expected Scaling} \\
\midrule
KK self-energy &
$\displaystyle\frac{\delta\omega^\Sigma}{\omega^{\rm GR}}
$$\sim$$\frac{g_{\rm eff}^2(\omega^{\rm GR})^2}{m_1^2}$
\\[2ex]
BH background &
$\displaystyle\frac{\delta\omega^{\rm bg}}{\omega^{\rm GR}}
$$\sim$$\mathcal O(\Upsilon)$\\
\bottomrule
\end{tabularx}
\end{table}

Thus, observable QNM modifications from warped extra dimensions arise essentially through the background route, controlled by $\Upsilon$, while the effective-operator contribution is strongly suppressed for a large KK gap. In~a minimal separable compactification with no zero-mode--KK mixing and no background deformation,
\begin{equation}
\Sigma_P=0\,,\qquad\delta g_{\mu\nu}^{\rm BH}=0\,,
\end{equation}
consistent with \cref{eq:no_mixing}; 
the~zero-mode QNM spectrum remains the GR spectrum. The~expressions derived above enter the Fisher analysis through $\Upsilon$ (\cref{sec8}). Unless~the dependence on compactification parameters is explicitly propagated, the~Fisher parameters should be interpreted as phenomenological QNM shifts rather than direct measurements of $m_\alpha$, $g_\alpha$, $\ell_{\rm AdS}$, or~$\Upsilon$.

\subsection{Nonlocal Propagation, Spectral Densities, and~Late-Time~Tails}
\label{sub:pheno-tails}

A continuous KK spectrum or a nonlocal self-energy modifies the analytic structure of the retarded Green functions, introducing branch cuts in the complex-frequency plane in addition to discrete poles. In~RS2, the~continuum generates nonanalytic frequency-space contributions that affect the transient response. The~time-domain waveform therefore cannot be fully described by prompt emission and discrete QNMs alone; it generically includes late-time power-law tails governed by backscattering from the extra-dimensional potential barrier or bulk continuum,
\begin{equation}
h(t)=h_{\rm prompt}(t)+h_{\rm QNM}(t)+h_{\rm tail}(t)\,.
\end{equation}

The tail component $h_{\rm tail}(t)\propto t^{-p}$ is sensitive to the spectral density $\rho(m)$ and the bulk curvature scale $\ell$. In~GR, a~multipole-$\ell$ perturbation of a Schwarzschild black hole has the Price-law tail $p=2\ell+3$, subdominant to the ringdown. For~the extra-dimensional scenarios considered here, neither the exponent $p$ nor the tail amplitude relative to the QNM signal has been derived. This channel is therefore listed but not quantified, and~no accessibility limit is assigned to~it.

\section{Scalar Sector, Polarizations, and~Compactification~Moduli}
\label{sec:scalar}

Beyond the tensor channel, a~compactification may add vector and scalar degrees of freedom that contribute to the detector output through $\delta\tilde h_{\rm sc}$ in \cref{eq:phenomenological_decomposition}. We discuss the polarization content of massive KK excitations, the~radion, compactification moduli and their dipole radiation, and~brane-bending~modes.

\subsection{Polarization Content and the Heavy-KK~Truncation}
\label{sub:polar_content}

A massive spin-2 excitation in four dimensions carries five physical polarizations: two tensor, two vector, and~one scalar. This is the polarization content of each individual massive KK mode in the single-extra-dimension warped cases, $N=1$, relevant to RS1 and~RS2.

For compactifications with $N>1$, such as the toroidal and Calabi--Yau cases of \cref{sec2}, the~internal components $h_{\mu a}$ and $h_{ab}$ generate additional vector and scalar sectors. Their number and dynamics depend on the internal geometry, the~harmonic decomposition, and~the constraints of the massive spin-2 system; they should not be counted as independent four-dimensional polarizations of a single KK level without solving the corresponding constrained~problem.

All direct KK polarization channels require a propagating massive KK excitation. Under~the heavy-KK assumption of \cref{sub:heavyKK}, the~massive tower is integrated out and no direct KK polarization is included in the detector waveform. The~scalar channel considered below arises instead from a light radion or modulus, which is not suppressed by the tensor KK~gap.

\subsection{Radion and Detector~Response}
\label{sub:radion}

A separate scalar channel may arise from a light radion or modulus, decoupled from the heavy tensor tower. In~RS1, the~radion satisfies schematically
\begin{equation}
\label{eq:radion_equation}
\left(\Box-m_{\rm rad}^2\right)\varphi=\alpha_{\rm rad}T\,,
\end{equation}
where $m_{\rm rad}$ is the radion mass, set by the stabilization potential and generically independent of the tensor KK gap $m_1$, while $\alpha_{\rm rad}$ is the radion-matter coupling, fixed by the analogue of the wavefunction-overlap coupling $g_\alpha$ in \cref{eq:KK_coupling}.

If sufficiently light and coupled to matter, subject to the no-hair suppression for black-hole sources discussed in \cref{sub:rs1}, the~radion may generate breathing and longitudinal responses. The~detector strain is
\begin{equation}
\label{eq:polarization_response}
h_I=F_I^+h_++F_I^\times h_\times+F_I^b h_b+F_I^L h_L\,,
\end{equation}
where $F_I^+$, $F_I^\times$, $F_I^b$, and~$F_I^L$ are the antenna-pattern functions of detector $I$ for the plus, cross, breathing, and~longitudinal polarizations. The~breathing and longitudinal components originate from the same scalar degree of freedom and are not independent modes: for $m_{\rm rad}\ll\omega$, the~longitudinal response is suppressed relative to the breathing one by $m_{\rm rad}^2/\omega^2$. Together, they constitute the $\delta\tilde h_{\rm sc}$ channel of \cref{eq:phenomenological_decomposition}.

Emission of a scalar of mass $m_{\rm rad}$ is kinematically open only for
\begin{equation}
m_{\rm rad}c^2<\hbar\omega\,,
\end{equation}
equivalently when its Compton wavelength $\lambda_{\rm rad}=\hbar/(m_{\rm rad}c)$ exceeds $c/\omega$. For~heavier radions, the~scalar channel reduces to a short-range Yukawa correction to the conservative force and produces no propagating polarization signal in the detector~band.

\subsection{Moduli and Dipole~Radiation}
\label{sub:moduli}

Compactifications with nontrivial internal geometry, such as $T^N$ and Calabi--Yau manifolds, generically contain volume and shape moduli. If~unstabilized, they are massless, couple to matter with gravitational strength, and~give rise to scalar--tensor phenomenology, including dipole radiation in binaries with different scalar charges $\alpha_{1,2}$.

The dipole contribution to the inspiral phase scales as
\begin{equation}
\label{eq:dipole_phase}
\delta\Psi_{\rm dip}\propto\frac{(\alpha_1-\alpha_2)^2}{\eta}
\left(\frac{\pi G_NMf}{c^3}\right)^{-7/3}\,,
\end{equation}
with a normalization fixed by the scalar theory and by the effective scalar charges. Since the leading GR phase scales as $v^{-5}$ and the dipole term as $v^{-7}$, the~relative correction is of order $v^{-2}$, i.e.,\ a $-1$PN~contribution.

The leading dipole channel is suppressed for BH--BH binaries by the no-hair theorems for static scalar charges and is therefore primarily relevant for binaries containing neutron stars. Massless or light moduli are also constrained by the Cassini measurement of the post-Newtonian parameter $\gamma$, which bounds the scalar coupling to matter; the corresponding constraints are discussed in \cref{sec:numerics}. Whether stabilization renders the moduli sufficiently heavy to decouple is model-dependent.

\subsection{Brane-Bending~Modes}
\label{sub:brane_bending}

Brane-bending modes, i.e.,\ fluctuations in the brane position, contribute to the brane-to-brane response and enter the tensor structure of the RS2 propagator through the static potential of \cref{sub:rs2_potential}. They do not constitute an independent radiating polarization in the detector band when the brane-bending degree of freedom is integrated out or when the brane is effectively rigid on the GW wavelength scale. A~light, dynamical brane-bending mode could instead generate an additional scalar channel, but~its mass, couplings, and~phenomenology are model-dependent and are not computed~here.

\section{Model-by-Model Realization and Quantitative~Accessibility}
\label{sec:numerics}
\label{sub:pheno-scope}

The screening developed in the previous sections covers the four compactifications of \cref{sec2}: the torus $T^N$, Ricci-flat/Calabi--Yau compactifications, RS1, and~RS2. It assigns quantitative limits only to the channels whose dependence on the microscopic parameters has been derived above; the remaining channels are identified as phenomenological channels with model-dependent normalizations. The~conclusions should not be extended to scenarios outside this set. These include Dvali--Gabadadze--Porrati (DGP) gravity~\cite{DGP} and cascading gravity~\cite{deRham}, in~which gravity is quasi-localized on the brane and a crossover scale controls the transition to higher-dimensional behavior, so \mbox{\cref{eq:dL_leakage}} and the propagation-speed bound are the primary tests. They also include RS models with a Gauss--Bonnet bulk term~\cite{Charmousis}, ADD-type large-extra-dimension scenarios~\cite{ADD} with brane-fluctuation modes or explicitly stabilized moduli, black strings~\cite{Chamblin,Gregory}, higher-dimensional black holes with horizon radius below $\ell$, which are relevant at most for primordial black holes, and~stochastic backgrounds with KK or bulk spectral features or nonlinear \mbox{GW~memory.}

We now specialize the analytic framework of \cref{sec7,sec5,sec:prop,sec5ring,sec:scalar} to each of the four compactifications, evaluating the explicit microscopic results of \cref{sec4,sec5} numerically and comparing them with current bounds and with the sensitivity of third-generation detectors. The~channels whose derivation from microscopic parameters is complete are identified in \cref{tab:channel_map}.

\subsection{Detector Benchmarks and Reference~Scales}
\label{sub:num-benchmarks}

Throughout this section, we adopt the following reference thresholds for ET. Waveform-parameter uncertainties from third-generation forecasts are $\sigma_{\rm PN}$$\sim$$10^{-3}$, the~phase-sensitivity threshold is $\sim$$10^{-2}$--$10^{-3}\,\mathrm{rad}$, and~the ringdown resolution is assumed to reach fractional frequency shifts of $\sim$$10^{-2}$--$10^{-3}$ for high-SNR events. These are benchmark values adopted for the screening and are not derived from the Fisher analysis of \cref{sec8}, which compares the relative performance of the two ET configurations. The~GW photon energy is $\hbar\omega=4.1\times10^{-13}\,\mathrm{eV}$ at $f=100\,\mathrm{Hz}$ and $\hbar\omega\simeq1.0\times10^{-12}\,\mathrm{eV}$ at $f_{220}\simeq250\,\mathrm{Hz}$, the~ringdown frequency of a $60\,M_\odot$ black~hole.

\subsection{Kinematic Accessibility of Propagating~Modes}
\label{sub:num-kinematics}

We first consider direct propagation of KK modes. At~$f=100\,\mathrm{Hz}$, \cref{eq:propagation_condition} with $\hbar\omega=4.1\times10^{-13}\,\mathrm{eV}$ requires a compactification size
\begin{equation}
R\gtrsim\frac{\hbar c}{\hbar\omega} \simeq5\times10^{5}\,\mathrm{m}
\end{equation}
for a single flat dimension. This exceeds the torsion-balance limit \hl{$R\lesssim44\,\upmu\mathrm{m}$} 
 \cite{Kapner} and the Yukawa-range bound below $38.6\,\upmu\mathrm{m}$ \cite{Lee} by more than ten orders of magnitude. Direct KK propagation in the detector band is therefore excluded for the compactifications considered here; the dispersive channel of \cref{sub:pheno-prop} is relevant only for an effectively ultralight graviton mass, independently bounded by GWTC-3 as $m_g\lesssim10^{-23}\,\mathrm{eV}/c^2$ \cite{LIGO}.

For scalar or radion emission, the~threshold $m_{\rm rad}c^2<\hbar\omega$ of \cref{sub:radion} corresponds at $100\,\mathrm{Hz}$ to
\begin{equation}
m_{\rm rad}c^2\lesssim4\times10^{-13}\,\mathrm{eV}\,,
\end{equation}
i.e.,\ a Compton wavelength above $5\times10^{5}\,\mathrm{m}$.

\subsection{Heavy KK~Towers}
\label{sub:num-heavyKK}
In this section, we will investigate the implications caused by the heavy component of the KK mass~tower.

\subsubsection{Toroidal Compactification, $M_4\times T^N$}
\label{sub:model-toroidal}

For a toroidal compactification with $N=2$, the~KK masses are given by \cref{eq:toroidal_masses}. Taking $M_D=5\,\mathrm{TeV}$ and $M_{\rm Pl}=2.435\times10^{18}\,\mathrm{GeV}$, the~relation $M_{\rm Pl}^2=M_D^{2+N}V_N$ gives
\begin{equation}
R \simeq3.1\,\upmu\mathrm{m}\,,\qquad m_1c^2 \simeq64.6\,\mathrm{meV}\,.
\end{equation}

At $f=100\,\mathrm{Hz}$, where $\hbar\omega=4.14\times10^{-13}\,\mathrm{eV}$, the~kinematic factor is
\begin{equation}
\left(\frac{\hbar\omega}{m_1c^2}\right)^2 \simeq4.1\times10^{-23}\,.
\end{equation}

With an $\mathcal O(1)$ Wilson coefficient, $|c_i|$$\sim$$ g^2S_4$ as in \cref{sub:eft_derivation}, the~waveform correction \cref{eq:eps_KK_def} is
\begin{equation}
\epsilon_{\rm KK} \simeq c_i\left(\frac{\hbar\omega}{m_1c^2}\right)^2
 \sim 4\times10^{-23}\,,
\end{equation}
up to the uncertainty on $c_i$. Comparing with $\sigma_{\rm PN}$$\sim$$10^{-3}$, the~heavy tensor KK tower lies about twenty orders of magnitude below the observational reach. Reaching the benchmark sensitivity would require $|c_i|\gtrsim2\times10^{19}$, far outside any plausible variation in the matching coefficient. The~toroidal tower is therefore inaccessible to direct interferometric~detection.

\subsubsection{Ricci-Flat and Calabi--Yau Compactification, $M_4\times X_N$}
\label{sub:model-CY}

For the tensor sector, the~KK spectrum is determined by \cref{eq:CY_spectrum}. For~$N=6$, with~$V_6=(2\pi R)^6$ and $M_D=5\,\mathrm{TeV}$, the~compactification scale is $R\simeq4.9\times10^{-16}\,\mathrm{m}$. Assuming a dimensionless internal eigenvalue $\hat\lambda_1$$\sim$$1$, the~estimate
\begin{equation}
m_1c^2 \sim 0.4\,\mathrm{GeV}
\end{equation}
is consistent with this dimensional benchmark, although~it is not a universal prediction for all Calabi--Yau manifolds. At~$f=300\,\mathrm{Hz}$,
\begin{equation}
\left(\frac{\hbar\omega}{m_1c^2}\right)^2
 \sim 10^{-41}\,,
\end{equation}
so the tensor KK tower is inaccessible to ET, provided that $\hat\lambda_1$$\sim$$\mathcal O(1)$ and no unstabilized light moduli~intervene.

\subsubsection{Randall--Sundrum~I}
\label{sub:model-rs1}

In RS1~\cite{Randall1}, using the wavefunction normalization of \cref{sec4}, the~tensor KK gap is
\begin{equation}
m_1=x_1\frac{k}{M_{\rm Pl}}\Lambda_\pi \simeq0.38\,\mathrm{TeV}
\end{equation}
for $\Lambda_\pi=1\,\mathrm{TeV}$ and $k/M_{\rm Pl}=0.1$. At~$f=300\,\mathrm{Hz}$,
\begin{equation}
\left(\frac{\hbar\omega}{m_1c^2}\right)^2 \sim 10^{-47}\,,
\end{equation}
excluding tensor KK states from ET observation. The~brane-induced mixing of \cref{subsec:exactRS1} enters through $c_{\rm mix}$ in \cref{eq:cmix} with the same suppression. The~conclusion therefore holds unless $c_{\rm mix}$ is enhanced by many orders of magnitude beyond values compatible with the effective~description.

Even a sub-eV KK scale, such as the toroidal benchmark $m_1c^2\simeq0.0646\,\mathrm{eV}$ of \mbox{\cref{sub:model-toroidal}}, remains inaccessible. Its Compton wavelength is $\lambda_1\simeq3.05\,\upmu\mathrm{m}$, so over binary separations of order kilometers, the massive-mode effects are exponentially suppressed, while the off-shell correction at $f=100\,\mathrm{Hz}$ is of order $4\times10^{-23}$. Such modes are therefore kinematically decoupled and do not yield observable Fisher constraints from the mixing parameters~alone.

For the RS1 radion, the~constraint depends strongly on its mass, interaction length, coupling convention, and~stabilization model, and~cannot be extended indiscriminately to every radion scenario. For~black-hole binaries, the~leading scalar charge is additionally suppressed by no-hair theorems~\cite{Hawking}.

\subsection{Randall--Sundrum II: Conservative and Bulk-Radiation~Channels}
\label{sub:model-rs2}

We evaluate \cref{eq:rs2_explicit_phase} for a binary neutron-star system with $M=2.8\,M_\odot$, $\eta\simeq0.25$, and~$f=100\,\mathrm{Hz}$, taking the laboratory bound $\ell\lesssim50\,\upmu\mathrm{m}$. Since $G_NM/c^2\simeq4.13\,\mathrm{km}$, one has $\alpha_{\rm RS2}\simeq1.5\times10^{-16}$ and $(\pi G_NMf/c^3)^{-1/3}\simeq6.1$. The~absolute conservative phase deformation is therefore
\begin{equation}
|\delta\Psi_{\rm RS2}^{\rm cons}|
\lesssim
6\times10^{-15}\,\mathrm{rad}\,,
\end{equation}
\textls[-15]{about twelve orders of magnitude below the ET phase-sensitivity threshold of \mbox{$\sim$$10^{-2}$--$10^{-3}\,\mathrm{rad}$}}. The~conservative static-potential 2PN correction is thus unobservable throughout the region allowed by laboratory bounds, and~$\ell$ cannot be measured through the conservative inspiral phase with ET. A~dedicated Fisher bound on $\alpha_{\rm RS2}$, confirming this conclusion, is derived in \cref{subsec:rs2_fisher_microscopic}.

Taking the optimistic bound of \cref{eq:rs2_optimistic_bound} at face value, the~curvature radius required for the deformation to reach the statistical uncertainty is $\ell\simeq12\,\mathrm{m}$ for the binary-neutron-star benchmark, and~up to several hundred meters for heavier binaries. This lies more than five orders of magnitude above the laboratory bound and outside the domain of validity of the weak-field expansion. The~allowed region of $\ell$ is therefore entirely below the detector~reach.

For bulk energy loss, with~$\ell\lesssim10^{-4}\,\mathrm{m}$ and $f\lesssim10^3\,\mathrm{Hz}$, one obtains the constraint:
\begin{equation}
\left(\frac{\omega\ell}{c}\right)^2
\lesssim5\times10^{-18}\,.
\end{equation}

Since the accumulated GR phase is of order $10^3\,\mathrm{rad}$, the~corresponding phase deformation is at most of order $10^{-14}\,\mathrm{rad}$, again far below the ET~thresholds.

The potentially testable RS2 channels are therefore the non-conservative and strong-field ones: bulk-graviton energy leakage, late-time waveform tails, and~QNM shifts controlled by the tidal-charge parameter $\Upsilon$ during merger and ringdown (\cref{sub:model-qnm-upsilon}). Only the last has a quantitative parametrization in this~work.

\subsection{Quasinormal-Mode Shifts and Tidal-Charge Parameter $\Upsilon$}
\label{sub:model-qnm-upsilon}

The KK tower contributes to the QNM frequencies through the
self-energy term of \cref{sub:qnm_selfenergy}. For~a
$60\,M_\odot$ BH, with~$f_{220}\simeq250\,\mathrm{Hz}$ and
$\hbar\omega\simeq1.0\times10^{-12}\,\mathrm{eV}$, a~sub-millimeter KK gap
$m_1c^2=\hbar c/(100\,\upmu\mathrm{m})=2.0\,\mathrm{meV}$ gives
\begin{equation}
\left(\frac{\hbar\omega}{m_1c^2}\right)^2
 \simeq3\times10^{-19}\,,
\end{equation}
while a TeV-scale gap gives $\sim10^{-48}$, up~to
dimensionless overlap factors. Off-shell KK self-energy
corrections are therefore negligible for direct QNM
spectroscopy unless the spectrum contains an accidentally
light or near-resonant~mode.

The tidal-charge background provides a distinct route to
spectral modifications. Here, $\Upsilon$ is the dimensionless
geometric parameter of \cref{eq:tidal_charge_lapse}, and~its
effect is not controlled by the tensor KK gap. A~third-order
WKB evaluation of the full $\Upsilon$-dependent problem of
\cref{sub:qnm_background} gives, to~first order around
$\Upsilon=0$, for~the axial $\ell=2$, $n=0$ mode,
\begin{equation}
\label{eq:RW_tidal_charge_numbers}
\frac{\delta\omega^{R}}{\omega^{R}} \simeq+0.17\,\Upsilon\,,\qquad
\frac{\delta\tau}{\tau} \simeq-0.056\,\Upsilon\,,
\end{equation}
i.e.,\ $\mathcal C_\omega \simeq+0.17$ and
$\mathcal C_\tau \simeq-0.056$. The~calculation reproduces the
Schwarzschild fundamental mode to $0.2\%$; this agreement checks the undeformed reference spectrum, rather than independently establishing the accuracy of the $\Upsilon$-derivatives. The~coefficients are nearly
identical for $\ell=3$, namely $+0.17$ and $-0.059$. Retaining only the first term of \cref{eq:RW_tidal_charge_QNM_shift} would change the
real-frequency coefficient by about $10\%$ but reverse the sign of the damping-time shift, which would become $\simeq+0.19\,\Upsilon$. For~the negative-charge branch motivated by the RS interpretation, the~oscillation frequency decreases and the damping time increases.
In fractional terms, the~damping time is about three times less sensitive to $\Upsilon$ than the~frequency.

Existing ringdown analyses of GWTC-3 have constrained tidal charge using QNM information and emphasize correlations with the other BH parameters~\cite{Mishra}. As~an indicative sensitivity scale, if~the fractional frequency uncertainty reaches the benchmark
$\sigma_{\delta\hat\omega} \sim 10^{-2}$--$10^{-3}$ of
\cref{sub:num-benchmarks}, then \cref{eq:RW_tidal_charge_numbers} gives
\begin{equation}
\label{eq:upsilon_qnm_sensitivity_scale}
\sigma_\Upsilon^{(\omega)} \simeq\frac{\sigma_{\delta\hat\omega}}
{|\mathcal C_\omega|} \sim 0.06\text{--}0.006\,.
\end{equation}

This is a conditional conversion of an assumed spectral precision and its interpretation requires that the adopted background and perturbation prescription apply to the BH parameters. A~combined use of frequency and damping time must also retain their~correlation.

The geometric parameter $\Upsilon$ must be distinguished from a coefficient introduced directly in the inspiral phase. The~background and QNM calculation do not, by~themselves, determine the waveform of a comparable-mass binary. To~represent an additional inspiral channel, we
therefore introduce a separate dimensionless coefficient
$A_\Upsilon$ through the ppE-type prescription:
\begin{equation}
\label{eq:upsilon_phen_waveform}
\widetilde h_I(f;\boldsymbol\lambda,A_\Upsilon)=
\widetilde h_{I,\rm GR}(f;\boldsymbol\lambda)
\exp\!\left[iA_\Upsilon K_\Upsilon(f;\boldsymbol\lambda)\right]\,,
\end{equation}
where
\begin{equation}
\label{eq:upsilon_phen_kernel}
K_\Upsilon(f;\boldsymbol\lambda)=\left(\pi\frac{G_N\mathcal M_z}{c^3}f
\right)^{-3/2}\,,\qquad\mathcal M_z=(1+z)\mathcal M_{\rm source}\,.
\end{equation}

The exponent and normalization define a phenomenological benchmark and are not derived here from the tidal-charge metric, its orbital properties, or~its QNM spectrum. The~prescription modifies only the phase of the detector-projected waveform. The~stationary-phase time
used to evaluate the detector response is retained at its GR~value.

Interpreting this inspiral coefficient as a geometric tidal charge requires an additional mapping, for~example:
\begin{equation}
\label{eq:upsilon_phen_mapping}
A_\Upsilon=\kappa\Upsilon\,.
\end{equation}

For a specified nonzero constant $\kappa$, a~local uncertainty would translate as $\sigma_\Upsilon=\sigma_{A_\Upsilon}/|\kappa|$.
Neither the value of $\kappa$ nor its possible dependence on source properties is derived here. Taking $\kappa=\mathcal O(1)$ is an optional benchmark assumption, not a prediction of the background model. Constraints on $A_\Upsilon$ and on the geometric $\Upsilon$ must
therefore remain distinct unless this mapping is explicitly~adopted.

Two limitations govern the physical interpretation. First, if~the
bulk-induced deformation scales with a positive power of $\ell/r_h$, the~natural value of $\Upsilon$ for stellar-mass BH would be strongly suppressed. Sensitivity to a free tidal-charge parameter must
therefore not be read as sensitivity to a specific compactification. Second, the~numerical QNM coefficients are conditional on the effective perturbation model of \cref{sub:qnm_background}. The~background metric alone does not determine the complete bulk-coupled perturbation dynamics. Within~these assumptions, tidal-charge spectral shifts remain a potentially accessible strong-field channel, whereas the heavy-KK
self-energy corrections remain parametrically~unobservable.

\subsection{Experimental Bounds on Non-Waveform~Channels}
\label{sub:num-bounds}

In the cumulative bulk-leakage sector, the~comparison between the GW distance of GW170817 and the electromagnetic distance of its host galaxy is consistent with $D=4$ and requires a screening scale
\begin{equation}
R_c\gtrsim20\,\mathrm{Mpc}
\end{equation}
for $D>4$ \cite{LIGO2}, constraining \cref{eq:dL_leakage}.

The joint observation of GW170817 and GRB~170817A bounds the propagation-speed parameter as
\begin{equation}
-3\times10^{-15}
\lesssim\epsilon_c
\lesssim7\times10^{-16}
\end{equation}
\hl{ref.} 
 \cite{LIGO2}. Using the lower bound in \cref{eq:mass_from_speed} gives
\begin{equation}
mc^2\lesssim8\times10^{-8}\,\hbar\omega\,,
\end{equation}
i.e.,\ $mc^2\lesssim3\times10^{-20}\,\mathrm{eV}$ at $100\,\mathrm{Hz}$.

The massless-moduli case of \cref{eq:dipole_phase} and the associated dipole radiation of \mbox{\cref{sub:moduli}} are independently constrained by the Cassini measurement of the post-Newtonian parameter,
\begin{equation}
|\gamma-1|\lesssim2.3\times10^{-5}
\end{equation}
\hl{ref.} \cite{Cassini}, which bounds the scalar coupling to~matter.

\subsection{Summary of Quantitative Accessibility Across~Channels}
\label{sub:model-summary-table}

\Cref{tab:final_accessibility} summarizes the channels, microscopic parameters, theoretical estimates, ET sensitivity thresholds, and~observational outcomes obtained in this~work.

\begin{table}[H]
\centering

\caption{Summary of quantitative accessibility for extra-dimensional signatures with the Einstein Telescope (ET). Labels in parentheses denote the status: (NO) non-observable, (O) observable forecast target, and~(B) bounded or constrained. The~status (O)$^\ast$ refers to $\Upsilon$ as a free phenomenological parameter, whose relation to the microscopic RS2 parameters is not derived (\cref{sub:model-qnm-upsilon}); its threshold is the assumed benchmark of \cref{sub:num-benchmarks}.}
\label{tab:final_accessibility}

  \begin{tabularx}{\textwidth}{LLLL}

\toprule
Channel/Model & Microscopic parameter & Theoretical estimate/scaling & ET sensitivity threshold \\
\midrule
Toroidal KK tower ($T^2$) & $m_1c^2\simeq64.6\,\mathrm{meV}$ & $\epsilon_{\rm KK}$$\sim$$4\times10^{-23}$ & $\sigma_{\rm PN}$$\sim$$10^{-3}$ (NO) \\\midrule
Calabi--Yau KK tower ($X_6$) & $m_1c^2$$\sim$$0.4\,\mathrm{GeV}$ & $(\hbar\omega/m_1c^2)^2$$\sim$$10^{-41}$ & $\sigma_{\rm PN}$$\sim$$10^{-3}$ (NO) \\\midrule
RS1 tensor KK tower & $m_1\simeq0.38\,\mathrm{TeV}$ & $(\hbar\omega/m_1c^2)^2$$\sim$$10^{-47}$ & $\sigma_{\rm PN}$$\sim$$10^{-3}$ (NO) \\\midrule
RS2 conservative inspiral & $\alpha_{\rm RS2}$, $\ell\lesssim50\,\upmu\mathrm{m}$ & $|\delta\Psi_{\rm RS2}^{\rm cons}|\lesssim6\times10^{-15}\,\mathrm{rad}$ & $\sim$$10^{-2}$--$10^{-3}\,\mathrm{rad}$ (NO) \\\midrule
Tidal-charge QNM shifts & $\Upsilon$ & $\delta\omega^R/\omega^R\simeq+0.17\,\Upsilon$ & $\sim$$10^{-2}$--$10^{-3}$ (O)$^\ast$ \\\midrule
Bulk leakage, $d_L^{\rm GW}\neq d_L^{\rm EM}$ & $R_c\gtrsim20\,\mathrm{Mpc}$ & \cref{eq:dL_leakage}, $D=4$ bound & GW--EM comparison (B by GW170817) \\\midrule
Moduli/dilaton dipole & Scalar charges $\alpha_{1,2}$ & \cref{eq:dipole_phase}, $-1$PN & $\sim$$10^{-3}$ (B by Cassini) \\
\bottomrule
\end{tabularx}
\end{table}

\section{Fisher Forecast for Future~Interferometers}
\label{sec8}
 
Having established through the analytic screening (\cref{sec7,sec5,sec:prop,sec5ring,sec:scalar}) and its numerical evaluation (\cref{sec:numerics}) that the heavy tensor KK towers ($T^N$, CY, RS1) contribute to the waveform at most at the level of
$(\hbar\omega/m_1c^2)^2$$\sim$$10^{-23}$--$10^{-22}$, and~that the conservative RS2 inspiral phase lies far below the ET threshold, we restrict the Fisher-matrix parameter-estimation analysis to the surviving phenomenological~channels:
\begin{enumerate}
    \item A continuum-inspired phase deformation
    $\beta_{\rm RS2}$, a~ppE-type benchmark whose
    normalization is not derived from the microscopic RS2
    parameters (\cref{sub:rs2_propagator});
    \item QNM frequency and damping-time shifts
    $\delta\hat\omega$ and $\delta\hat\tau$, which also carry
    the tidal-charge channel through
    $\delta\hat\omega=\mathcal C_\omega\Upsilon$ and
    $\delta\hat\tau=\mathcal C_\tau\Upsilon$
    (\cref{eq:RW_tidal_charge_numbers});
    \item A ppE-type inspiral phase amplitude $A_\Upsilon$, introduced
    in \cref{eq:upsilon_phen_waveform} and treated as a free
    phenomenological coefficient. In~the absolute-sensitivity forecast, it is identified with the dimensionless tidal-charge parameter $\Upsilon$ as a modeling choice (\cref{subsec:upsilon_mc});
    \item A scalar polarization amplitude $a_{\rm sc}$, associated with the light scalar (radion-type) mode of the compactification and relevant for binaries containing non-BH compact objects.
\end{enumerate}
 
The RS2 analysis contains two logically distinct phenomena. The~parameter $\alpha_{\rm RS2}$ describes the conservative weak-field correction derived from the static RS2 potential and is used only for an absolute analytic detectability estimate. By~contrast, $\beta_{\rm RS2}$ denotes a
separately prescribed phase kernel and is included in the Monte Carlo analysis solely as a phenomenological benchmark. No direct relation between $\alpha_{\rm RS2}$ and $\beta_{\rm RS2}$ is~assumed.
 
Following this bottom-up, theory-guided phenomenological parametrization~\cite{Torri}, this section compares two candidate ET geometries: a double-L network ($2L$) with $15\,\mathrm{km}$ arms and a triangular network ($\triangle$) with $10\,\mathrm{km}$ arms. These nominal configurations are used for the main results, in~particular, for \cref{tab:R_summary} \hl{and} 
 \cref{fig:R_beta_RS2,fig:R_omega_QNM,fig:R_Upsilon,fig:R_Upsilon_z,fig:R_radion_sc}. Because~the arm length changes the noise curve, this nominal pair mixes the effect of the arm length with that of the geometry. We therefore repeat every comparison with a triangle of $15\,\mathrm{km}$ arms (equal-arm control), for~which the two networks share the same noise curve and the geometric contribution is isolated. The~extra-dimensional scenarios motivate the functional forms of the deformations, while the forecast does not constrain compactification parameters. The~analysis is exploratory, uses idealized noise and response assumptions, and~its limitations are collected in \cref{subsec:summary_tradeoffs}.
 
\subsection{Detector Response and Network Inner~Product}
\label{subsec:inner_product}
 
For detector $I$, the~waveform is expanded around a fiducial GR signal as
\begin{equation}
\label{eq:Fisher_linearized_waveform}
\tilde h_I(f;\boldsymbol\theta)=\tilde h_I^{\rm GR}(f;\boldsymbol\theta_{\rm GR})+
\delta\tilde h_I(f;\boldsymbol\theta_{\rm extra})\,,
\end{equation}
where $\boldsymbol\theta_{\rm GR}$ denotes the standard compact-binary parameters and
\begin{equation}
\boldsymbol\theta_{\rm extra}=\{\beta_{\rm RS2},\,
\delta\hat\omega_{\ell mn},\,\delta\hat\tau_{\ell mn},\,A_\Upsilon,\,
a_{\rm sc}\}
\end{equation}
are the accessible non-GR channels. Each channel is analyzed separately. For~every channel, the Fisher matrix contains the GR (or, for~the ringdown, the~amplitude and phase) nuisance parameters plus a single deformation parameter (two for the QNM channel; see below), and~no joint multi-channel forecast is~attempted.
 
For a detector network, the~noise-weighted inner product is
\begin{equation}
\label{eq:inner_product}
(a|b)=4\,{\rm Re}\sum_I\int_{f_{\min}}^{f_{\max}}df\,
\frac{\tilde a_I(f)\tilde b_I^*(f)}{S_{n,I}(f)}\,,
\end{equation}
where $S_{n,I}(f)$ is the one-sided noise power spectral
density of detector $I$, and~the noise is assumed as uncorrelated between detectors. The~sum runs over the interferometric channels of the configuration, i.e.,~$I=\{1,2,3\}$ for the triangular configuration, consisting of three co-located $60^\circ$ underground interferometers, and~$I=\{1,2\}$ for the $2L$ configuration, consisting of two $90^\circ$ interferometers located in different sites, separated by about $1200\,\mathrm{km}$.
 
Detector positions are computed on a spherical Earth of radius $6371\,\mathrm{km}$, and~the detector tensors are $D=\tfrac12(\hat u\hat u-\hat v\hat v)$, with~$\hat u,\hat v$ the arm directions, so that the factor $\sin60^\circ$ of the triangular interferometers is contained in the response. The~arm azimuths, measured from the local north, are $(0^\circ,60^\circ)$, $(120^\circ,180^\circ)$, $(240^\circ,300^\circ)$ for the triangle, and,~respectively, $(0^\circ,90^\circ)$ and $(45^\circ,135^\circ)$ for the two different sites for the $2L$. These orientations are inputs of the analysis and antenna patterns are computed in the long-wavelength limit. The~arm length enters through the noise model, while the site separation enters through the arrival-time~delays.
 
Each configuration uses the public ET amplitude spectral densities (ASDs) of Ref.~\cite{Branchesi} for the HFLF-cryogenic design, in~the version distributed with the \texttt{gwfast} package, with~$S_n={\rm ASD}^2$ for one nested interferometer. The~triangle uses the ASD of the $10\,\mathrm{km}$ design for each of its three interferometers, the~$2L$ uses the $15\,\mathrm{km}$ design for each of its two, and~the equal-arm control triangle uses the $15\,\mathrm{km}$ design. The~curves are used exactly as supplied, with~no rescaling or interpolation between arm lengths. The~same frequency grid, source realizations, waveform models and analysis settings are used for all~configurations. 
 
For inspiral channels, the~detector response and arrival-time delays are evaluated at the stationary-phase time--frequency map
\begin{equation}
\label{eq:spa_time}
t(f)=t_c+\frac{1}{2\pi}\frac{d\Psi_{\rm PN}}{df}\,,
\end{equation}
with detector positions and response tensors rotated in the Earth-fixed frame by $\omega_\oplus t(f)$. This is required for the long signals considered here, with~$t(f)-t_c$$\sim$$10^3$--$10^4\,\mathrm{s}$ at
$f_{\min}=5\,\mathrm{Hz}$ for the binary-neutron-star benchmark, and~breaks exact degeneracies of the restricted quadrupole waveform in the extrinsic sector $\{D_L,\phi_c,\iota,\psi,\theta,\phi\}$. For~the short, narrowband ringdown signals, Earth rotation is negligible and static responses are~used.
 
The Fisher and covariance matrices are
\begin{equation}
\label{eq:Fisher_matrix}
\Gamma_{ab}=\left(\partial_a h\middle|\partial_b h\right)\,,\qquad
C^{ab}=(\Gamma^{-1})^{ab}\,,\qquad\sigma_a=\sqrt{C^{aa}}\,.
\end{equation}

For the inspiral channels, marginalization over the GR nuisance parameters $\boldsymbol\lambda=\{t_c,\allowbreak\phi_c,\allowbreak\ln\mathcal M_z,\allowbreak\eta,\allowbreak\ln D_L,\allowbreak\cos\iota,\allowbreak\psi,\allowbreak\theta,\allowbreak\phi\}$ (with $\mathcal M_z$ the detector-frame chirp mass) is performed on the full $10\times10$ matrix, i.e.,\ the nine GR parameters plus one deformation parameter. For~the ringdown channels, the nuisance parameters are the amplitude and the phase of the mode, so the matrix is $4\times4$ (\cref{subsec:mc_results}).
 
The Fisher formalism~\cite{Flanagan} is a linearized-signal approximation, reliable when the SNR is high enough for the likelihood to be Gaussian around the fiducial point~\cite{Vallisneri}. For~parameters with physical boundaries, such as $\alpha_{\rm RS2}\ge0$, or~at low SNR, the~Fisher uncertainties should be read as local sensitivity scales rather than calibrated confidence intervals. We write $\Gamma=J^{T}J$, with~$J$ the whitened Jacobian in which real and imaginary parts are stacked and the columns are normalized, and~obtain $\sigma_a$ from its singular value decomposition. The~condition number $\kappa$ of the column-normalized matrix is $\kappa=(s_{\max}/s_{\min})^2$. If~the matrix is rank-deficient or $\kappa\geq10^{12}$, no uncertainty is assigned to the event and the event is excluded from the paired statistics. No pseudo-inverse is used, since discarding singular directions can underestimate $\sigma$ when the null space is not orthogonal to the deformation direction. The~excluded events are listed in \cref{subsec:mc_results}.
 
\subsection{Comparison~Metric}
\label{subsec:comparison_metric}
 
For each extra parameter $\vartheta_A$, a~$2L$ network and a triangular network are compared through the ratio of marginalized uncertainties over an ensemble of simulated events,
\begin{equation}
\label{eq:R_general}
R_A\equiv\frac{\sigma_{\vartheta_A}^{(2L)}}
{\sigma_{\vartheta_A}^{(\triangle)}}\,.
\end{equation}

Thus, $R_A<1$ favors the $2L$ configuration, while $R_A>1$ favors the triangular one. The~nominal comparison is $2L$ ($15\,\mathrm{km}$) versus triangle ($10\,\mathrm{km}$). The~equal-arm control, $2L$ ($15\,\mathrm{km}$) versus $\triangle$ ($15\,\mathrm{km}$), is denoted by the superscript (eq). Since $\sigma_{\vartheta_A}\propto\rho^{-1}$ at fixed waveform shape, $R_A$ at the natural SNR mixes the difference in
network SNR with the geometric breaking of parameter degeneracies. We therefore also report
\begin{equation}
\label{eq:R_fixed_snr}
R_A^{\rm fixed}\equiv R_A\frac{\rho_{2L}}{\rho_\triangle}\,,
\end{equation}
which compares the configurations at the same network SNR. This quantity removes the overall SNR scale, but~not the different spectral shape of the noise curves. Indeed, it isolates the geometric contribution only when the two networks share the same ASD, i.e.,\ for the equal-arm pair. For~the nominal pair, it also contains the effect of the different shapes of the $10\,\mathrm{km}$ and $15\,\mathrm{km}$ curves. In~the figures, the~upper panel shows $R_A$ at natural SNR and the lower panel shows $R_A^{\rm fixed}$. For~each channel, we quote the mean $\langle R_A\rangle$, the~median $\widetilde R_A$ and the fraction of events with $R_A<1$.
 
\subsection{Conservative RS2 Parameter: Fisher Formulation and Absolute~Detectability}
\label{subsec:rs2_fisher_microscopic}
 
The conservative phase correction derived in \cref{sub:rs2_conservative_phase} permits a Fisher analysis in which the deformation parameter is related to the microscopic AdS curvature radius $\ell$. \mbox{We define}
\begin{equation}
\alpha_{\rm RS2}=\left(\frac{\ell c^2}{G_NM}\right)^2\,,
\qquad\alpha_{\rm RS2}\geq0\,.
\end{equation}

Within the conservative circular-orbit model, with~the leading four-dimensional quadrupole flux evaluated on the corrected orbit, the~stationary-phase correction is (\cref{eq:rs2_explicit_phase})
\begin{equation}
\delta\Psi_{\rm RS2}^{\rm cons}(f)=\alpha_{\rm RS2}\,Q(f;\boldsymbol\lambda)\,,
\label{eq:rs2_conservative_phase_corrected}
\end{equation}
with
\begin{equation}
Q(f;\boldsymbol\lambda)=-\frac{25}{16\eta}
\left(\frac{\pi G_NMf}{c^3}\right)^{-1/3}\,.
\label{eq:rs2_fisher_kernel}
\end{equation}

This is a relative 2PN correction in the adopted approximation. Its coefficient is not a prediction for the full five-dimensional inspiral, since bulk-radiative, strong-field and source-dependent effects have not been computed. We distinguish the absolute measurability of this specific conservative contribution from the relative performance of the idealized detector~geometries.
 
The frequency-domain waveform of detector $I$ is
\begin{equation}
\widetilde h_I(f;\boldsymbol\lambda,\alpha_{\rm RS2})=
\widetilde h_I^{\rm GR}(f;\boldsymbol\lambda)
\exp\!\left[i\alpha_{\rm RS2}Q(f;\boldsymbol\lambda)\right]\,,
\label{eq:rs2_fisher_waveform}
\end{equation}
so at the GR fiducial point $\alpha_{\rm RS2}=0$,
\begin{equation}
\left.\frac{\partial\widetilde h_I}
{\partial\alpha_{\rm RS2}}\right|_{\alpha_{\rm RS2}=0}
=iQ(f;\boldsymbol\lambda)
\widetilde h_I^{\rm GR}(f;\boldsymbol\lambda)\,.
\label{eq:rs2_fisher_derivative}
\end{equation}

The network Fisher matrix is
\begin{equation}
\Gamma_{ab}=4\,{\rm Re}\sum_{I,J}\int_{f_{\min}}^{f_{\max}}df\,
\frac{\partial\widetilde h_I^*}{\partial\theta^a}
\left[S_n^{-1}(f)\right]_{IJ}
\frac{\partial\widetilde h_J}{\partial\theta^b}\,,
\label{eq:rs2_network_fisher}
\end{equation}
where $\boldsymbol\theta=(\boldsymbol\lambda,\alpha_{\rm RS2})$
and $S_n(f)$ is the network noise cross-spectral-density matrix, which reduces to the usual sum weighted by $1/S_{n,I}(f)$ for uncorrelated detector channels. After~marginalization over the standard parameters, the~effective information on
$\alpha_{\rm RS2}$ is
\begin{equation}
\mathcal I_\alpha=\Gamma_{\alpha\alpha}-\Gamma_{\alpha\lambda}
\Gamma_{\lambda\lambda}^{-1}\Gamma_{\lambda\alpha}\,,
\label{eq:rs2_schur}
\end{equation}
with
\begin{equation}
\sigma_\alpha=\mathcal I_\alpha^{-1/2}\,.
\label{eq:rs2_sigma_alpha}
\end{equation}

This expression is meaningful provided that the relevant Fisher blocks are well-conditioned and that the adopted parameter set captures the dominant~correlations.
 
Since the deformation is quadratic in $\ell$, a~Fisher expansion directly in $\ell$ around $\ell=0$ is singular. A~one-sided upper limit on the non-negative $\alpha_{\rm RS2}$ can instead be converted into a bound on $\ell$,
\begin{equation}
\ell_{\rm upper}=\frac{G_NM}{c^2}\sqrt{\alpha_{\rm upper}}\,.
\label{eq:rs2_ell_bound}
\end{equation}

The conversion from $\sigma_\alpha$ to $\alpha_{\rm upper}$ must account for the physical boundary $\alpha_{\rm RS2}\geq0$. A~Gaussian Fisher standard deviation is not by itself a calibrated one-sided confidence or credible limit. For~a specified physically allowed radius $\ell_\star$,
\begin{equation}
\alpha_\star=\left(\frac{\ell_\star c^2}{G_NM}\right)^2\,,
\end{equation}
a useful local sensitivity diagnostic is
\begin{equation}
\mathcal S_{\rm RS2}=\frac{\alpha_\star}{\sigma_\alpha}\,.
\label{eq:rs2_sensitivity}
\end{equation}

This ratio estimates the Fisher-scale distinguishability of the conservative deformation from the GR fiducial waveform. It is neither a detection significance nor a substitute for Bayesian model selection. The~detector-geometry comparison is performed through the ratio
\begin{equation}
R_\alpha=\frac{\sigma_\alpha^{\rm 2L}}{\sigma_\alpha^\triangle}\,,
\label{eq:rs2_geometry_ratio}
\end{equation}
with identical source realizations and consistent noise and response~assumptions.
 
With the waveform of \cref{eq:rs2_fisher_waveform}, the~unmarginalized network Fisher information at the GR fiducial point is
$\Gamma_{\alpha\alpha}=\rho^2\langle Q^2\rangle_\rho$, where $\rho$ is the network SNR and the brackets denote the normalized, positive, noise-weighted frequency average. Since $|Q(f)|$ decreases monotonically for $f\geq f_{\min}$, $\Gamma_{\alpha\alpha}\leq\rho^2|Q(f_{\min})|^2$, and~the marginalized uncertainty satisfies the detector-independent optimistic bound
\begin{equation}
\sigma_\alpha\geq\frac{1}{\rho\,|Q(f_{\min})|}\,.
\label{eq:rs2_optimistic_bound}
\end{equation}

This bound requires no particular noise curve, but~it is not an ET sensitivity forecast. Indeed, it neglects nuisance-parameter correlations and assumes a prescribed network~SNR.
 
For $\ell=14\,\upmu\mathrm{m}$, $\eta=1/4$, $f_{\min}=3\,\mathrm{Hz}$ (a lower cutoff than the $5\,\mathrm{Hz}$ used in the Monte Carlo, chosen to make the bound maximally optimistic)
and an intentionally optimistic network SNR $\rho=10^3$, the~equal-mass benchmarks $M=(2.8,\,30,\,60)\,M_\odot$ give
$\alpha_{\rm RS2}=(1.15\times10^{-17},\,9.99\times10^{-20},\,
2.50\times10^{-20})$. The~corresponding lower bounds on
$\sigma_\alpha$ are $(8.10\times10^{-6},\,1.79\times10^{-5},\,2.25\times10^{-5})$,
and the local Fisher-scale ratios $\alpha_{\rm RS2}/\sigma_\alpha$
are bounded above by $(1.42\times10^{-12},\,5.6\times10^{-15},\,1.1\times10^{-15})$.
For the largest laboratory-allowed radius, $\ell=50\,\upmu\mathrm{m}$ (\cref{sub:model-rs2}), one has $\alpha_{\rm RS2}\simeq1.5\times10^{-16}$ for $M=2.8\,M_\odot$, and~the ratio remains bounded by $\simeq2\times10^{-11}$. Even this idealized bound excludes the measurability of the conservative correction over the region allowed by laboratory~bounds.
 
The formal curvature radii obtained by setting
$\alpha_{\rm RS2}=\sigma_\alpha$ range from about ten to several hundred meters. They lie outside the domain of validity of the weak-field expansion and must not be reported as physical RS2 constraints. Moreover, the~relative 2PN frequency dependence can correlate with the GR intrinsic
parameters, making the actual marginalized uncertainties larger than the optimistic bound. The~phase kernel $\beta_{\rm RS2}$ used in the Monte Carlo comparison is a separate phenomenological benchmark and cannot be interpreted as a constraint on $\ell$ through
\cref{eq:rs2_conservative_phase_corrected}. A~fully predictive RS2 forecast would require the source-dependent retarded response and the missing radiative corrections. The~present treatment is therefore microscopically parametrized only within the conservative weak-field~approximation.
 
\subsection{Monte Carlo Setup, Numerical Stability and~Reproducibility}
\label{subsec:mc_results}
 
\begin{figure}[htpb]

\includegraphics[width=1.05\textwidth]{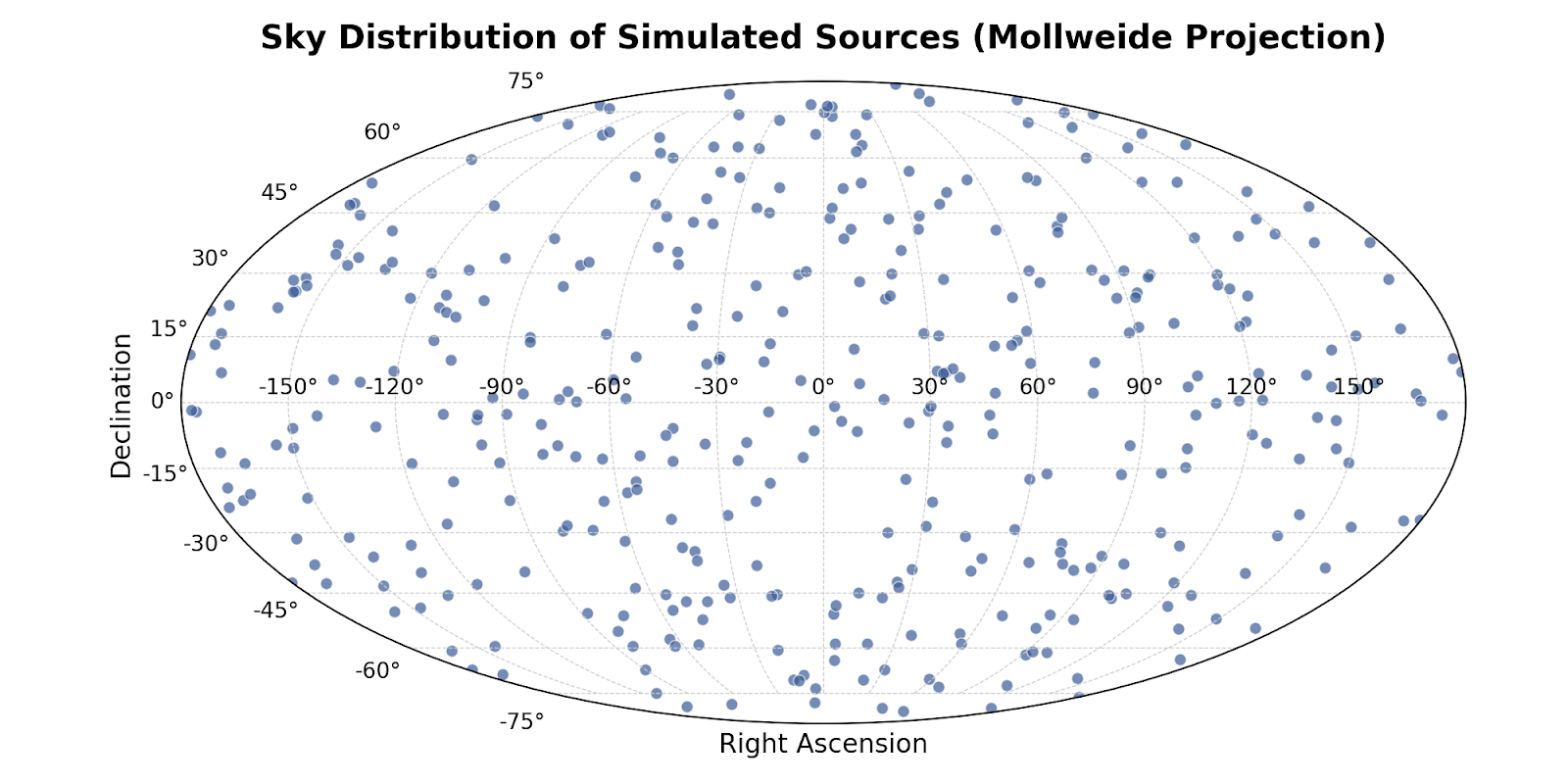}
\caption{\hl{Sky} distribution, in~a Mollweide projection, of~the $N=400$ simulated sources used for the comparison between the nominal $2L$ configuration with $15\,\mathrm{km}$ arms and the nominal triangular configuration with $10\,\mathrm{km}$ arms. The~same $400$ sky locations, together with the corresponding orientation realizations, are common to the two networks and to the equal-arm control, so the ratios $R_A$ and $R_A^{\rm fixed}$ are evaluated event-by-event on identical sources. The~positions are drawn isotropically on the sphere, and~the right ascension and the declination are shown on the horizontal and vertical \mbox{axes, respectively.}}
\label{fig:sky_distribution}
\end{figure}

The inspiral channels ($\beta_{\rm RS2}$, $A_\Upsilon$, $a_{\rm sc}$) use a common ensemble of $N=400$ independent binary-objects realizations, reused for every channel and for every configuration, so that the comparison is event-by-event \cref{fig:sky_distribution}. The~full simulation setup is summarized in \cref{tab:setup}. The~source-frame component masses are independent and uniform in $[1.1,2.5]\,M_\odot$ (chirp masses $1.0$--$2.1\,M_\odot$, total masses up to $4.9\,M_\odot$), all sources are placed at $z=0.1$ ($D_L=475.6\,\mathrm{Mpc}$ for flat $\Lambda$CDM with $H_0=67.7\,\mathrm{km\,s^{-1}\,Mpc^{-1}}$ and $\Omega_m=0.31$), and~the detector-frame chirp mass is $\mathcal M_z=(1+z)\mathcal M_c$. Sky position is isotropic in the Earth-fixed frame at the coalescence time, $\cos\iota$ is uniform in $[-1,1]$, $\psi$ and $\phi_c$ are uniform in $[0,2\pi)$, and~$t_c=0$. No SNR threshold is applied. The~median network SNR of the GR signal is $67.5$ for $\triangle$ ($10\,\mathrm{km}$), $89.9$ for $2L$ ($15\,\mathrm{km}$) and $93.4$ for $\triangle$ ($15\,\mathrm{km}$).

The GR signal in detector $I$ is
\begin{equation}
\tilde h_I^{\rm GR}=\mathcal A(f)\left[F_I^+\frac{1+\cos^2\iota}{2}-iF_I^\times\cos\iota\right]e^{i\Psi_{\rm GR}(f)}\,e^{2\pi if\tau_I(f)}\,,
\end{equation}
\textls[-15]{with the Newtonian amplitude $\mathcal A(f)=\sqrt{5/24}\,\pi^{-2/3}\,(c/D_L)\,(G\mathcal M_z/c^3)^{5/6}f^{-7/6}$, the~arrival-time delay} $\tau_I(f)$ and the nonspinning 1.5PN stationary-phase phase
\begin{equation}
\Psi_{\rm GR}=2\pi ft_c-\phi_c-\frac{\pi}{4}+\frac{3}{128\,\eta\,v^5}\left[1+\left(\frac{3715}{756}+\frac{55\,\eta}{9}\right)v^2-16\pi v^3\right]\,,
\end{equation}
where $v=(\pi GMf/c^3)^{1/3}$ and $M=\mathcal M_z\eta^{-3/5}$. The~band is $5$--$1000\,\mathrm{Hz}$ on $400$ logarithmically spaced points with trapezoidal weights. No ISCO cutoff is applied, and,~for the heaviest binaries, the detector-frame ISCO frequency is about $800\,\mathrm{Hz}$, so the upper part of the band lies beyond the validity of the inspiral waveform for those~events.
 
The ringdown channels use a fixed $(2,2,0)$ benchmark with $f_{220}=251\,\mathrm{Hz}$ and $\tau_{220}=4\,\mathrm{ms}$, with~the sky location and orientation of the same $N=400$ realizations. The~mode is described by the single-pole response
\begin{equation}
\tilde H(f)=\frac{1}{\tau^{-1}+2\pi i(f-f_{220})}\,,
\end{equation}
over $5$--$2000\,\mathrm{Hz}$, multiplied by the static network response $\sum_I|F_I^+(1+\cos^2\iota)/2-iF_I^\times\cos\iota|^2$ in power. The~Fisher matrix has four parameters ($\delta\hat\omega$, $\delta\hat\tau$, amplitude, phase) and is computed once per noise curve. The~uncertainty of each event is the common value divided by the square root of its network antenna power. Sky position, inclination and polarization are therefore not marginalized in the ringdown channel, and~the amplitude is normalized to an arbitrary fixed value, so only ratios between configurations, and~not absolute uncertainties or SNRs, are~meaningful.
 
For the inspiral channels, the~derivatives with respect to $\phi_c$, $\ln D_L$, and~the phase part of the derivatives with respect to $t_c$, $\ln\mathcal M_z$ and $\eta$ are analytic, as~are the derivatives with respect to the deformation parameters. The~amplitude and response contributions to the derivatives with respect to $t_c$, $\ln\mathcal M_z$, $\eta$, $\cos\iota$, $\psi$, $\theta$ and $\phi$ are computed by central finite differences, with~steps $10^{-7}$ for $\ln\mathcal M_z$ and $\eta$ and $10^{-5}$ for the other parameters. For~the sky-averaged ratios, the same $N$ realizations are used for every configuration. The~reported uncertainties are marginalized over the full set of GR source parameters. The~fiducial deformation parameter is zero for all~channels.
 
\begin{table}[H]
\centering

\caption{\hl{Summary} of the simulation setup. All the inputs, the~ASD file hashes and the library versions are stored together with the results in a configuration~file.}
\label{tab:setup}

\begin{tabularx}{\textwidth}{lL}
\toprule
Ensemble & $N=400$ BNS, seed $20260825$, $m_{1,2}\in[1.1,2.5]\,M_\odot$ uniform, $z=0.1$, isotropic sky, uniform $\cos\iota,\psi,\phi_c$, no SNR cut \\\midrule
Waveform & Newtonian amplitude, nonspinning 1.5PN SPA phase, Earth rotation through $t(f)$, no tides, spins or merger \\\midrule
Band and grid & $5$--$1000\,\mathrm{Hz}$ (inspiral), $5$--$2000\,\mathrm{Hz}$ (ringdown), log-spaced grids of $400$ and $4000$ points \\\midrule
Noise & Ref.~\cite{Branchesi} HFLF-cryo ASD, $10\,\mathrm{km}$ and $15\,\mathrm{km}$, one nested interferometer per channel, uncorrelated \\\midrule
Networks & $\triangle$: $3\times60^\circ$ at Sos Enattos; $2L$: Sos Enattos and Meuse--Rhine, azimuths $(0^\circ,90^\circ)$ and $(45^\circ,135^\circ)$ \\\midrule
Parameters & $9$ GR parameters $+1$ deformation (inspiral); $4$ (ringdown) \\\midrule
Solver & SVD of the normalized whitened Jacobian; exclusion if rank-deficient or $\kappa\geq10^{12}$ \\\midrule
Pairs & $2L$ ($15\,\mathrm{km}$)/$\triangle$ ($10\,\mathrm{km}$) nominal; $2L$ ($15\,\mathrm{km}$)/$\triangle$ ($15\,\mathrm{km}$) equal-arm \\
\bottomrule
\end{tabularx}
\end{table}
 
The rank and condition number $\kappa$ of the whitened Jacobian matrices were evaluated for all event-channel combinations. 
Due to geometric degeneracies in nearly face-on orientations ($\cos\iota \approx \pm 1$), where $\psi$ and $\phi_c$ become ill-defined, a~small number of realizations were excluded. 
For $\beta_{\rm RS2}$ and $A_\Upsilon$, the~excluded realization is event index $165$; for $a_{\rm sc}$, the~excluded realizations are event indices $165$ and $248$. 
These numbers identify individual source realizations in the common $N=400$ ensemble and do not denote sample sizes. 
The corresponding Fisher matrices have rank $9$ and $\kappa$$\sim$$10^{16}$. 
All remaining events satisfy $\kappa \le 7.2 \times 10^{11}$ (median $\kappa $$\sim$$ 10^5$--$10^6$), making our selection invariant to any condition threshold between $10^{12}$ and $10^{15}$. 
Ringdown matrices are uniformly well-conditioned ($\kappa \simeq 7$, rank 4) with zero exclusions. 
Consequently, the~paired sample size is $N=399$ for $\beta_{\rm RS2}$ and $A_\Upsilon$, $N=398$ for $a_{\rm sc}$, and~$N=400$ for QNM parameters. 
Given $N=400$, the~binomial sampling uncertainty on fractions near $50\%$ is $\simeq 2.5\%$. 
Thus, configuration preferences reported in \cref{tab:R_summary} are only meaningful beyond this threshold and should be interpreted as population-level sensitivities rather than event-level~predictions.

The numerical pipeline passed multiple internal consistency checks: antenna responses derived from the detector tensor match explicit geometric formulations to machine precision ($\sim$$10^{-16}$), and~analytic derivatives agree with finite-difference approximations to within a relative error of $1.1 \times 10^{-5}$ across all GR parameters. Grid-convergence tests (varying the frequency grid between 200, 400, and~1000 points) yield median ratio shifts below $0.03\%$ across all channels. Confidence bounds on medians are evaluated via 2000 bootstrap resamplings (\cref{subsec:channel_by_channel}).

\begin{table}[H]
\centering

\caption{Monte Carlo Fisher comparison for waveform-level benchmark parameters. Upper block: nominal comparison, $2L$ network with $15\,\mathrm{km}$ arms versus triangular network with $10\,\mathrm{km}$ arms. Lower block: equal-arm control, $2L$ with $15\,\mathrm{km}$ arms versus triangle with $15\,\mathrm{km}$ arms. The~statistics use $399$ events for $\beta_{\rm RS2}$ and $A_\Upsilon$, $398$ for $a_{\rm sc}$ and $400$ for the QNM parameters (\cref{subsec:mc_results}). Left: ratio $R_A$ at the natural network SNR. Right: ratio $R_A^{\rm fixed}$ at fixed network SNR, \cref{eq:R_fixed_snr}. $R_A<1$ favors the $2L$ configuration and $R_A>1$ favors the triangular configuration. The~continuum parameter $\beta_{\rm RS2}$ and the tidal-charge phase amplitude $A_\Upsilon$ are phenomenological
benchmarks. For~the QNM parameters ($^\dagger$), $R_\omega=R_\tau$ to numerical precision; the fixed-SNR ratio is constant over the events ($1.0006$ for the nominal pair and $1$ for the equal-arm pair), so a fraction of events below unity is not defined.}
\label{tab:R_summary}

  \begin{tabular}{lcccccc}

\toprule
 & \multicolumn{3}{c}{\textbf{Natural SNR}} & \multicolumn{3}{c}{\textbf{Fixed Network SNR}} \\
\cmidrule{2-7}
\textbf{\hl{Channel} 
} & \boldmath{$\langle R_A\rangle$} & \boldmath{$\widetilde R_A$} & \textbf{frac. \boldmath{$<1$}} & \boldmath{$\langle R_A^{\rm fixed}\rangle$} & \boldmath{$\widetilde R_A^{\rm fixed}$} & \textbf{frac.} \boldmath{$<1$} \\
\midrule
\multicolumn{7}{l}{\emph{\hl{Nominal}
: $2L$ ($15\,\mathrm{km}$) vs. $\triangle$ ($10\,\mathrm{km}$)}} \\
$\beta_{\rm RS2}$ (continuum-inspired) & $0.810$ & $0.805$ & $99.5\%$ & $1.063$ & $1.069$ & $2.8\%$ \\
$A_\Upsilon$ (tidal-charge ppE-type) & $0.747$ & $0.751$ & $98.7\%$ & $0.980$ & $0.997$ & $62.9\%$ \\
$\delta\hat\omega,\ \delta\hat\tau$ (QNM) $^\dagger$ & $0.782$ & $0.780$ & $99.5\%$ & $1.0006$ & $1.0006$ & -- \\
$a_{\rm sc}$ (scalar breathing) & $0.709$ & $0.652$ & $81.9\%$ & $0.928$ & $0.849$ & $61.3\%$ \\
\midrule
\multicolumn{7}{l}{\emph{\hl{Equal-arm control:} $2L$ ($15\,\mathrm{km}$) vs. $\triangle$ ($15\,\mathrm{km}$)}} \\
$\beta_{\rm RS2}$ (continuum-inspired) & $1.047$ & $1.040$ & $26.3\%$ & $0.993$ & $0.998$ & $73.9\%$ \\
$A_\Upsilon$ (tidal-charge ppE-type) & $1.032$ & $1.037$ & $33.3\%$ & $0.979$ & $0.994$ & $73.4\%$ \\
$\delta\hat\omega,\ \delta\hat\tau$ (QNM) $^\dagger$ & $1.054$ & $1.050$ & $20.0\%$ & $1.000$ & $1.000$ & -- \\
$a_{\rm sc}$ (scalar breathing) & $1.032$ & $0.953$ & $52.8\%$ & $0.977$ & $0.896$ & $56.8\%$ \\
\bottomrule
\end{tabular}
\end{table}

\subsection{Channel-by-Channel Fisher~Sensitivity}
\label{subsec:channel_by_channel}
 
The results for the accessible channels are summarized in \cref{tab:R_summary} and in \cref{fig:R_beta_RS2,fig:R_omega_QNM,fig:R_Upsilon,fig:R_Upsilon_z,fig:R_radion_sc}. For~each channel, the upper panel of the figure quantifies the overall relative performance, combining network sensitivity, spectral shape and degeneracy breaking, while the lower panel removes the overall SNR scale (\cref{eq:R_fixed_snr}). The~results are illustrative, model-dependent sensitivity comparisons for the proposed~geometries.

\subsubsection{Continuum-Inspired Phase~Deformation}
 
\begin{figure}[htbp]

\includegraphics[width=1\textwidth]{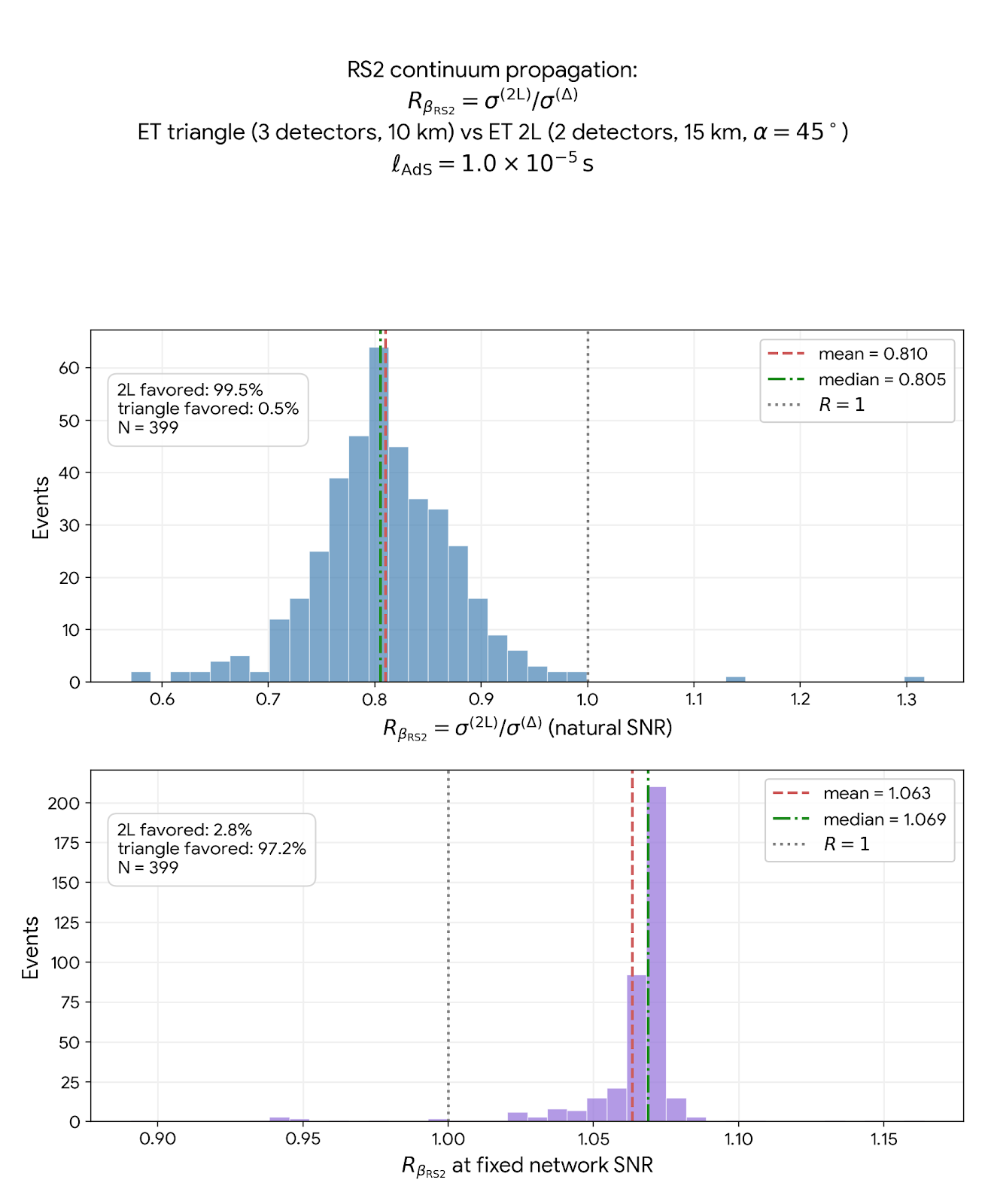}
\caption{\hl{Distribution} of $R_{\beta_{\rm RS2}}=\sigma_{\beta_{\rm RS2}}^{(2L)}/\sigma_{\beta_{\rm RS2}}^{(\triangle)}$ over $N=399$ common source realizations (one event with a rank-deficient Fisher matrix is excluded), comparing the nominal $2L$ configuration with $15\,\mathrm{km}$ arms to the nominal triangular configuration with $10\,\mathrm{km}$ arms, for~the RS2 continuum-propagation phase kernel $\delta\Psi=\beta_{\rm RS2}(2\pi f\ell_{\rm AdS})^2 \ln(2\pi f\ell_{\rm AdS})$ with $\ell_{\rm AdS}=10^{-5}\,\mathrm{s}$, corresponding to a light-crossing time of about $3\,\mathrm{km}$. \emph{Top:} ratio at natural network SNR (mean $0.810$, median $0.805$, $2L$ favored in $99.5\%$ of events). \emph{Bottom:} ratio at fixed network SNR, \cref{eq:R_fixed_snr} (mean $1.063$, median $1.069$, $2L$ favored in $2.8\%$ of events); the shift from unity reflects the different spectral shapes of the $10\,\mathrm{km}$ and $15\,\mathrm{km}$ noise curves (the equal-arm median is $0.998$). The~Fisher uncertainties are marginalized over ten waveform parameters,
with the real ECEF detector geometry, public ET sensitivity curves, and~Earth-rotating stationary-phase response.}
\label{fig:R_beta_RS2}
\end{figure}

The parameter $\beta_{\rm RS2}$ is a phenomenological coefficient describing a prescribed frequency-dependent deformation of the GW signal. The~forecast quantifies the network sensitivity to this deformation and does not constrain the AdS curvature radius. For~the nominal pair, \cref{fig:R_beta_RS2} shows that the $2L$ configuration yields the tighter constraint in $99.5\%$ of the events at natural SNR, with~median ratio $0.805$ ($95\%$ bootstrap interval $0.800$--$0.812$). This reflects mainly the higher network SNR of the $2L$ with $15\,\mathrm{km}$ arms (median $89.9$ against $67.5$ for the $10\,\mathrm{km}$ triangle). At~fixed SNR, the ratio becomes $1.069$, with~the triangle favored in $97.2\%$ of the events. This shift is not geometric: for the equal-arm pair, the fixed-SNR median is $0.998$ and the natural-SNR median $1.040$, the~latter corresponding to the slightly higher SNR of the $15\,\mathrm{km}$ triangle ($93.4$). The~fixed-SNR ratio of the nominal pair is therefore due to the different spectral shapes of the $10\,\mathrm{km}$ and $15\,\mathrm{km}$ ASD, whose ratio varies between about $1.25$ and $1.43$ between $20$ and $500\,\mathrm{Hz}$ and weights differently the frequency range probed by the kernel. At~equal arm length, neither geometry is preferred beyond $\simeq4\%$.

\vspace{-3pt}
\unskip
 
\subsubsection{Black-Hole Ringdown and Quasinormal~Modes}
 
\begin{figure}[htbp]
\centering
\includegraphics[width=1\textwidth]{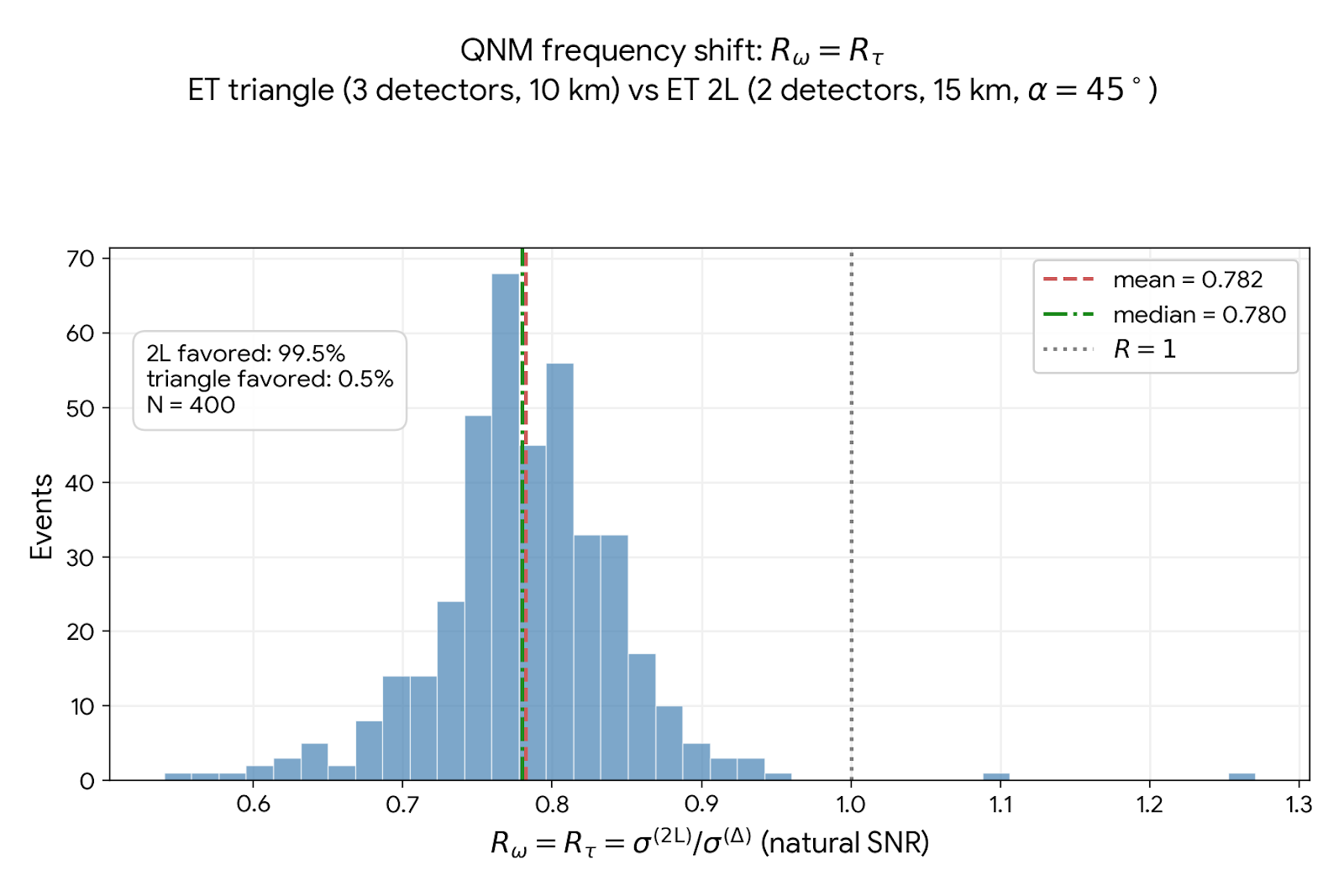}
\caption{Distribution of
$R_\omega=\sigma_{\delta\hat\omega}^{(2L)}/\sigma_{\delta\hat\omega}^{(\triangle)}$, identical to that of $R_\tau=\sigma_{\delta\hat\tau}^{(2L)}/\sigma_{\delta\hat\tau}^{(\triangle)}$, for~$N=400$ sky-location and orientation realizations of a fixed
$(2,2,0)$ ringdown benchmark with $f_{220}=251\,\mathrm{Hz}$ and $\tau_{220}=4\,\mathrm{ms}$,
comparing the nominal $2L$ configuration with $15\,\mathrm{km}$ arms to the nominal triangular configuration with $10\,\mathrm{km}$ arms. \emph{Top:} ratio at natural network SNR (mean $0.782$, median $0.780$,
$2L$ favored in $99.5\%$ of events). The~ratio at fixed network SNR, \cref{eq:R_fixed_snr}, takes the same value, $1.0006$, for~all the events. The~two ratios coincide because $\sigma_{\delta\hat\tau}\propto\sigma_{\delta\hat\omega}$ with a constant factor in this benchmark, which is the same for the two networks and cancels in the $2L/\triangle$ ratio. The~uncertainties are marginalized over the amplitude and the phase of the mode (four-parameter Fisher matrix); sky position, inclination and polarization are not marginalized. The~detector response is static because Earth rotation is negligible over the signal duration.}
\label{fig:R_omega_QNM}
\end{figure}

The ringdown modification enters through $\delta\hat\omega$ and $\delta\hat\tau$. For~the nominal pair, \cref{fig:R_omega_QNM} shows that the $2L$ configuration yields tighter constraints in $99.5\%$ of the events at natural SNR, with~median ratio $0.780$ ($95\%$ interval $0.774$--$0.789$). This is the product of two effects: the equal-arm ratio, $\widetilde R^{\rm (eq)}=1.050$, with~the triangle favored in $80\%$ of the events and close to the sky-averaged antenna-power ratio $\sqrt{9/8}\simeq1.06$, as~expected for a short, static ringdown signal whose uncertainty scales as the inverse network SNR, and the~lower 
noise of the $15\,\mathrm{km}$ curve at $251\,\mathrm{Hz}$, whose ASD is smaller than the $10\,\mathrm{km}$ one by a factor $\simeq1.35$ ($1.05/1.35\simeq0.78$). The~triangular preference at equal arm length therefore reflects only its larger network antenna power, and~the $2L$ preference in the nominal comparison only the better noise~curve.

At fixed network SNR, the two networks are equivalent; that is, the ratio is $1$ for the equal-arm pair and $1.0006$, for~every event, for~the nominal pair, the~residual being due to the different spectral shapes of the two ASDs across the width of the line. The~benchmark thus carries no geometric information. The~equality $R_\omega=R_\tau$ (verified to $7\times10^{-16}$ for any pair) follows from the structure of the single-pole model. Since $\partial_{\ln f}\tilde H=i\,2\pi f_{220}\tau_{220}\,\partial_{\ln\tau}\tilde H$ and the nuisance space spanned by the complex amplitude is invariant under multiplication by $i$, the~two parameters are uncorrelated after marginalization and, for~any noise curve, $\sigma_{\delta\hat\tau}=2\pi f_{220}\tau_{220}\,\sigma_{\delta\hat\omega}\simeq6.3\,\sigma_{\delta\hat\omega}$.

\unskip
 
\subsubsection{Tidal-Charge Parameter and Phase Amplitude $A_\Upsilon$}
\label{subsec:upsilon_mc}
 
\begin{figure}[htbp]

\includegraphics[width=1\textwidth]{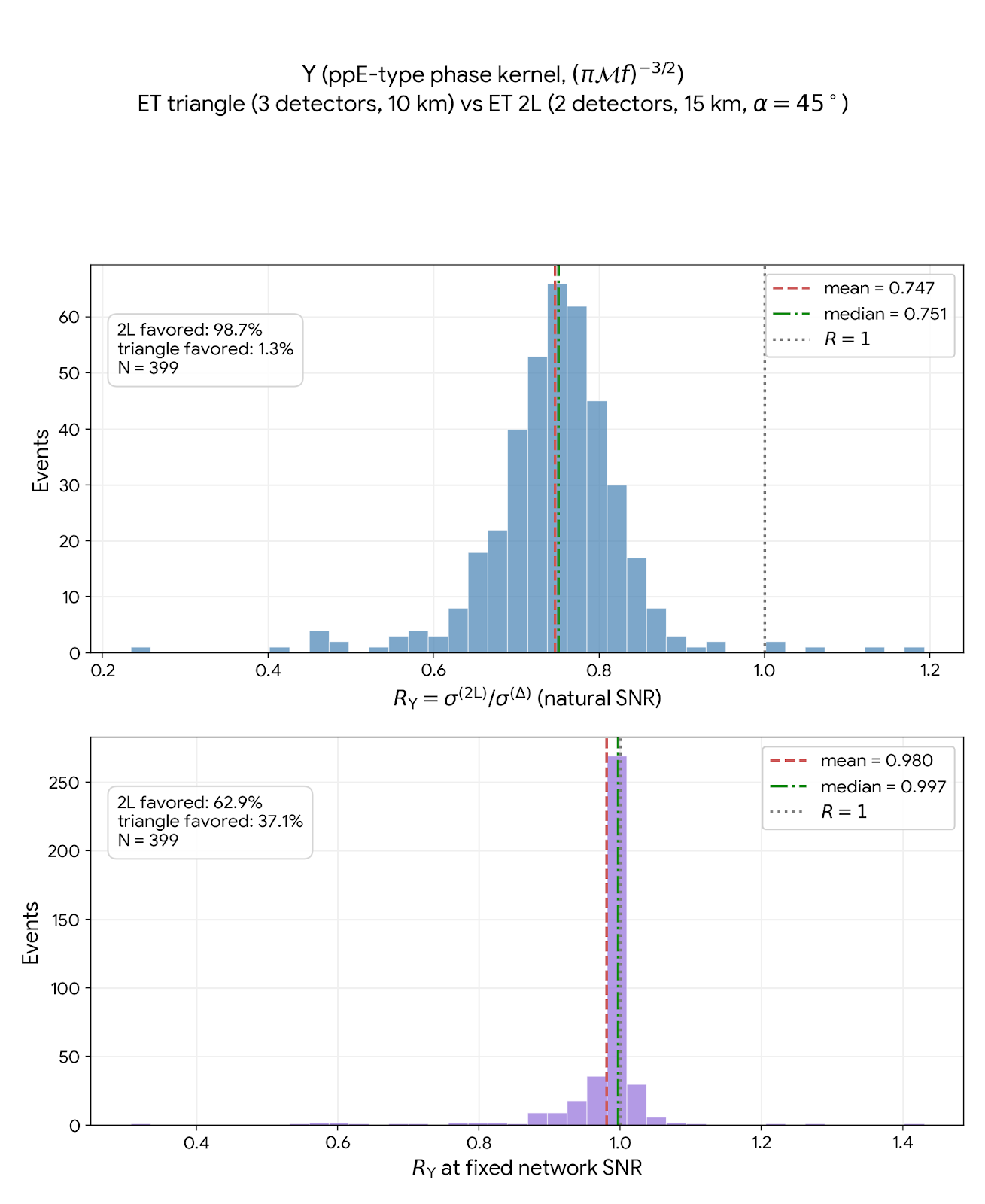}
\caption{Distribution of
$R_{A_\Upsilon}=\sigma_{A_\Upsilon}^{(2L)}/\sigma_{A_\Upsilon}^{(\triangle)}$ for the tidal-charge ppE-type phase amplitude $A_\Upsilon$, treated as a free phenomenological amplitude ($N=399$ events), comparing the nominal $2L$ configuration with $15\,\mathrm{km}$ arms to the nominal triangular
configuration with $10\,\mathrm{km}$ arms. \emph{\hl{Top:}
} ratio at natural network SNR (mean $0.747$, median $0.751$, $2L$ favored in $98.7\%$ of events). \emph{\hl{Bottom}:} ratio at fixed network SNR, \cref{eq:R_fixed_snr} (mean $0.980$, median $0.997$, $2L$ favored in $62.9\%$ of events). The~uncertainties are marginalized over ten waveform parameters, with~the real ECEF detector geometry, public ET sensitivity
curves, and~Earth-rotating stationary-phase response. The~fixed-SNR result is consistent with equivalent networks, the~residual being at the level of the different spectral shapes of the two noise curves.}
\label{fig:R_Upsilon}
\end{figure}

\begin{figure}[htbp]

\includegraphics[width=1\textwidth]{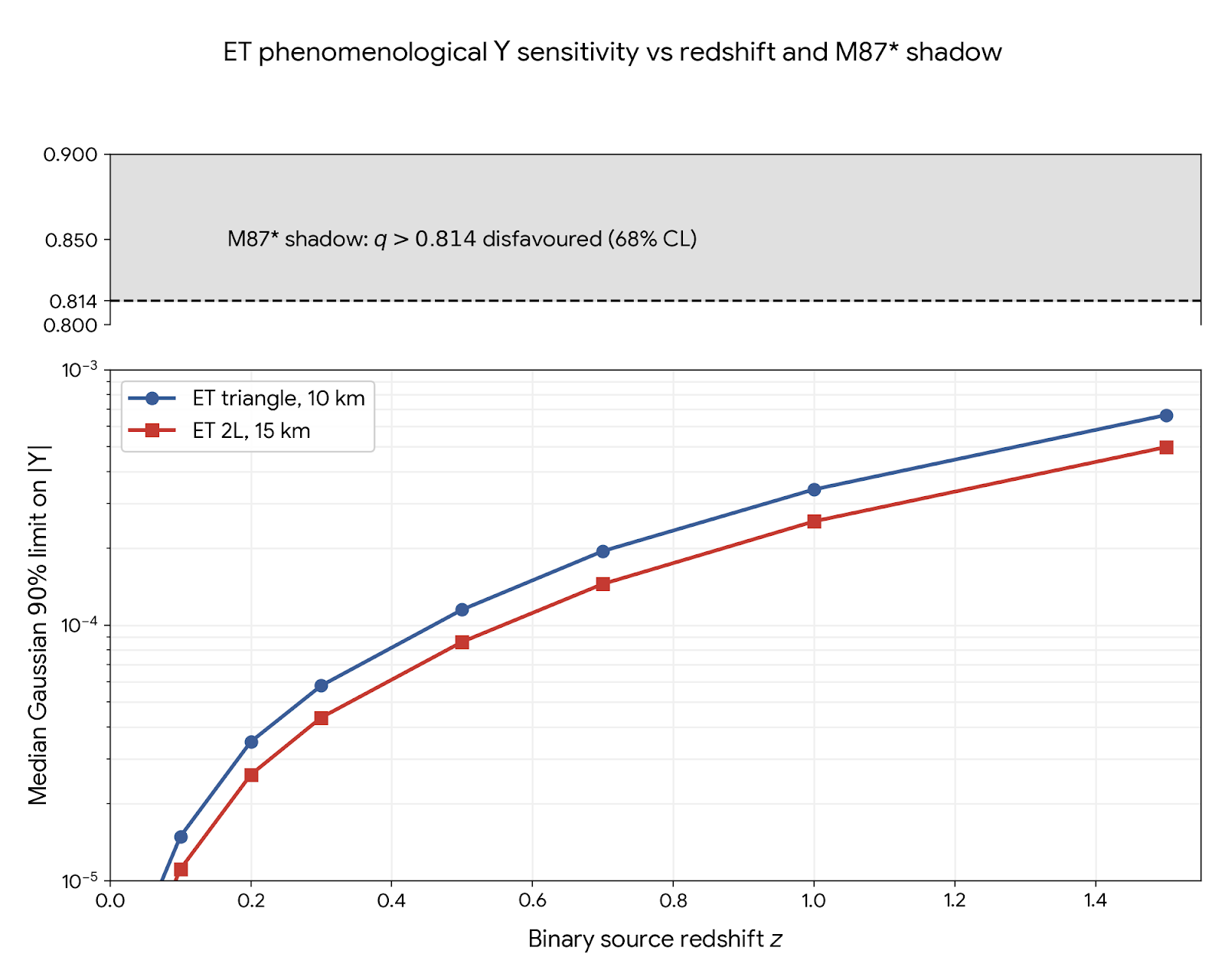}
\caption{\hl{Median} Gaussian $90\%$ upper limit on the dimensionless tidal-charge parameter $|\Upsilon|$, obtained from the inspiral phase deformation under the benchmark identification $A_\Upsilon\equiv\Upsilon$, as~a function of the binary source redshift $z$, for~the nominal triangular configuration with $10\,\mathrm{km}$ arms (blue circles) and the nominal $2L$ configuration with $15\,\mathrm{km}$ arms (red squares). At~each redshift, the same ensemble of $N=400$ binary realizations and sky orientations is reused, with~the observed masses and luminosity distances updated consistently. The~limits are medians over the simulated events at each redshift, marginalized over the GR parameters, with the real ECEF geometry, public ET sensitivity curves and Earth-rotating stationary-phase response. The~grey band in the upper panel marks the region $q>0.814$ disfavored at $68\%$ CL by the M87* shadow for the dimensionless tidal-charge parameter $q$ \cite{Sahoo,Banerjee}. Because~M87* is a fixed source with an already-performed measurement, the~band is horizontal. The~redshift on the horizontal axis refers to the ET binaries, not to M87*. The~vertical axis is broken to display simultaneously the ET limits and the shadow threshold. The~comparison is conditional on the benchmark identification $A_\Upsilon\equiv\Upsilon$, which is a modeling hypothesis rather than a relation derived from the tidal-charge metric. The~two limits are also quoted at different confidence levels ($68\%$ versus $90\%$). No detection threshold is applied: at high redshift, these are
formal Fisher forecasts, not validated likelihood-based limits.}
\label{fig:R_Upsilon_z}
\end{figure}

In the Monte Carlo, the~inspiral channel is the ppE-type phase amplitude $A_\Upsilon$ of \cref{eq:upsilon_phen_waveform}, which multiplies the kernel $K_\Upsilon$ of \cref{eq:upsilon_phen_kernel},
\begin{equation}
\delta\Psi_{A_\Upsilon}(f)=A_\Upsilon K_\Upsilon(f)=A_\Upsilon\left(\frac{\pi G\mathcal M_zf}{c^3}\right)^{-3/2}\,,
\end{equation}
where $\mathcal M_z=(1+z)\mathcal M_c$ is the detector-frame chirp mass, for~binary-neutron-star-like sources, with~the real ECEF detector geometry, the~Earth-rotating stationary-phase response and the public ET sensitivity curves. In~terms of the velocity variable $v$ of the GR phase, the~kernel scales as $v^{-9/2}$. The~kernel is not derived from the tidal-charge BH of \cref{sub:qnm_background}, whose effect acts on the ringdown through $\delta\hat\omega=\mathcal C_\omega\Upsilon$ and $\delta\hat\tau=\mathcal C_\tau\Upsilon$
(\cref{eq:RW_tidal_charge_numbers}). \Cref{fig:R_Upsilon}
therefore measures the network response to a generic phase
deformation of this type. \Cref{fig:R_Upsilon_z} expresses the same response as an absolute limit, and~reads it as a limit on $\Upsilon$ only through the identification $A_\Upsilon\equiv\Upsilon$ adopted below, which is a modeling~choice.
 
For a fixed circular polarization, the strain in detector $k$ is
\begin{equation}
\tilde h_k\propto(F_k^++iF_k^\times)A(f)e^{i\Psi(f)}
e^{2\pi if\tau_k}\,.
\end{equation}

The Fisher information on a parameter entering only through the common phase $\Psi(f)$ is proportional to the network antenna power
\begin{equation}
P(\theta,\phi,\psi)=\sum_k\left[(F_k^+)^2+(F_k^\times)^2\right]\,.
\end{equation}

Averaged over sky position and polarization angle, the~triangle gives $\langle P\rangle_\triangle=9/10$, while the $2L$ configuration gives $\langle P\rangle_{2L}=4/5$. If~the deformation were exactly orthogonal to all marginalized parameters and the noise curves were equal, the~natural-SNR ratio would follow the per-event network-SNR ratio, whose sky average is
\begin{equation}
\sqrt{\langle P\rangle_\triangle/\langle P\rangle_{2L}} =\sqrt{9/8} \simeq1.06\,,
\end{equation}
as found for the ringdown channels at equal arm~length.
 
In the full ten-parameter analysis, however, the~kernel correlates with the coalescence time, chirp mass and symmetric mass ratio, and~Earth rotation makes the antenna response frequency-dependent, so the marginalized ratio is not fixed by the antenna powers alone. For~the nominal pair, we find $\widetilde R_{A_\Upsilon}=0.751$ at natural SNR ($95\%$ interval $0.744$--$0.760$), with~$98.7\%$ of events favoring the $2L$ configuration, and~$\widetilde R_{A_\Upsilon}^{\rm fixed}=0.997$ at fixed SNR ($0.996$--$0.998$), with~$62.9\%$ favoring the $2L$ configuration (\cref{fig:R_Upsilon}). The~$2L$ advantage at natural SNR is therefore entirely accounted for by its higher network SNR, and~the geometric contribution at fixed SNR is below $1\%$. This is confirmed by the equal-arm control, for~which the medians are $\widetilde R^{\rm (eq)}_{A_\Upsilon}=1.037$ at natural SNR (the triangle with $15\,\mathrm{km}$ arms has a $4\%$ higher SNR than the $2L$) and $\widetilde R^{\rm (eq),fixed}_{A_\Upsilon}=0.994$ at fixed SNR. Therefore, when both networks have the same arm length, their sensitivities to $A_\Upsilon$ are substantially~equivalent.

\Cref{fig:R_Upsilon_z} shows the absolute sensitivity to the dimensionless tidal-charge parameter $\Upsilon$ as a function of the source redshift, for~
$z\in\{0.05,\,0.1,\,0.2,\,0.3,\,0.5,\,0.7,\,1.0,\,1.5\}$.
At each redshift, the same ensemble of $N=400$ binary realizations and sky orientations is reused, with~the observed (redshifted) masses and luminosity distances updated~consistently.
 
The inspiral phase deformation is modeled as
\begin{equation}
\Delta\Psi(f)=A_\Upsilon
\left(\pi\frac{G\mathcal M_z}{c^3}f\right)^{-3/2}\,.
\label{eq:A_Upsilon_kernel}
\end{equation}

In this forecast, the ppE phase amplitude is identified with the dimensionless tidal-charge parameter, $A_\Upsilon\equiv\Upsilon$ (denoted $q$ in Ref.~\cite{Sahoo}), adopting the convention $\kappa=1$ in the relation $A_\Upsilon=\kappa\Upsilon$. This identification defines the normalization of $\Upsilon$ in the inspiral channel and is a modeling hypothesis, not a relation derived from the
tidal-charge metric. Under~this convention, the~Fisher constraint on the phase amplitude is read as a constraint on $\Upsilon$, with~$\sigma_\Upsilon\equiv\sigma_{A_\Upsilon}$, and~the plotted quantity is the median Gaussian $90\%$ upper limit on $|\Upsilon|$,
\begin{equation}
|\Upsilon|^{90\%}=1.64485\,\sigma_\Upsilon\,,
\end{equation}
marginalized over the GR parameters, i.e.,\ the half-width of the symmetric $90\%$ interval. The~factor $1.64485$ is the two-sided $90\%$ quantile of a standard normal distribution, equivalently its one-sided $95\%$ quantile.

The limit degrades monotonically with $z$, from~
$\simeq(1$--$2)\times10^{-5}$ at $z=0.1$ (we find $1.11\times10^{-5}$ for the $2L$ and $1.49\times10^{-5}$ for the triangle) to $\simeq(5$--$7)\times10^{-4}$ at $z=1.5$. This reflects primarily the decrease in the network SNR with luminosity distance. In~addition, the~redshifting of the chirp mass entering the kernel introduces an extra degradation scaling as $(1+z)^{3/2}$ at fixed observed frequency. The~$2L$ configuration yields a limit lower than that of the triangle by a nearly constant factor, $|\Upsilon|^{90\%}_{2L}/|\Upsilon|^{90\%}_{\triangle}\simeq0.75$
over the whole range. This factor coincides with the population-median ratio $\widetilde R_{A_\Upsilon}=0.751$ of \cref{tab:R_summary} (at $z=0.1$, the ratio of the median limits is $0.747$), and~has the same origin. Since $\sigma_{A_\Upsilon}\propto\rho^{-1}$ at fixed waveform shape and the SNR ratio of the two networks does not depend on the redshift, it is the SNR advantage of the $15\,\mathrm{km}$ noise curve and not a geometric effect. For~triangle and $2L$ with equal arm length, the two limits would differ by a few percent (\cref{tab:R_summary}).

\textls[-15]{The ppE-type inspiral result constrains the phase amplitude $A_\Upsilon$ entering \mbox{\cref{eq:upsilon_phen_waveform}}.} Under~the benchmark identification $A_\Upsilon\equiv\Upsilon$ ($\kappa=1$), the~inspiral limit shown in \cref{fig:R_Upsilon_z} is read as a limit on $\Upsilon$, with~$\sigma_\Upsilon\equiv\sigma_{A_\Upsilon}$. More generally, if~a linear mapping $A_\Upsilon=\kappa\Upsilon$ with $\kappa\neq1$ is adopted, the~local uncertainty on the geometric parameter translates as
\begin{equation}
\sigma_\Upsilon^{\rm insp}=\frac{\sigma_{A_\Upsilon}}{|\kappa|}\,.
\end{equation}

The identification $\kappa=1$ adopted here is a modeling
choice, and~the numerical value of the limit depends on the
normalization and on the frequency index of the kernel
$K_\Upsilon$. The~inspiral forecast should therefore be read as a
constraint on the phase deformation under this identification,
while the geometric parameter is accessed independently,
through the background and perturbation prescription of
\cref{sub:qnm_background}, by~the ringdown~channel.

The sensitivity to the geometric tidal-charge parameter is also carried by the ringdown. Since $\Upsilon$ shifts frequency and damping time simultaneously, for~a single parameter with $\delta\hat\omega=\mathcal C_\omega\Upsilon$ and $\delta\hat\tau=\mathcal C_\tau\Upsilon$, the Fisher information is
\begin{equation}
\Gamma_{\Upsilon\Upsilon}=\mathcal C_\omega^2\Gamma_{\hat\omega\hat\omega}
+2\mathcal C_\omega\mathcal C_\tau\Gamma_{\hat\omega\hat\tau}
+\mathcal C_\tau^2\Gamma_{\hat\tau\hat\tau}\,,
\end{equation}
where the $\Gamma$ are the marginalized Fisher blocks of the ringdown parameters. In~the present benchmark, $\Gamma_{\hat\omega\hat\tau}=0$ exactly (\cref{subsec:channel_by_channel}), so
\begin{equation}
\sigma_\Upsilon^{-2}=\left(\frac{\mathcal C_\omega}{\sigma_{\delta\hat\omega}}\right)^2+\left(\frac{\mathcal C_\tau}{\sigma_{\delta\hat\tau}}\right)^2\,.
\end{equation}

With $|\mathcal C_\tau|\simeq\mathcal C_\omega/3$ and $\sigma_{\delta\hat\tau}\simeq6.3\,\sigma_{\delta\hat\omega}$, the~damping-time term contributes only $\simeq$$0.3\%$ of the information, and~the frequency channel dominates. Since $R_\omega=R_\tau$, the~relative performance for $\Upsilon$ through the ringdown follows the common QNM ratio, $\widetilde R=0.780$ at natural SNR and $R^{\rm fixed}=1.0006$ for the nominal pair ($1.050$ and $1$ at equal arm length), i.e.,\ it is determined by the noise curve and not by the geometry, and~it differs from the ppE-type inspiral value only through the spectral shape of the~noise.

The shaded region in \cref{fig:R_Upsilon_z} is an external, model-dependent reference interval for the dimensionless tidal-charge parameter, obtained from shadow-based analyses~\cite{Sahoo,Banerjee}. It is not an ET forecast, and~it has no redshift dependence merely because it is displayed alongside a redshift-dependent inspiral forecast. The~M87* constraint refers to one fixed supermassive BH, and~its applicability to stellar-mass binary remnants requires the additional assumption that the same effective tidal-charge parametrization and mapping apply across very different source masses and~environments.

The band is therefore an order-of-magnitude reference. It does not constitute a joint constraint, a~Bayesian prior, or~a prediction that the same numerical interval applies to the binary remnant. The~comparison is illustrative; it relies on the modeling identification $A_\Upsilon\equiv\Upsilon$, on~the normalization and frequency index of the kernel $K_\Upsilon$, on~the transfer of the parametrization between very different source masses, and~on limits quoted at different confidence levels ($68\%$ for the shadow, $90\%$ for ET). 
 
\subsubsection{Scalar~Polarization}
 
\begin{figure}[htbp]

\includegraphics[width=0.9\textwidth]{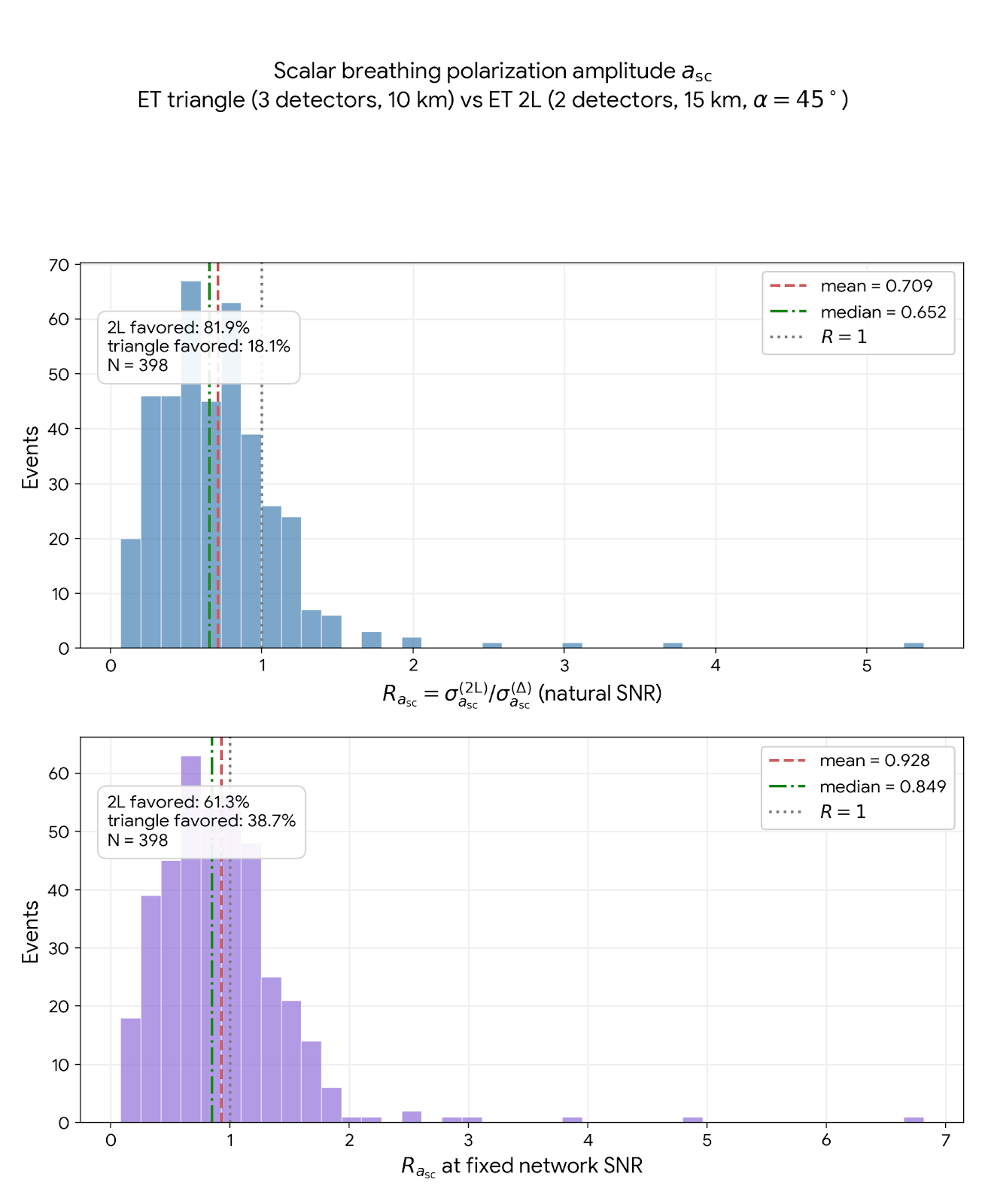}
\caption{\hl{Distribution}
 of $R_{a_{\rm sc}}=\sigma_{a_{\rm sc}}^{(2L)}/
\sigma_{a_{\rm sc}}^{(\triangle)}$ over $N=398$ common source realizations of NS-containing binaries with a scalar breathing component (two nearly face-on events with a rank-deficient Fisher matrix are excluded), comparing the nominal $2L$ configuration with $15\,\mathrm{km}$ arms to the nominal triangular configuration with $10\,\mathrm{km}$ arms.
\emph{\hl{Top}:} ratio at natural network SNR (mean $0.709$, median $0.652$, $2L$ favored in $81.9\%$ of events). \emph{\hl{Bottom}:} ratio at fixed network SNR, \cref{eq:R_fixed_snr} (mean $0.928$, median $0.849$,
$2L$ favored in $61.3\%$ of events). The~uncertainties are marginalized over ten waveform parameters, with~the real ECEF detector geometry, public ET sensitivity curves, and~Earth-rotating stationary-phase response.}
\label{fig:R_radion_sc}
\end{figure}

For binaries containing non-BH compact objects, such as BNS or NSBH systems, a~light scalar degree of freedom introduces a transverse breathing polarization component $a_{\rm sc}$. In~the Monte Carlo, $a_{\rm sc}$ multiplies a normalized breathing-mode template added to the tensor inspiral waveform,
\begin{equation}
\delta\tilde h_I=a_{\rm sc}\,\mathcal A(f)\,F_I^{b}(f)\,\sin^2\iota\;e^{i\Psi_{\rm GR}(f)}\,e^{2\pi if\tau_I(f)}\,,
\end{equation}
with $F_I^b=D_I:(\hat p\hat p+\hat q\hat q)$ the breathing response of detector $I$, the~same amplitude and phase as the tensor signal and a unit coefficient. Therefore, $a_{\rm sc}$ is the ratio of the breathing amplitude, with~this angular dependence, to~the Newtonian tensor amplitude. The~amplitude is real, and~its Fisher derivative is the projected breathing response of the network, including Earth rotation and arrival-time delays. For~the nominal pair, \cref{fig:R_radion_sc} shows that the $2L$ configuration yields the tighter marginalized constraint in $81.9\%$ of the events at natural SNR, with~$\widetilde R=0.652$ ($95\%$ interval $0.600$--$0.705$), and~in $61.3\%$ at fixed SNR, with~$\widetilde R^{\rm fixed}=0.849$ ($0.782$--$0.913$). The~distributions are broad, and~a $15\%$ median advantage at fixed SNR corresponds to a preference for the $2L$ in only $61\%$ of the events. For~equal arm length, the medians are $0.953$ at natural SNR and $0.896$ at fixed SNR, with~the $2L$ favored in $52.8\%$ and $56.8\%$ of the events. The~scalar amplitude is thus the only channel where the equal-arm comparison retains a preference larger than $5\%$.
 
A set of controlled tests on $40$ events randomly selected from the ensemble (nominal pair) isolates the origin of the fixed-SNR preference. With~the GR parameters held fixed, the~median ratio at fixed SNR is $1.03$ and the $2L$ is favored in only $42.5\%$ of the events, i.e.,\ no preference. With~the GR parameters marginalized, the median is $0.75$, with~the $2L$ favored in $72.5\%$ of the events (the corresponding median over the full ensemble is $0.849$, the~difference being compatible with the sampling uncertainty of a $40$-event subset), and~for a complex scalar amplitude (amplitude and phase free), it is $0.42$, with~the $2L$ favored in $90\%$ of the events. With~a static detector response and marginalized GR parameters, the~Fisher matrix is rank-deficient for all the $40$ events and both networks. The~residual advantage of the $2L$ in the scalar channel thus arises from a different degeneracy between the breathing response and the GR extrinsic parameters, which is broken only by the rotation of the Earth. We note that, for the triangle, the three responses to any polarization, the~breathing one included, sum to zero (a null stream), so the breathing response lies in the same two-dimensional space as those of the tensor~polarizations.

\subsection{Limitations}
\label{subsec:summary_tradeoffs}

The results hold within the following assumptions of the simulations.
The public $10\,\mathrm{km}$ and $15\,\mathrm{km}$ ET sensitivity curves are used for the triangular and $2L$ configurations, respectively, in~the version distributed with \texttt{\hl{gwfast}
}.
Antenna patterns are computed in the long-wavelength limit, the~noise is assumed independent across detectors with the same ASD for the three interferometers of the triangle, and~neither site-specific noise nor duty cycle is included.
The $2L$ orientation is a fixed input and is not optimized.
Each channel is analyzed separately, with~one deformation parameter at a time.
The Fisher formalism assumes a linearized signal and a high SNR (the median network SNR is $68$--$93$ at $z=0.1$, and~lower at the larger redshifts of \cref{fig:R_Upsilon_z}), and~does not provide calibrated confidence intervals near physical boundaries.
The population results use a single redshift, $z=0.1$, and~a single representative mass distribution, with~a restricted nonspinning 1.5PN waveform, without~ISCO cutoff and without tidal effects.
The ringdown is a single-mode benchmark with fixed amplitude, in~which the extrinsic parameters are not marginalized.
Finally, the~pipeline has not been validated against the standard-science results of Ref.~\cite{Branchesi}, and~no scan has been performed in the numerical settings listed in \cref{subsec:mc_results}.

\section{Conclusions}
\label{sec:conclusions}

We studied the GW phenomenology of extra-dimensional compactifications (toroidal, Calabi--Yau, and~Randall--Sundrum RS1 and RS2) in two steps: a feasibility screening connecting microscopic parameters to waveform observables, and~a Fisher-matrix study of the channels that remain potentially accessible. The~status of each channel (derived from microscopic parameters, parametric scaling, or~phenomenological benchmark) is summarized in \cref{tab:channel_map}.

The screening shows that the direct effects of heavy KK tensor towers are strongly suppressed. For~the toroidal, Calabi--Yau, and~RS1 cases, the waveform corrections scale as $(\hbar\omega/m_1c^2)^2$$\sim$$10^{-23}$--$10^{-22}$, about twenty orders of magnitude below the ET threshold sensitivity, and~direct propagation of KK modes in the detector band is excluded by laboratory bounds. For~the minimal RS2 model, the~conservative inspiral-phase correction of \cref{sub:rs2-theory} stays below the ET measurability threshold throughout the region allowed by laboratory bounds: at the largest allowed radius, $\ell=50\,\mu{\rm m}$, it gives $|\delta\Psi^{\rm cons}_{\rm RS2}|\lesssim2\times10^{-14}\,{\rm rad}$ for an equal-mass binary neutron star of $2.8\,M_\odot$ (the maximum, reached at $f_{\min}=3\,\mathrm{Hz}$), and~the detector-independent Fisher bound of \cref{subsec:rs2_fisher_microscopic} confirms that it is unmeasurable even at a network SNR of $10^3$.

The accessible channels are therefore phenomenological benchmarks, not predictions fixed by the compactification parameters: a continuum-inspired phase kernel ($\beta_{\rm RS2}$), a~tidal-charge inspiral kernel ($A_\Upsilon$), QNM frequency and damping-time shifts controlled by the effective tidal-charge parameter $\Upsilon$, and~a scalar breathing polarization ($a_{\rm sc}$). In~particular, $\Upsilon$ remains a phenomenological parameter: it is not directly derived from a specific compactification model or from a unique microscopic relation, but~it can nevertheless be treated and tested experimentally as a free parameter in GW searches. In~the inspiral channel, $\Upsilon$ is accessed only through the benchmark identification $A_\Upsilon\equiv\Upsilon$, so the Fisher uncertainties quantify the detectability of the assumed deformations, not constraints on compactification~parameters.

The Fisher analysis compares a $2L$ network with $15\,\mathrm{km}$ arms and a triangular network with $10\,\mathrm{km}$ arms (nominal comparison), plus a triangle with $15\,\mathrm{km}$ arms as equal-arm control, on~$N=400$ common source realizations. For~each channel, we report the uncertainty ratio $R_A$ at the natural network SNR and the SNR-normalized ratio $R_A^{\rm fixed}$, which removes the overall SNR scale; only the equal-arm control isolates the~geometry.

In the nominal comparison, the~$2L$ with $15\,\mathrm{km}$ arms outperforms the triangle with $10\,\mathrm{km}$ arms in all channels.
At natural SNR, it is favored in $99.5\%$, $98.7\%$ and $99.5\%$ of the events for the continuum-inspired kernel, the~tidal-charge kernel $A_\Upsilon$ (i.e.,\ the tidal parameter $\Upsilon$) and the QNM shifts, with~median ratios $0.805$, $0.751$ and $0.780$, and~in $81.9\%$ of the events for the scalar amplitude, with~median $0.652$.
This advantage follows from the higher network SNR of the $15\,\mathrm{km}$ noise curve (median $89.9$ against $67.5$), not from the geometry: for the equal-arm pair, the medians at natural SNR are $1.04$, $1.04$ and $1.05$ for the same three channels, slightly favoring the triangle, and~at fixed SNR, they are within $0.6\%$ of unity.
The ringdown benchmark carries no geometric information, and~the $7\%$ fixed-SNR shift of the continuum-inspired kernel in the nominal comparison reflects the different spectral shapes of the two noise curves. Only the scalar breathing amplitude retains, at~equal arm length and fixed SNR, a~broad preference for the $2L$ (median $0.896$, $56.8\%$ of the events), arising from the degeneracy with the GR parameters broken by the Earth's rotation. Overall, the~differences between the configurations are modest (at most $\sim$$35\%$ in the median uncertainty) and should not be read as a ranking of the final ET~designs.

The same advantage holds for the absolute sensitivity to $\Upsilon$ (\cref{fig:R_Upsilon_z}): the median $90\%$ limits grow from $\simeq$$(1$--$1.5)\times10^{-5}$ at $z=0.1$ to $\simeq$$(5$--$7)\times10^{-4}$ at $z=1.5$, and~the $2L$ limit is lower than the triangle one by a nearly constant factor $\simeq0.75$, which is the SNR advantage of the $15\,\mathrm{km}$ curve. These limits depend on the normalization and frequency index of the kernel $K_\Upsilon$, which are not derived from the tidal-charge metric, so they do not establish whether ET reaches the range of $\Upsilon$ of physical interest. The~M87* shadow band in the figure is an external, model-dependent order-of-magnitude reference and not a joint constraint, so the comparison does not indicate whether ET is more or less sensitive than the shadow~measurement.

Although the tidal-charge deformation is expected to affect the ringdown through shifts in the QNM frequency and damping time, the~absolute sensitivity to $\Upsilon$ from the ringdown at a specified network SNR remains to be calculated. The~present analysis quantifies only the relative response of the considered ET configurations and the parametric dependence of the QNM shifts on $\Upsilon$. A~dedicated forecast of the absolute ringdown constraint on $\Upsilon$, for~a specified network SNR and source population, is left for future~work.

\vspace{6pt}
\funding{This research received no external funding.}

\dataavailability{\hl{The} detector sensitivity curves used in this work are publicly available through the \texttt{gwfast} package~\cite{Iacovelli1, Iacovelli2}. The~10~km triangular configuration and the 15~km 2L configuration employ the Einstein Telescope HFLF-cryogenic amplitude spectral densities distributed in \texttt{gwfast} under \path{gwfast/data/psds/ET_designs_comparison_paper/HFLF_cryo/ETLength10km/} and \path{gwfast/data/psds/ET_designs_comparison_paper/HFLF_cryo/ETLength15km/}, respectively. These curves, as~well as the proposed detector geometries, originate from Branchesi~et~al.~\cite{Branchesi}. 
}

\acknowledgments{The author would like to acknowledge networking support by the COST Action~23130.}

\conflictsofinterest{The authors declare no conflicts of interest.}

\begin{adjustwidth}{-\extralength}{0cm}

\PublishersNote{}
\end{adjustwidth}
\end{document}